\input lanlmac
\def\href#1#2{{#2}}
\def\hhref#1{{#1}}
\input epsf.tex

\overfullrule=0mm

\newcount\figno
\figno=0
\def\fig#1#2#3{
\par\begingroup\parindent=0pt\leftskip=1cm\rightskip=1cm\parindent=0pt
\baselineskip=11pt
\global\advance\figno by 1
\midinsert
\epsfxsize=#3
\centerline{\epsfbox{#2}}
\vskip 12pt
{\bf Fig.\ \the\figno:} #1\par
\endinsert\endgroup\par
}
\def\figlabel#1{\xdef#1{\the\figno}}
\def\encadremath#1{\vbox{\hrule\hbox{\vrule\kern8pt\vbox{\kern8pt
\hbox{$\displaystyle #1$}\kern8pt}
\kern8pt\vrule}\hrule}}


\def\IR{\relax{\rm I\kern-.18em R}}
\font\cmss=cmss10 \font\cmsss=cmss10 at 7pt

\def\q#1{\left[#1\right]_x}

\font\cmss=cmss10 \font\cmsss=cmss10 at 7pt
\def\IZ{\relax\ifmmode\mathchoice
{\hbox{\cmss Z\kern-.4em Z}}{\hbox{\cmss Z\kern-.4em Z}}
{\lower.9pt\hbox{\cmsss Z\kern-.4em Z}}
{\lower1.2pt\hbox{\cmsss Z\kern-.4em Z}}\else{\cmss Z\kern-.4em Z}\fi}
\def\IN{\relax{\rm I\kern-.18em N}}
\def\b{\circ}
\def\n{\bullet}

\def\gbbbb{\Gamma_4^{\hbox{$\scriptstyle \b \b$}\kern -8.2pt
\raise -4pt \hbox{$\scriptstyle \b \b$}}}
\def\gnnnn{\Gamma_4^{\hbox{$\scriptstyle \n \n$}\kern -8.2pt  
\raise -4pt \hbox{$\scriptstyle \n \n$}}}
\def\gnnnnnn{\Gamma_6^{\hbox{$\scriptstyle \n \n \n$}\kern -12.3pt
\raise -4pt \hbox{$\scriptstyle \n \n \n$}}}
\def\gbbbbbb{\Gamma_6^{\hbox{$\scriptstyle \b \b \b$}\kern -12.3pt
\raise -4pt \hbox{$\scriptstyle \b \b \b$}}}
\def\gbbbbc{\Gamma_{4\, c}^{\hbox{$\scriptstyle \b \b$}\kern -8.2pt
\raise -4pt \hbox{$\scriptstyle \b \b$}}}
\def\gnnnnc{\Gamma_{4\, c}^{\hbox{$\scriptstyle \n \n$}\kern -8.2pt
\raise -4pt \hbox{$\scriptstyle \n \n$}}}
\def\Rbud#1{{\cal R}_{#1}^{-\kern-1.5pt\blacktriangleright}}
\def\Rleaf#1{{\cal R}_{#1}^{-\kern-1.5pt\vartriangleright}}
\def\Rbudb#1{{\cal R}_{#1}^{\circ\kern-1.5pt-\kern-1.5pt\blacktriangleright}}
\def\Rleafb#1{{\cal R}_{#1}^{\circ\kern-1.5pt-\kern-1.5pt\vartriangleright}}
\def\Rbudn#1{{\cal R}_{#1}^{\bullet\kern-1.5pt-\kern-1.5pt\blacktriangleright}}
\def\Rleafn#1{{\cal R}_{#1}^{\bullet\kern-1.5pt-\kern-1.5pt\vartriangleright}}
\def\Wleaf#1{{\cal W}_{#1}^{-\kern-1.5pt\vartriangleright}}
\def\Cleaf{{\cal C}^{-\kern-1.5pt\vartriangleright}}
\def\Cbud{{\cal C}^{-\kern-1.5pt\blacktriangleright}}
\def\Crleaf{{\cal C}^{-\kern-1.5pt\circledR}}


\magnification=\magstep1
\baselineskip=12pt
\hsize=6.3truein
\vsize=8.7truein
 at 8truept
 at 8truept
 at 10truept

\font\bigrm=cmr12 at 14pt
\centerline{\bigrm Distance statistics in quadrangulations with a boundary,}
\medskip
\centerline{\bigrm or with a self-avoiding loop}

\bigskip\bigskip

\centerline{J. Bouttier and E. Guitter}
  \smallskip
  \centerline{Institut de Physique Th\'eorique}
  \centerline{CEA, IPhT, F-91191 Gif-sur-Yvette, France}
  \centerline{CNRS, URA 2306}
\centerline{\tt jeremie.bouttier@cea.fr}
\centerline{\tt emmanuel.guitter@cea.fr}

  \bigskip

     \bigskip\bigskip

     \centerline{\bf Abstract}
     \smallskip
     {\narrower\noindent
We consider quadrangulations with a boundary and derive explicit expressions 
for the generating functions of these maps with either a marked vertex at a 
prescribed distance from the boundary, or two boundary vertices at a prescribed
mutual distance in the map. For large maps, this yields explicit formulas for 
the bulk-boundary and boundary-boundary correlators in the various encountered 
scaling regimes: a small boundary, a dense boundary and a critical boundary 
regime. The critical boundary regime is characterized by a one-parameter family
of scaling functions interpolating between the Brownian map and the Brownian 
Continuum Random Tree. We discuss the cases of both generic and self-avoiding 
boundaries, which are shown to share the same universal scaling limit. We 
finally address the question of the bulk-loop distance statistics in the 
context of planar quadrangulations equipped with a self-avoiding loop. Here 
again, a new family of scaling functions describing critical loops is 
discovered.
\par}

     \bigskip

\nref\QGRA{V. Kazakov, {\it Bilocal regularization of models of random
surfaces}, Phys. Lett. {\bf B150} (1985) 282-284; F. David, {\it Planar
diagrams, two-dimensional lattice gravity and surface models},
Nucl. Phys. {\bf B257} (1985) 45-58; J. Ambj\o rn, B. Durhuus and J. Fr\"ohlich,
{\it Diseases of triangulated random surface models and possible cures},
Nucl. Phys. {\bf B257} (1985) 433-449; V. Kazakov, I. Kostov and A. Migdal
{\it Critical properties of randomly triangulated planar random surfaces},
Phys. Lett. {\bf B157} (1985) 295-300.}
\nref\TUT{W. Tutte,
{\it A Census of planar triangulations} Canad. J. of Math. {\bf 14} (1962) 21-38;
{\it A Census of Hamiltonian polygons} Canad. J. of Math. {\bf 14} (1962) 402-417;
{\it A Census of slicings}, Canad. J. of Math. {\bf 14} (1962) 708-722;
{\it A Census of Planar Maps}, Canad. J. of Math. {\bf 15} (1963) 249-271.
}
\nref\BIPZ{E. Br\'ezin, C. Itzykson, G. Parisi and J.-B. Zuber, {\it Planar
Diagrams}, Comm. Math. Phys. {\bf 59} (1978) 35-51.}
\nref\DGZ{for a review, see: P. Di Francesco, P. Ginsparg and 
J. Zinn--Justin, {\it 2D Gravity and Random Matrices},
Physics Reports {\bf 254} (1995) 1-131.}
\nref\Pol{A. M. Polyakov, {\it Quantum geometry of bosonic strings},
Phys. Lett. B {\bf 103}, 207-210 (1981); {\it Quantum geometry of fermionic 
strings}, Phys. Lett. B {\bf 103}, 211-213 (1981).}
\nref\AW{J. Ambj\o rn and Y. Watabiki, {\it Scaling in quantum gravity},
Nucl.Phys. {\bf B445} (1995) 129-144, arXiv:hep-th/9501049.}
\nref\AJW{J. Ambj\o rn, J. Jurkiewicz and Y. Watabiki,
{\it On the fractal structure of two-dimensional quantum gravity},
Nucl.Phys. {\bf B454} (1995) 313-342, arXiv:hep-lat/9507014.}
\nref\ADJ{J. Ambj\o rn, B. Durhuus and T. Jonsson, {\it Quantum Geometry:
A statistical field theory approach}, Cambridge University Press, 1997.}
\nref\MS{M. Marcus and G. Schaeffer, {\it Une bijection simple pour les
cartes orientables} (2001), available at 
\hhref{http://www.lix.polytechnique.fr/Labo/Gilles.Schaeffer/Biblio/};
see also G. Schaeffer, {\it Conjugaison d'arbres
et cartes combinatoires al\'eatoires}, PhD Thesis, Universit\'e 
Bordeaux I (1998) and G. Chapuy, M. Marcus and G. Schaeffer, 
{\it A bijection for rooted maps on orientable surfaces}, 
arXiv:0712.3649 [math.CO].}
\nref\MOB{J. Bouttier, P. Di Francesco and E. Guitter. {\it 
Planar maps as labeled mobiles},
Elec. Jour. of Combinatorics {\bf 11} (2004) R69, arXiv:math.CO/0405099.}
\nref\FOMAP{J. Bouttier, P. Di Francesco and E. Guitter. {\it Blocked edges 
on Eulerian maps and mobiles: Application to spanning trees, hard particles 
and the Ising model}, 	J. Phys. A: Math. Theor. {\bf 40} (2007) 7411-7440, 
arXiv:math.CO/0702097.}
\nref\GEOD{J. Bouttier, P. Di Francesco and E. Guitter, {\it Geodesic
distance in planar graphs}, Nucl. Phys. {\bf B663}[FS] (2003) 535-567, 
arXiv:cond-mat/0303272.}
\nref\CS{P. Chassaing and G. Schaeffer, {\it Random Planar Lattices and 
Integrated SuperBrownian Excursion}, 
Probability Theory and Related Fields {\bf 128(2)} (2004) 161-212, 
arXiv:math.CO/0205226.}
\nref\MW{G. Miermont and M. Weill, {\it Radius and profile of random planar
maps with faces of arbitrary degrees}, Electron. J. Probab. 
{\bf 13} (2008) 79-106, arXiv:0706.3334 [math.PR].}
\nref\MARMO{J. F. Marckert and A. Mokkadem, {\it Limit of normalized
quadrangulations: the Brownian map}, Annals of Probability {\bf 34(6)}
(2006) 2144-2202, arXiv:math.PR/0403398.}
\nref\LEGALL{J. F. Le Gall, {\it The topological structure of scaling limits 
of large planar maps}, invent. math. {\bf 169} (2007) 621-670,
arXiv:math.PR/0607567.}
\nref\LGP{J. F. Le Gall and F. Paulin,
{\it Scaling limits of bipartite planar maps are homeomorphic to the 2-sphere},
Geomet. Funct. Anal. {\bf 18}, 893-918 (2008), arXiv:math.PR/0612315.}
\nref\MierS{G. Miermont, {\it On the sphericity of scaling limits of 
random planar quadrangulations}, Elect. Comm. Probab. {\bf 13} (2008) 248-257, 
arXiv:0712.3687 [math.PR].}
\nref\STATGEOD{J. Bouttier and E. Guitter, {\it Statistics of
geodesics in large quadrangulations}, J. Phys. A: Math. Theor. {\bf 41} 
(2008) 145001 (30pp), arXiv:0712.2160 [math-ph].}
\nref\Mier{G. Miermont, {\it Tessellations of random maps of arbitrary
genus}, Ann. Sci. \'Ec. Norm. Sup\'er., to appear, arXiv:0712.3688 [math.PR]}
\nref\LEGALLGEOD{J.-F. Le Gall, {\it Geodesics in large planar maps and 
in the Brownian map}, Acta Math., to appear, arXiv:0804.3012 [math.PR].}
\nref\THREEPOINT{J. Bouttier and E. Guitter, {\it The three-point function 
of planar quadrangulations}, J. Stat. Mech. (2008) P07020, 
arXiv:0805.2355 [math-ph].}
\nref\LOOP{J. Bouttier and E. Guitter, {\it Confluence of geodesic paths and 
separating loops in large planar quadrangulations}, 
J. Stat. Mech. (2009) P03001, arXiv:0811.0509 [math-ph].}
\nref\PDF{P. Di Francesco, {\it 2D Quantum Gravity, Matrix Models and Graph 
Combinatorics}, Lecture notes given at the summer school ``Applications of 
random matrices to physics'', Les Houches, June 2004, arXiv:math-ph/0406013.}
\nref\ALDOUS{D. Aldous, {\it Tree-Based Models for Random Distribution of Mass},
J. Stat. Phys. {\bf 73} 625-641 (1993).}
\nref\KO{I. Kostov, {\it O(n) vector model on a planar random lattice: 
spectrum of anomalous dimensions}, Mod. Phys. Lett. A  {\bf 4}  217-226 (1988).}
\nref\GJ{I. P. Goulden and D. M. Jackson, {\it Combinatorial enumeration},
John Wiley, New York, 1983.}
\nref\DFG{P. Di Francesco and E. Guitter, {\it Integrability of graph 
combinatorics via random walks and heaps of dimers}, 
J. Stat. Mech. (2005) P09001, arXiv:math/0506542 [math.CO].}

\newsec{Introduction}
Understanding the properties of random maps is a fundamental question for 
both the mathematical and the physical community. In mathematics, maps raise 
at a discrete level beautiful and rather involved combinatorial problems 
while, at a continuous level, they give rise to new probabilistic objects, 
like the Brownian map, whose construction is still under investigation. 
In physics, maps are used as discretizations for fluctuating surfaces in 
various domains ranging from low energy physics, for instance in the context 
of fluid membrane statistics, to high energy physics in the fields of string 
theory or of two-dimensional quantum gravity \QGRA. 

The first incursion into these problems dealt mainly with global properties 
of random maps. At the combinatorial level, this amounted to a precise 
enumeration of various families of maps by several methods developed by 
mathematicians or physicists. These include in particular the original approach
through recursive decomposition, developed by Tutte in a series of papers 
\TUT, and the approach through random matrix integrals, which provide a 
systematic and powerful machinery for enumeration [\xref\BIPZ,\xref\DGZ]. 
With this latter technique, the study was extended to maps with possible extra 
statistical degrees of freedom such as spins or particles (see \DGZ\ for a 
review). At a continuous level, many results such as exponents characterizing
global properties of the maps were obtained heuristically via the so-called 
Liouville model \Pol.  Beside global properties, one then addressed the more 
refined question of the actual dependence of correlations on the 
{\it distance} along the map. A first expression was obtained in Refs.~\AW\ 
and \AJW\ (see also \ADJ) for the so-called {\it two-point function}, which 
gives the law for the distance between two points on the map and more generally
for the ``loop-loop propagator'', measuring the distance between two 
boundary loops at the extremities of a cylindrical map. These results
were obtained in the context of triangulations via Tutte's recursive 
decomposition approach, but at the price of heuristic arguments which, although
non-rigorous at the discrete level, led eventually to the correct continuous
correlators. Apart from this result, little remained known for quite a
while on the statistics of distances in maps or, equivalently, on the metric 
structure of the Brownian map, probably because neither the recursive 
decomposition nor the matrix integral approach, nor even the Liouville model,
are well-suited to address questions on the distance. 

Fortunately, a completely new enumeration technique was then discovered, where
the distance plays a central role. It uses bijections to code the maps
by much simpler objects such as the so-called {\it well-labeled} trees, where
the labels precisely retain some of the distances in the map. This bijective
approach was initiated by Schaeffer for quadrangulations (maps with 
tetravalent faces only) \MS, and later 
extended to maps with arbitrary prescribed 
face valences \MOB, Eulerian maps and maps with particles or spins \FOMAP. 
As far as distance statistics is concerned, a first application dealt 
again with the two-point function, giving the law
for the distance between two vertices on the map. This law was derived exactly 
at the discrete level in Ref.~\GEOD\ in the case of quadrangulations of fixed 
area, and was shown to converge to a continuous universal scaling function, 
giving a 
rigorous proof of the expression of Refs.~\AW\ and \AJW. A related quantity, the
radius, was discussed in Refs.~[\xref\CS,\xref\MW]. The universal continuous
two-point function is an intrinsic characteristic of the Brownian map, 
a more fundamental object
toward which many families of random planar maps (falling in the universality 
class of the so-called 2D pure gravity, such as maps with arbitrary bounded 
face degrees or maps coupled to non-critical statistical models) are expected
to converge in the scaling limit where the area of the maps is large 
and scales as the fourth power of the distance [\xref\MARMO,\xref\LEGALL].
Transposing the Schaeffer bijection or its extensions at the continuous level 
allowed to construct this Brownian map as a random metric space, 
which was shown in particular to be homeomorphic to the two-dimensional sphere 
[\xref\LGP,\xref\MierS]. Other properties of the Brownian map could be derived 
by first understanding their discrete counterparts and then taking a continuous 
scaling limit. For instance, the statistics of geodesics (i.e.\ paths of 
shortest length) between two points was considered in Ref.~\STATGEOD\ and it 
was shown that for typical points, all geodesics coalesce into a unique 
macroscopic geodesic path in the scaling limit [\xref\Mier,\xref\LEGALLGEOD]. 
The so-called {\it three-point function}, which measures the joint probability 
distribution for the pairwise distances between three uniformly chosen random 
vertices, was computed exactly in Ref.~\THREEPOINT\ in the case of 
quadrangulations of fixed area, and its universal scaling limit was analyzed.
A remarkable property of confluence was discovered by Le Gall \LEGALLGEOD,
stating that the three geodesic paths joining three typical points on
the Brownian map merge by pairs before reaching their endpoints. 
Again, a complete characterization of the geometry of a geodesic triangle 
could be obtained from an exact solution at the discrete level \LOOP.

So far, all the above rigorous results on the distance statistics obtained
via the bijective approach dealt with closed planar maps, i.e with the 
topology of the sphere. In this paper, we extend these results to the more 
general context of maps {\it with a boundary}. More precisely, we are 
interested in the
{\it bulk-boundary correlator}, which gives the law for the distance to
the boundary of a vertex drawn at random in the bulk of the map,
and in the {\it boundary-boundary correlator}, measuring the distance
between two vertices on the boundary. Note that, in some sense, the 
bulk-boundary correlator lies half-way between the two-point function of
planar maps and the loop-loop propagator of cylindrical maps. Our main 
results are explicit
exact expressions for these correlators, already at a discrete level,
in the particular case of quadrangulations. For quadrangulations of large area
$n$, several regimes are obtained according to whether the boundary is small 
(with a length which remains finite), dense (with a length of order $n$)
or critical (with a length of order $n^{1/2}$). We derive from our
discrete results explicit 
expressions for the scaling limit of the bulk-boundary and boundary-boundary
correlators in all these regimes. Most, but not all, of these scaling limits 
are universal and we recover in particular some results of Refs.~\AW\ and \AJW,
here via a rigorous discrete enumeration. Maps with a critical boundary are
characterized by a one-parameter family of universal scaling functions, 
corresponding to a new probabilistic limiting object interpolating between 
the Brownian map and the Brownian Continuum Random Tree.
We first derive our expressions in the context of quadrangulations with 
{\it generic boundaries}, i.e boundaries which may have ``pinch-points''
separating the map into several components. Our results for the bulk-boundary 
correlator are then extended, both at the discrete level and in the various 
continuous regimes, to the case of {\it self-avoiding boundaries} where 
pinch-points on the boundary are forbidden. For large maps, 
generic and self-avoiding boundaries lead to the same scaling regimes.
We finally address the problem of planar quadrangulations equipped with a 
self-avoiding loop. We give exact discrete and continuous expressions for the 
{\it bulk-loop correlator}, which is the law for the distance to the loop of 
a vertex drawn at random in the map. Here again, for large maps, a regime of 
critical self-avoiding loop is found, described by a new one-parameter
family of scaling functions.

The paper is organized as follows: Section 2 presents our main results,
which are then proved in Sections 3 to 6.  We give in Sect.~2.1 
a precise definition of quadrangulations with a boundary and 
present in Sect.~2.2 a number of explicit discrete formulas for various 
generating functions encoding the bulk-boundary and boundary-boundary 
correlators. The case of both generic and self-avoiding boundaries are
discussed. We then turn in Sect.~2.3 to the statistics of distances in
large maps, for which several scaling regimes are found, depending on
whether the boundary is small, dense, or critical. We give in 
particular explicit expressions for various scaling functions characterizing 
these regimes. We finally discuss in Sect.~2.4 the bulk-loop correlator for
quadrangulations with a self-avoiding loop, both at the discrete and 
continuous levels. Section 3 gives a precise derivation of our
results for quadrangulations with a generic boundary at the discrete
level. We present in Sect.~3.1 a bijection relating these quadrangulations
to cyclic sequences of well-labeled trees. This property is used in
Sect.~3.2 to derive explicit formulas for associated generating functions.
Physically, these correspond to the bulk-boundary 
and boundary-boundary correlators in a grand canonical ensemble 
with a boundary of fluctuating length, conjugate to a fugacity
parameter $z$. We then deduce in Sect.~3.3 the corresponding 
fixed-length generating functions, corresponding to an ensemble 
of quadrangulations with a boundary of fixed length. Section 4
is devoted to the scaling limit of large maps with a generic boundary.
We analyze in Sect.~4.1 the singularities of our discrete generating 
functions, which control the large map properties. Three different scaling
regimes are found: a small boundary regime for $z<1/8$, discussed in details
in Sect.~4.2, a dense boundary regime for $z>1/8$, discussed in details
in Sect.~4.3 and finally a ``critical'' regime for $z\sim 1/8$. This latter 
regime is best analyzed in Sect.~4.4 in the fixed length ensemble 
by considering large quadrangulations whose boundary has a length 
proportional to the square-root of their area. This gives rise to
our one-parameter family of scaling functions.
Section 5 deals with quadrangulations with a self-avoiding boundary. 
We first show in Sect.~5.1 how to obtain an explicit formula
for the corresponding bulk-boundary correlator from the expression of its 
generic boundary counterpart. This property is 
used in Sect.~5.2 to analyze the large map scaling limit, which is
shown to present essentially the same three scaling regimes as above.
Finally, Section 6 is devoted to quadrangulations with a self-avoiding
loop. We show how to construct such quadrangulations by concatenating
two quadrangulations with self-avoiding boundaries of the same length. 
This property is used to derive explicit expressions for the discrete
and continuous bulk-loop correlators. In particular, a new family of 
scaling functions is found, describing maps with a critical self-avoiding
loop. We end this paper by a few concluding remarks gathered in Section 7.

\newsec{Main results}

This section presents a panorama of our results, which are listed
without derivation. We will then explain in the next sections how
those can be obtained.

\subsec{Quadrangulations with a boundary: definitions}

\fig{An example (a) of quadrangulation with a boundary of area $n=17$ and 
perimeter $2p=34$, which is the length of its contour (dashed green line). 
All faces are tetravalent, except for the external one which has degree $2p$.
Upon splitting the boundary at its separating vertices, we obtain (b) 
irreducible components (here six) which are either single edges or 
quadrangulations with a self-avoiding boundary.}{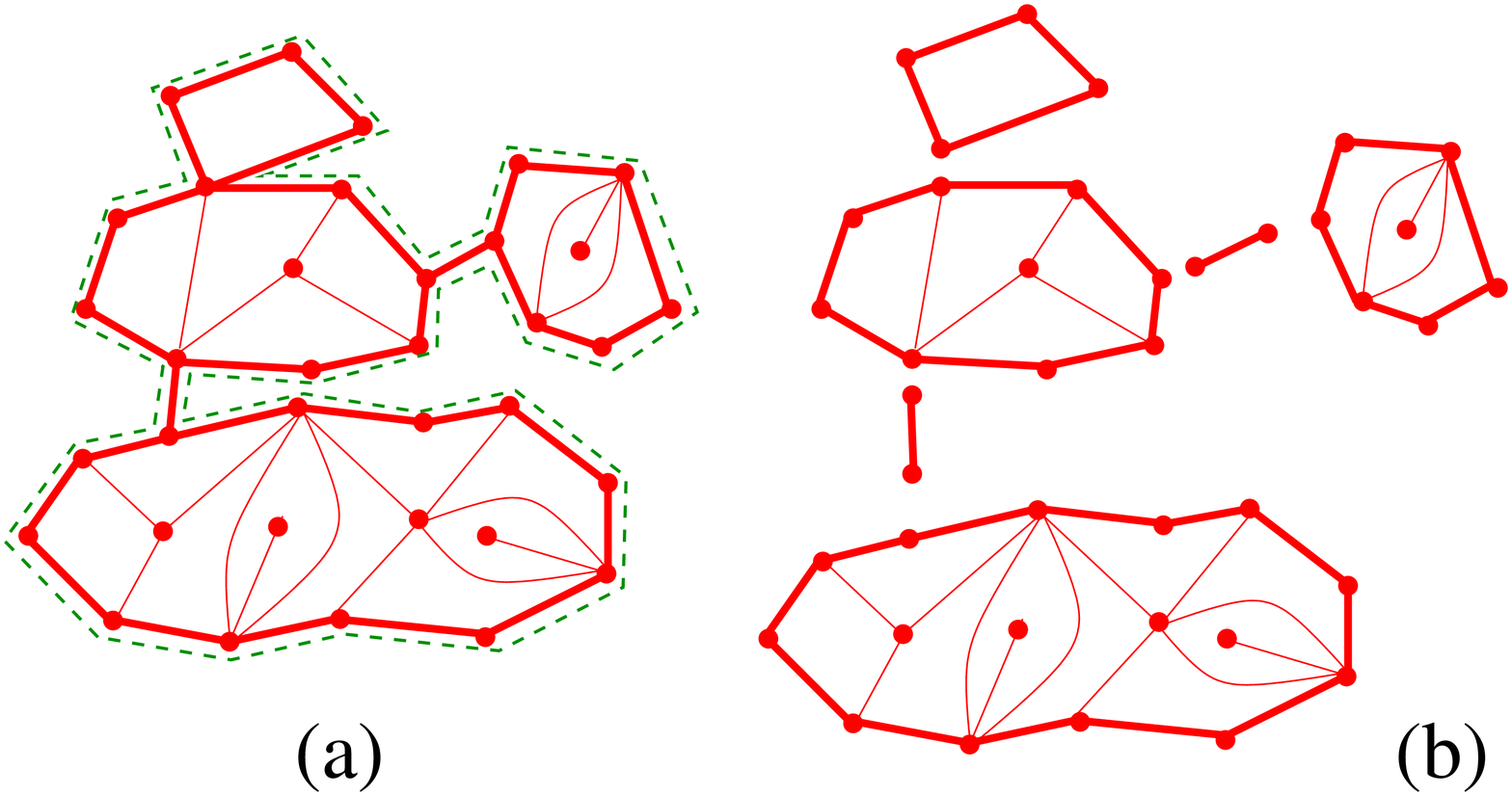}{12.cm}
\figlabel\quadwithb

Here and throughout the paper, we define a {\it quadrangulation with
a boundary} as a planar map with a distinguished face such that all the 
other faces of the map have degree $4$. Such maps are sometimes called
``pseudo-quadrangulations'' in the literature. We use the convention of 
representing the map in the plane with the point at infinity in the 
distinguished face (see Fig.~\quadwithb), which we call 
the {\it external face} accordingly. Note that the degree of the external 
face is necessarily even. 

We call the {\it boundary} of the quadrangulation the set of edges and
vertices incident to the external face. The actual sequence of edges
followed when going around the external face counterclockwise in the
plane is called the {\it contour}. The {\it perimeter} is the (even)
length of this sequence, which is nothing but the degree of the
external face. The {\it bulk} of the quadrangulation is the complement
of the external face in the plane, and the number of {\it inner} faces
it contains is called the {\it area}.

As in the usual terminology, a {\it pointed} map is a map with a
marked vertex, which is referred to as the {\it origin}, while a {\it
rooted} map is a map with a marked oriented edge, which is referred to
as the {\it root edge}. Here, we will make the useful convention that
a {\it rooted quadrangulation with a boundary} has always its root
edge on the boundary, with the external face incident to its right.
We also consider in the following maps which are both pointed and
rooted, or maps with are doubly rooted, in which cases all root edges
are boundary edges with the appropriate orientation.

Note that we consider here general maps which may possibly contain
separating vertices or edges (a vertex or an edge is separating if its
deletion disconnects the map).  In particular, the boundary may
contain such separating vertices or edges, which are those encountered
several times along the contour.  When the boundary has no separating
vertices nor edges, it is said to be {\it self-avoiding}.  Upon
splitting at the separating vertices of the boundary, a
quadrangulation with a {\it generic} boundary is naturally decomposed
into several ``irreducible'' components, that are quadrangulations
with a self-avoiding boundary, or possibly single edges, arranged in a
tree-like structure (see Fig.~\quadwithb). This decomposition will
later allow us to study quadrangulations with a self-avoiding boundary
using results obtained for generic boundaries, which we present first.
\bigskip

\subsec{Discrete results}

We wish to enumerate quadrangulations with a (generic) boundary having
prescribed area and perimeter. In the following, these two quantities
will be usually denoted respectively by $n$ and $2p$. As customary,
the results are best expressed via a {\it generating function},
corresponding to a sum over all quadrangulations with a boundary, a
given quadrangulation with area $n$ and perimeter $2p$ having a
contribution $g^n z^p$. We will thus consider power series in two
variables $g$ and $z$. Again, our results are here stated without
derivation, which can be found in Sections 3 (for generic boundaries)
and 5 (for self-avoiding boundaries).

Arguably the simplest generating function is
that for rooted quadrangulations with a boundary, already computed for instance
in Ref.~\PDF, and which may be written as 
\eqn\rootquadboun{W_0 = W (1 - g R^2 (W-1))}
where $R$ and $W$ are the unique power series satisfying the
algebraic equations
\eqn\RWeqs{\eqalign{R &= 1 + 3 g R^2 \cr W &= 1 + z R W^2.}}
The particular form of this generating function yields, by two
applications of the Lagrange inversion formula, an explicit expression
for the number of rooted quadrangulations with a boundary having area
$n$ and perimeter $2p$:
\eqn\Wzerozpgnpre{W_0\vert_{g^n z^p}= {3^n (2p)! \over p!(p-1)!}
{(2n+p-1)!\over n!(n+p+1)!}\ }
where $\cdot \vert_{g^n z^p}$ denotes the extraction of the coefficient of
$g^n z^p$ in the series.

In this paper, we are interested in refined quantities involving the
graph distance. Our main results are exact expressions for the
bulk-boundary correlator and the boundary-boundary correlator which
are defined as follows.

The {\it bulk-boundary correlator} is the generating function for
pointed quadrangulations with a boundary, where the origin is at a
prescribed distance, say $d$, from the boundary. It reads
\eqn\bulkbouncorr{G_d=\log\left({W_d\over W_{d-1}}\right)}
where
\eqn\Wdexplit{W_d \equiv W\ {1-(W-1) f_{d+1}\over 1-(W-1) f_d}, 
\qquad f_d \equiv x {(1-x^d) \over (1-x^{d+2})}}
and $x$ is the power series satisfying
\eqn\xeq{x = g R^2 (1 + x + x^2).}
A few remarks are in order. The maps considered here might have
non-trivial symmetries, and the generating function $G_d$
includes the corresponding usual inverse symmetry factor. $W_d$ has
itself a combinatorial interpretation, as the generating function for
pointed rooted quadrangulations with a boundary, such that the origin
is at distance {\it smaller than or equal to} $d$ from the boundary,
and such that the root edge starts from a (boundary) vertex closest to
the origin. Such maps have no symmetries. Finally,
the expression \rootquadboun\ is consistent with \Wdexplit\ for $d=0$,
while $W_d \to W$ for $d \to \infty$, so that $W$ is the generating
function for pointed rooted quadrangulations with a boundary.

The {\it boundary-boundary correlator} is the generating function for
doubly-rooted quadrangulations with a boundary, such that the two root
edges start from (boundary) vertices at a distance $d$ from each other
in the map. It reads
\eqn\Tdexplit{T_d=W^2 (W-1)^d f_1 \left({1\over
f_{d+1}}-2(W-1)+f_{d+1}(W-1)^2\right).}
No symmetry factors are involved.

Let us now consider quadrangulations with a self-avoiding boundary. A
classical combinatorial argument shows that the generating function
$\tilde{W_0}$ for rooted quadrangulations with a self-avoiding
boundary is related to $W_0$ via
\eqn\wwtildepre{{\tilde W}_0(g,Z) = W_0(g,z) \quad \hbox{with}
\quad Z=z W_0^2(g,z)}
where we emphasize that $\tilde{W_0}$ is a power series in two
variables $g$ and $Z$, $Z$ being the variable conjugated to the
half-perimeter of the self-avoiding boundary and $g$ still being
conjugated to the area. Algebra yields
\eqn\rootquadsaboun{\tilde{W}_0 = \tilde{W} (1 - g R^2 (\tilde{W}-1))}
where $R$ is still given by \RWeqs\ while $\tilde{W}$ obeys
\eqn\tildeWeq{ 
\tilde{W} = 1 + Z R / (1 -g R^2 (\tilde{W}-1))^2.}
By double Lagrange inversion we find the number of rooted
quadrangulations with a self-avoiding boundary having area $n$ and
perimeter $2p$
\eqn\expzerotildezpgnpre{{\tilde W}_0\vert_{g^n Z^p}=3^{n-p}
{(3p)!\over p!(2p-1)!}{(2n+p-1)!\over (n-p+1)!(n+2p)!}\ .}
Note that for self-avoiding boundaries we have the constraint $1\leq
p\leq n+1$. The bulk-boundary correlator, defined in the same way as
for generic boundaries, reads
\eqn\sabulkbouncorr{\tilde{G}_d=\log\left({\tilde{W}_d-\tilde{W}_0+1\over
\tilde{W}_{d-1}-\tilde{W}_0+1}\right) \quad {\rm for} \quad d \geq 1,
\qquad \tilde{G}_0=\tilde{W}_0-1}
where
\eqn\tildeWdexplit{\tilde{W}_d ={\tilde{W} \over \tilde{W_0}} \times
{1-({\tilde W}-1) f_{d+1}\over 1-({\tilde W}-1) f_d} +{\tilde W}_0-1.}
Again the generating function $\tilde{G}_d$ involves symmetry factors,
unlike $\tilde{W}_d$ which has the same combinatorial interpretation
as $W_d$, now in the context of quadrangulations with a self-avoiding
boundary. We have not been able to find a compact expression for the
boundary-boundary correlator for self-avoiding boundaries.
\bigskip

\subsec{Distance statistics in quadrangulations with a boundary}

From these above exact expressions, we may now derive statistical
information on distances in quadrangulations with a boundary. More
precisely, we consider random quadrangulations with a
boundary having prescribed area $n$ and perimeter $2p$, where each
sample map appears with a probability proportional to its inverse
symmetry factor. We are particularly interested in the large $n$
limit, for which we expect to find asymptotically the same results as
with the uniform measure.  Until further notice we consider generic
(possibly non self-avoiding) boundaries.

It proves convenient to consider first the {\it fixed $z$ ensemble}
where the area remains fixed equal to $n$ but the perimeter
fluctuates, and each sample map with perimeter $2p$ appears with
probability proportional to $z^p$ (besides the symmetry factor). This
model is well-defined (has a finite partition function for all $n$)
for $0 \leq z \leq 1/4$. The above generating functions correspond to
observables related to the distance. The bulk-boundary correlator
encodes the distance between the boundary and a random vertex
uniformly drawn in the bulk. The probability that this {\it
bulk-boundary distance} be $d$ is $( G_d |_{g^n} ) / ( \log W |_{g^n}
)$, while the probability that it be less than or equal to $d$ is
nothing but $( \log W_d |_{g^n} ) / ( \log W |_{g^n} )$.  The
boundary-boundary correlator encodes the distance between (the origins
of) two edges uniformly chosen on the boundary. The probability that
this {\it boundary-boundary distance} be $d$ is $( T_d |_{g^n} ) / ( 2
z {d \over dz} W_0 |_{g^n})$.

For $n \to \infty$, the model exhibits a phase transition at
$z=1/8$, which might be seen simply by analyzing $W_0$:
\item{-} when $z < 1/8$ ({\it subcritical regime}), the perimeter
remains finite as $n \to \infty$,
\item{-} when $z > 1/8$ ({\it supercritical regime}), the perimeter is
of order $n$ and, up to Gaussian fluctuations of order $n^{1/2}$, it
concentrates around its mean value $2 \langle p \rangle_n(z)$ with
\eqn\supcritperim{\langle p \rangle_n(z) \sim n { 8z-1 \over 1-4z }}
\item{-} when $z \sim 1/8$ ({\it critical regime}), the perimeter is of
order $n^{1/2}$.
\par 
\noindent We will be especially interested in the critical regime, and
for this case only we will perform the translation back to the fixed
perimeter ensemble, considering quadrangulations with a boundary
having fixed (large) area $n$ and (large) perimeter $2p$, keeping
the renormalized half-perimeter $P \equiv p \cdot n^{-1/2}$
finite. Let us now discuss the manifestations of the transition on the
distance statistics, as seen by analyzing $G_d$ and $T_d$.
The following table gives a qualitative summary of the asymptotic 
behaviors for the
perimeter, the bulk-boundary distance and the boundary-boundary distance 
in the various regimes.

\def\tvi{\vrule height 10pt depth 6pt width 0pt}
\def\tv{\tvi\vrule}
$$\vbox{\font\bidon=cmr8 \bidon
\offinterlineskip
\halign{\tv\ \hfill # \hfill\ \tv &\ \hfill # \hfill\ \tv
&\ \hfill # \hfill \ \tv &\ \hfill # \hfill \ \tv 
& \ \hfill # \hfill\  \tv \cr 
\noalign{\hrule}
 & perimeter & $\matrix{\hbox{bulk-boundary}\cr \hbox{distance}}$ & 
$\matrix{\hbox{boundary-boundary}\cr \hbox{distance}}$ & $\matrix{
\hbox{universality}\cr \hbox{class}}$ \cr
\noalign{\hrule}
$z<1/8$ & finite & $n^{1/4}$ & finite & Brownian map\cr  
\noalign{\hrule}
$z>1/8$ & $n$ & finite & $n^{1/2}$ & Brownian CRT\cr  
\noalign{\hrule}
$z\sim 1/8$ & $n^{1/2}$ & $n^{1/4}$ & $n^{1/4}$ & $\matrix{\hbox{Brownian map
 with}\cr   
\hbox{a boundary}}$ \cr  
\noalign{\hrule}
}}$$

In the subcritical regime, the bulk-boundary distance is of order
$n^{1/4}$ and admits a continuous limit law which does not depend on $z$
(provided it is in the subcritical range $]0,1/8[$), and whose (cumulative)
distribution function reads
\eqn\subcritbubo{\Phi(D) \equiv \lim_{n
\to \infty} { \left. \log W_{\lfloor D n^{1/4} \rfloor} \right|_{g^n}
\over \left. \log W \right|_{g^n} } = {2\over \sqrt{\pi}}
\int_{-\infty}^\infty d\xi\ {\rm i} \xi\ e^{-\xi^2}{\cal F}(D;-\xi^2)}
where
\eqn\defcalFint{{\cal F}(D;\mu) \equiv \sqrt{\mu}\left(
1+{3\over \sinh^2 \left(\sqrt{{3\over 2}}\, \mu^{1/4} D \right)
}\right)\ .}
We recognize the universal two-point function of pure 2D gravity [\xref\AW,
\xref\AJW] 
also obtained in the case of quadrangulations without a boundary
\GEOD.  This result agrees with the physical intuition: upon
rescaling distances by a factor $n^{-1/4}$, the boundary reduces to a
point which behaves no different from a typical point in a large
random quadrangulation.  Mathematically, the metric space obtained in
the scaling limit is expected to be the Brownian map. In contrast,
the boundary-boundary distance remains finite and admits a
non-universal discrete limit law with a computable, albeit complicated,
expression.

In the supercritical regime, the bulk-boundary distance remains finite
at large $n$ and admits a non-universal discrete limit law, whose
cumulative distribution function reads
\eqn\supcritbubo{\phi_z(d) \equiv \lim_{n
\to \infty} { \left. \log W_d \right|_{g^n}
\over \left. \log W \right|_{g^n} } = 
{(1-(x_{\rm crit}(z))^{d+1})
(1-(x_{\rm crit}(z))^{d+2})\over (1+(x_{\rm crit}(z))^{d+1})
(1+(x_{\rm crit}(z))^{d+2})}}
where
\eqn\expanfourint{x_{\rm crit}(z)\equiv{16 z-1-\sqrt{3((8z)^2-1)} \over
2(1-4z)}\ .}
The boundary-boundary distance is of order $n^{1/2}$ and admits a
continuous limit law. Using the square root of the mean half-perimeter
\supcritperim\ as distance unit, we find the Rayleigh probability
density function
\eqn\supcritbobo{\tilde{\rho}_{\rm bound.}(\delta) \equiv \lim_{n \to
\infty} \left( { \left. T_{\lfloor \delta \cdot \sqrt{\langle p \rangle_n(z)}
\rfloor} \right|_{g^n} \over \left. 2 z {d \over dz} W_0
\right|_{g^n}} \cdot \sqrt{\langle p \rangle_n(z)} \right) = 2 \delta \
e^{-\delta^2}}
which coincides with the two-point function for the Brownian Continuum
Random Tree \ALDOUS. The (Brownian) CRT is expected to be the limiting
metric space obtained when rescaling distances by a factor
$n^{-1/2}$. The physical interpretation is that, in the supercritical
phase, the boundary becomes ``dense'' in the quadrangulation, and
folds onto itself, creating a branched structure.

\fig{Plots of the cumulative distribution function ${\bar \Phi}(D,P)$ 
as a function of $D$, for $P=0.01$, $0.1$, $0.5$, $1.0$, $2.0$ and $5.0$ 
(thin lines from bottom to top). We also indicate in thick red line 
the (integrated) two-point function $\Phi(D)$.}{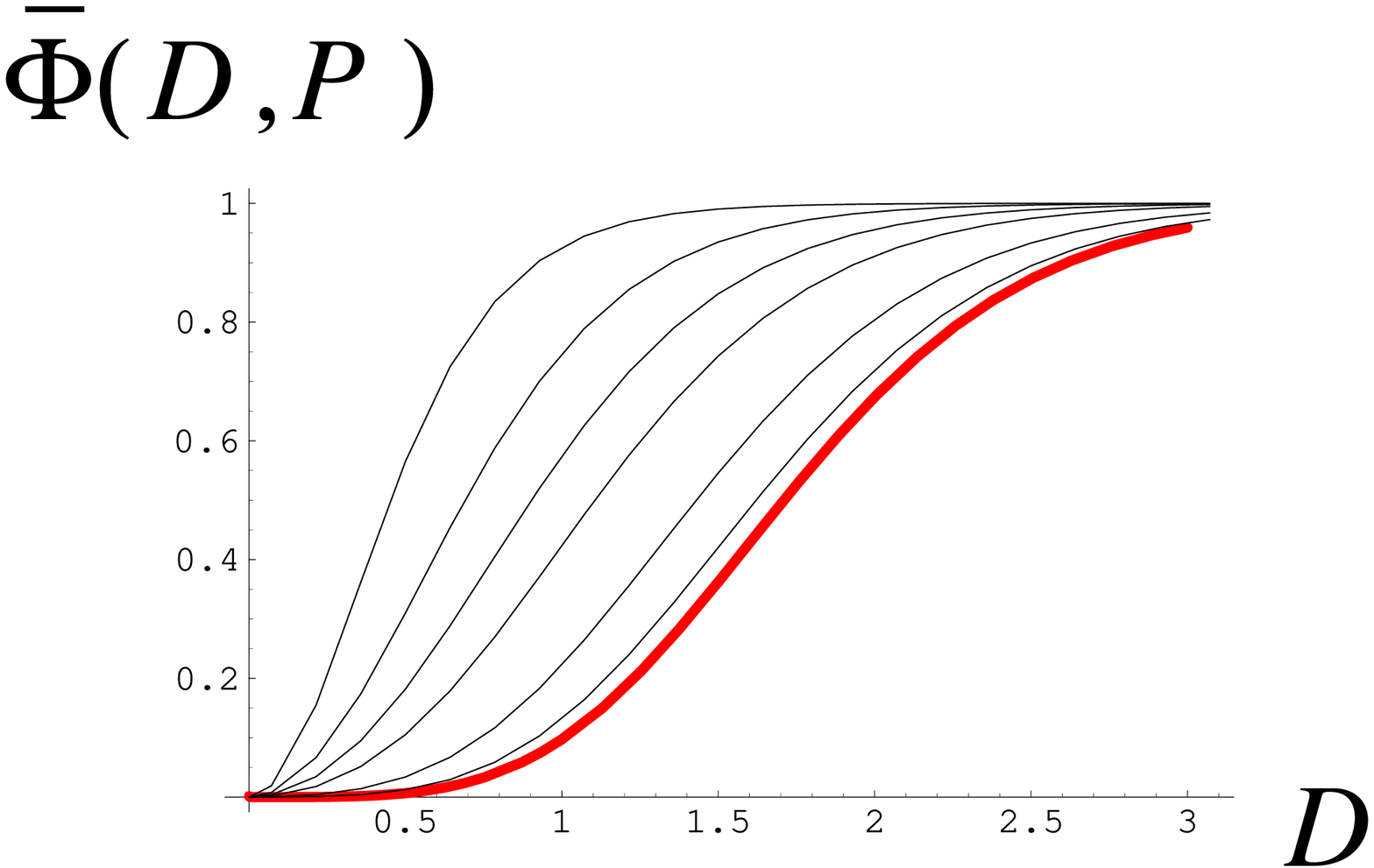}{9.cm}
\figlabel\PhiofP

\fig{Plots of the probability density $\tilde{\rho}_{\rm bound.}(\delta,P)$
as a function of the rescaled distance $\delta=D/\sqrt{P}$ for 
$P=0.5$, $1.0$, $1.5$, $2.0$, $3.0$, $5.0$, $10.0$ (thin lines from
bottom to top). We also indicate the Rayleigh law of Eq.~\supcritbobo (upper
blue thick line) corresponding to the limit $P\to\infty$ and the non-trivial 
law of Eq.~(2.25) (lower red thick line) corresponding to the limit
$P\to 0$.
}{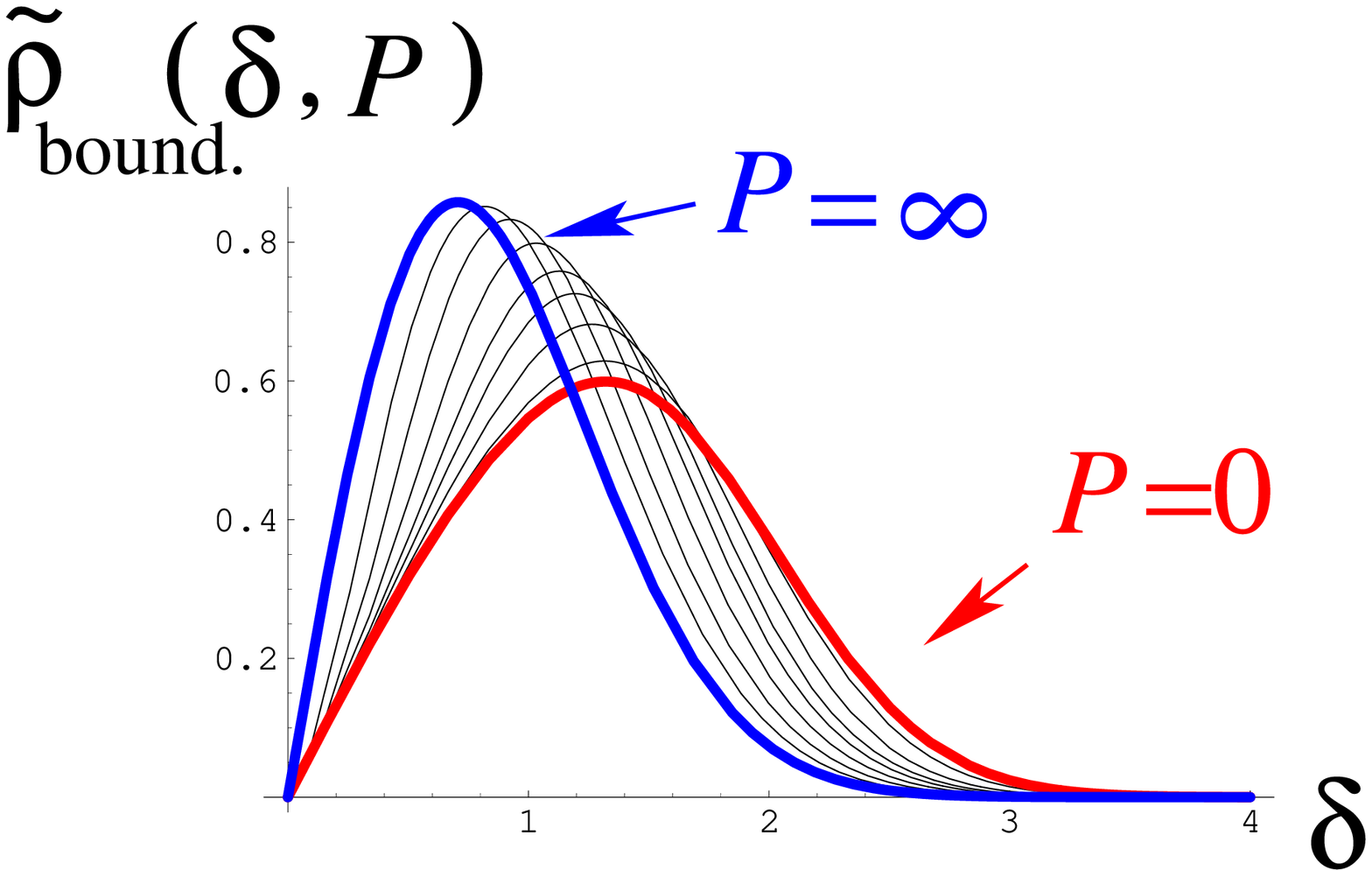}{10.cm}
\figlabel\rhotildeboundP

In the critical regime, both the bulk-boundary and boundary-boundary
distances are of order $n^{1/4}$. Both admit continuous limit laws,
which are best expressed in the (critical) fixed perimeter
ensemble. On the one hand, the bulk-boundary distance cumulative
distribution function reads
\eqn\critbubo{\eqalign{{\bar \Phi}(D,P) & \equiv \lim_{n \to \infty,\
p \sim P n^{1/2}} { \left. \log W_{\lfloor D n^{1/4} \rfloor}
\right|_{g^n z^p} \over \left. \log W \right|_{g^n z^p} } \cr &
=2\sqrt{P}\ e^{P^2/4}\int_{-\infty}^{\infty} d\xi\ {\xi\over {\rm i}}\
e^{-\xi^2}\ {\bar{\cal H}}(D,P;-\xi^2)\cr} }
where
\eqn\barcalHandf{\eqalign{\bar{\cal H}(D,P;\mu) &\equiv
{e^{-\sqrt{\mu} P}\over \sqrt{\pi} P^{3/2}} \left\{1+
\left(3\sqrt{\mu}-2 f^2(D;\mu)\right)\ \int_0^\infty dK\, e^{-{K^2
\over P}-2 f(D;\mu) K} 2K \right\} \cr f(D;\mu) &\equiv \sqrt{{3\over
2}}\, \mu^{1/4}\ \coth \left( \sqrt{{3\over 2}}\, \mu^{1/4} D \right).
\cr}}
It is plotted in Fig~\PhiofP\ as a function of $D$, for $P=0.01$,
$0.1$, $0.5$, $1.0$, $2.0$ and $5.0$. As we will show later, for $P
\to 0$ we recover the two-point function $\Phi(D)$ (shown in red on
Fig~\PhiofP), while for $P \to \infty$ the rescaled bulk-boundary
distance is of order $P^{-1/2}$ and ${\bar \Phi}(D,P)$ takes the
simple scaling form:
\eqn\PhilimlargePpre{{\bar{\Phi}}(D,P) \sim \tanh^2\left({\sqrt{3}
\over 2} D \ \sqrt{P}\right)}
consistent with the supercritical law \supcritbubo\ for $z \to 1/8^+$.
On the other hand, the boundary-boundary distance probability density
function reads
\eqn\critbobo{\eqalign{{\bar \rho}_{\rm bound.}(D,P) & \equiv \lim_{n
\to \infty,\ p \sim P n^{1/2}} \left( { \left. T_{\lfloor D n^{1/4}
\rfloor} \right|_{g^n z^p} \over \left. 2 z {d \over dz} W_0
\right|_{g^n z^p}} \cdot n^{1/4} \right) \cr & = {4\over 3 P^4} e^{-D^2/P}
\left\{(2D^3\!-\!3DP)\!+\!(4D^2P\!-2\!P^2) \sigma_1(D,P)\!+2\!DP^2
\sigma_2(D,P) \right\}\cr}}
where
\eqn\sigmadef{\eqalign{ \sigma_1(D,P) &= {2 e^{P^2/4} \over \sqrt{\pi}
P} \int_{-\infty}^{\infty}d\xi\ {\xi\over {\rm i}}\ e^{-\xi^2+{\rm
i}\, \xi\, P} f(D;-\xi^2) \cr \sigma_2(D,P) &= {2 e^{P^2/4} \over
\sqrt{\pi} P} \int_{-\infty}^{\infty}d\xi\ {\xi\over {\rm i}}\
e^{-\xi^2+{\rm i}\, \xi\, P} f^2(D;-\xi^2) \cr}}
and $f(D,\mu)$ is given by \barcalHandf. It is also natural to measure
the boundary-boundary distance in units of $\sqrt{p}$ which amounts to
introducing the variable $\delta = D / \sqrt{P}$. The corresponding
probability density $\tilde{\rho}_{\rm bound.}(\delta,P)$ is plotted
in Fig.~\rhotildeboundP\ for $P=0.5$, $1.0$, $1.5$, $2.0$, $3.0$,
$5.0$, $10.0$. When $P \to \infty$ we precisely recover the Rayleigh
density \supcritbobo, while for $P \to 0$ we have the particularly
simple but non-trivial expression:
\eqn\prdeltasmallP{{\tilde \rho}_{\rm bound.}(\delta,P)
{\buildrel {P\to 0} \over \to} {2\over 105}
e^{-\delta^2} (35\delta+28 \delta^3+12 \delta^5+3 \delta^7)\ .}

We emphasize that all these expressions are expected to be universal
(up to a possible rescaling of the distance and perimeter) and are
intrinsic to the metric space obtained in the scaling limit. More
precisely, up to a change of the distance scale, we have a
one-parameter family of random metric spaces indexed by the
(renormalized) perimeter $P$. It might be called the {\it Brownian map
with a boundary}. Note that $P$ is not homogeneous to a distance but
to its square, an indication that the fractal dimension of the
boundary is two.  The Brownian map with a boundary interpolates
smoothly between the Brownian map, recovered for $P=0$, and the
Brownian Continuum Random Tree, recovered in the limit $P \to \infty$.
Note that when $P \to 0$, we observe a deviation from the Brownian map
statistics for small distances of order $\sqrt{P}$ (corresponding to
finite values of $\delta$ above).

Let us now briefly mention the results for quadrangulations with a
self-avoiding boundary. Observe that the generating functions
$\tilde{W}_0$ and $\tilde{G}_d$ are related to random quadrangulations
with a self-avoiding boundary in exactly the same way as $W_0$ and
$G_d$ are related to random quadrangulations with a generic
boundary. A parallel approach can thus be followed. Here $Z$ denotes
the activity per unit of half-perimeter. There is now a phase
transition at $Z=2/9$ which is expected to be in the same universality
class as the above. From the exact expressions for $\tilde{W}_0$ and
$\tilde{G}_d$ we can show that:
\item{-} for $Z<2/9$, the perimeter remains finite as the area $n$
tends to infinity, while the bulk-boundary distance is of order
$n^{1/4}$ with the same limit law \subcritbubo,
\item{-} for $Z>2/9$, the perimeter is of order $n$ and concentrates
around its non-universal mean value, while the bulk-boundary
distance is finite and has a non-universal discrete limit law,
\item{-} for $Z \sim 2/9$, the perimeter is of order $n^{1/2}$, while
the bulk-boundary distance is of order $n^{1/4}$.
\par
\noindent
In this latter case, we may as well consider the critical fixed
perimeter ensemble, and compute the bulk-boundary distance cumulative
distribution function. We find the same expression as in \critbubo\ up
to a factor 3 in the renormalized perimeter $P$ (see details in
Sect.~5.2).  This is a first non-trivial check of the universality of
our analytical expressions.

\subsec{Application to self-avoiding loops}

Another interesting application of our exact discrete results is that
they allow us to study the statistics of distances in {\it
quadrangulations with a self-avoiding loop}. More precisely, a
self-avoiding loop is a closed path made of consecutive edges of the
quadrangulation, which is {\it simple}, i.e visits any vertex at
most once. We consider planar quadrangulations with a distinguished oriented
self-avoiding loop (and no boundary: all faces have degree 4). The
area is the total number of faces, while the loop length is
necessarily even. Again we consider a statistical model where the area
$n$ is fixed, where the loop length $2p$ may either be fixed or be
controlled by a weight $y^p$, and where in all rigor we need to
incorporate the inverse symmetry factor, irrelevant for $n \to
\infty$. This is a particular instance of the so-called ${\bf O}({\cal
N}=0)$ model on a random lattice, slightly different from the ones
studied with matrix model techniques \KO\ where the loops would run
on the dual map.

The connection with quadrangulations with a boundary is easily seen.
Upon cutting along the loop, a quadrangulation with a self-avoiding
loop yields two quadrangulations with a self-avoiding boundary,
constrained to have the same perimeter. The orientation of the
loop allows to distinguish these two pieces as left and right, and we
clearly have a bijection preserving the total area. We can therefore
express a number of generating functions for this problem in terms of
the generating functions found above. For instance, the generating
function for quadrangulations with a self-avoiding loop and a marked
vertex on the loop reads 
\eqn\rootdlooppre{\Gamma_0(g,y)=\sum_{p\geq
1}y^p \left({\tilde W}_0(g,Z)\vert_{Z^p}\right)^2}
where $g$ is the weight per face while $\sqrt{y}$ is the weight per
edge of the loop. More generally, we may consider the generating
function for quadrangulations with a self-avoiding loop and a marked
vertex at distance $d$ from the loop and lying on its right. It reads
\eqn\pointlooppre{\Gamma_d(g,y)= \sum_{p\geq 1} y^p {\tilde
W}_0(g,Z)\vert_{Z^p} {\tilde G}_d(g,Z)\vert_{Z^p}}
if the configurations are counted with their inverse symmetry
factor. Constraining the marked vertex to be on the right of the loop
ensures that both expressions are consistent for $d=0$, and by
symmetry it causes no loss of generality. Statistically, $\Gamma_d$
encodes the {\it bulk-loop distance}, i.e the distance between the
loop and a random vertex uniformly drawn in the bulk. In the following
sections, we provide more explicit (yet slightly involved) expressions
for ${\tilde W}_0\vert_{Z^p}$ and ${\tilde G}_d\vert_{Z^p}$, easing
the task of deducing the bulk-loop distance statistics for maps of
large fixed size $n$.  To sum up our results, we find a phase
transition at $y=4/81$.
\item{-} For $y<4/81$, the loop length remains finite as $n \to
\infty$, and the bulk-loop distance is of order $n^{1/4}$ with a
distribution again characterized by the two-point function
\subcritbubo. The physical interpretation is that the loop remains
microscopic and is thus irrelevant in the scaling limit, where
distances are rescaled by a factor $n^{-1/4}$, and which is still
described by the Brownian map.
\item{-} For $y>4/81$, the loop length is of order $n$ and the
bulk-loop distance is finite. The physical interpretation is that the
loop becomes dense in the quadrangulation. We however lack evidence
that the scaling limit (on a scale $n^{1/2}$) is still described by
the Brownian CRT, though this hypothesis is plausible.
\item{-} For $y \sim 4/81$, the loop length is of order $n^{1/2}$ and
the bulk-loop distance is of order $n^{1/4}$.
\par
\noindent 
\fig{The cumulative distribution function ${\hat \Phi}(D,P)$ as a function 
of the the bulk-loop distance for a self-avoiding loop of (rescaled)
half-length $P=0.01$, $0.1$, $0.2$, $0.5$ and $1.0$ (thin lines from
bottom to top). We also indicated (thick red line) the limiting 
two-point function $\Phi(D)$.
}{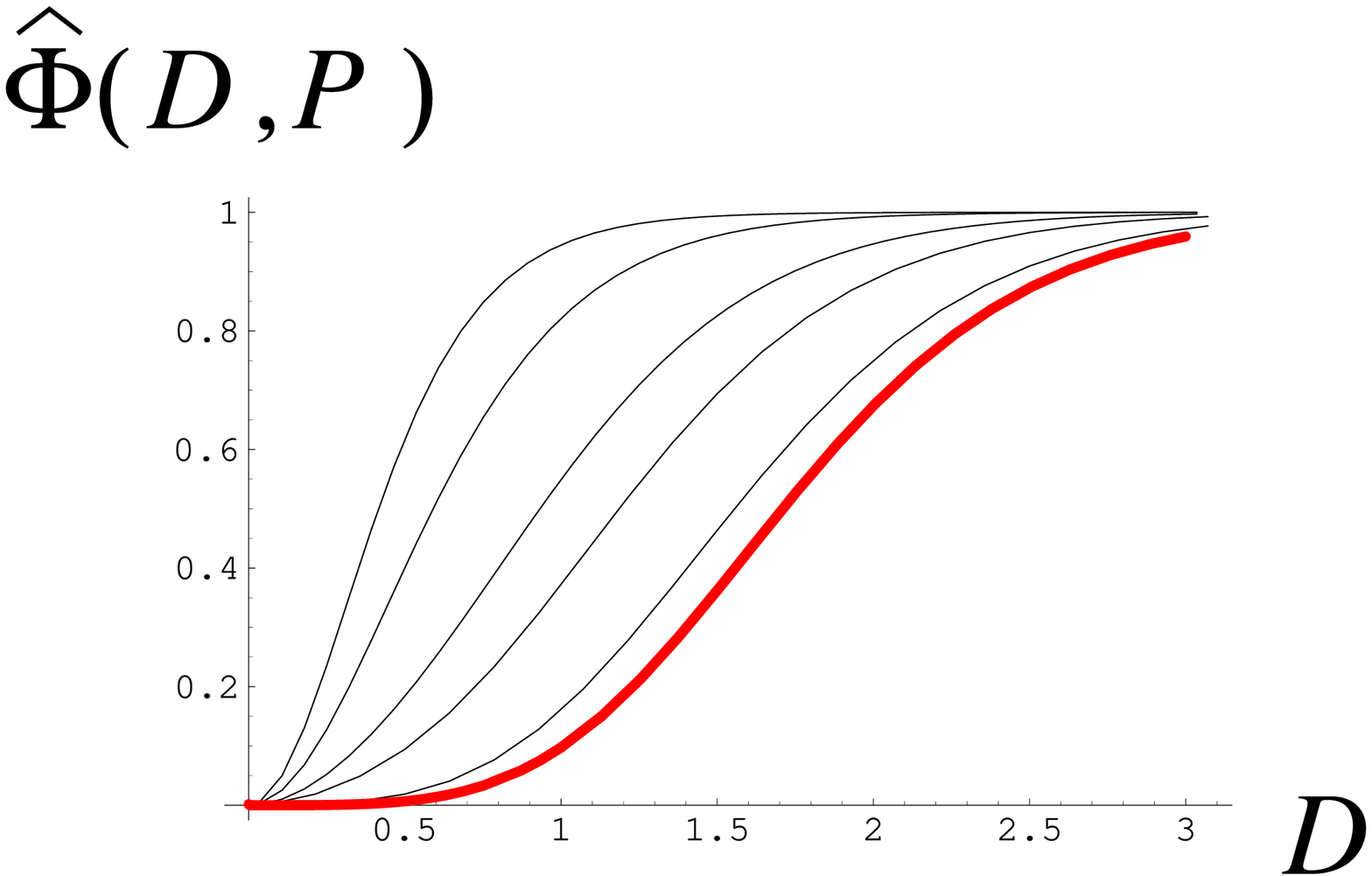}{9.cm}
\figlabel\hatPhiofP
Here again we are most interested in this critical case, and
the results are best expressed in the ensemble where both the area $n$
and the loop length $2p$ are prescribed, and jointly taken to be large
keeping the ratio $P = p \cdot n^{-1/2}$ finite. The scaling law for
the bulk-loop distance cumulative distribution function reads
\eqn\cumdihatpre{\eqalign{{\hat \Phi}(D,P) & \equiv \lim_{n \to
\infty,\ p \sim P n^{1/2}} { \left. \sum_{k=0}^{\lfloor D n^{1/4}
\rfloor} \Gamma_k \right|_{g^n y^p} \over \left. \sum_{k=0}^{\infty}
\Gamma_k \right|_{g^n y^p} } \cr & ={18 P^3 \sqrt{\pi} \over 1+18
P^2}\ e^{9 P^2}\int_{-\infty}^{\infty} d\xi\ {\xi\over {\rm i}}\
e^{-\xi^2}\ {\hat{\cal H}}(D,P;-\xi^2)\ ,\cr}}
where 
\eqn\hath{\hat{\cal H}(D,P;\mu) \equiv 3 {e^{-6 \sqrt{\mu} P} \over
\pi (3P)^4} (1\!+\!3P\sqrt{\mu}) \left\{1\!+\!
\left(3\sqrt{\mu}\!-\!2 f^2(D;\mu)\right)\ \int_0^\infty dK\, e^{-{K^2
\over 3P}-2 f(D;\mu) K} 2K \right\}. }
It is plotted in Fig~\hatPhiofP\ for $P=0.01$, $0.1$, $0.2$, $0.5$ and
$1.0$. 
When $P\to 0$, ${\hat \Phi}(D,P)\to \Phi(D)$ as expected while,
when $P\to \infty$, $D$ is of order $P^{-1/2}$ and we have
the scaling form
\eqn\hatPhilimlargePbis{{\hat{\Phi}}(D,P) \sim 
\tanh^2\left(\sqrt{9 \over 2} D \ \sqrt{P}\right) \ .}
The scaling function ${\hat \Phi}(D,P)$ is expected to be universal, and
characteristic of a model of self-avoiding loop on a Brownian map.

\newsec{Quadrangulations with a boundary: combinatorics}
We now come to the derivation of the expressions given in Sect.~2.2 for
the various generating functions concerning quadrangulations with a generic
boundary. Our approach is based on a bijection with simpler objects, namely
sequences of well-labeled trees, as discussed just below. 
\subsec{Bijection} 
\fig{An example (a) of pointed quadrangulation with a boundary. We have 
labeled each vertex by its graph distance from the origin (vertex with label
$0$). Adding (b) in each face an unlabeled vertex and connecting it
to those labeled vertices followed by a smaller label clockwise within the 
face, we end up (c) with a particular labeled mobile whose unlabeled vertices 
all have degree $2$, except for that associated with the external face, which 
has a degree equal to half the perimeter of the quadrangulation. The labels
around this vertex satisfy the property (P) of the text.}{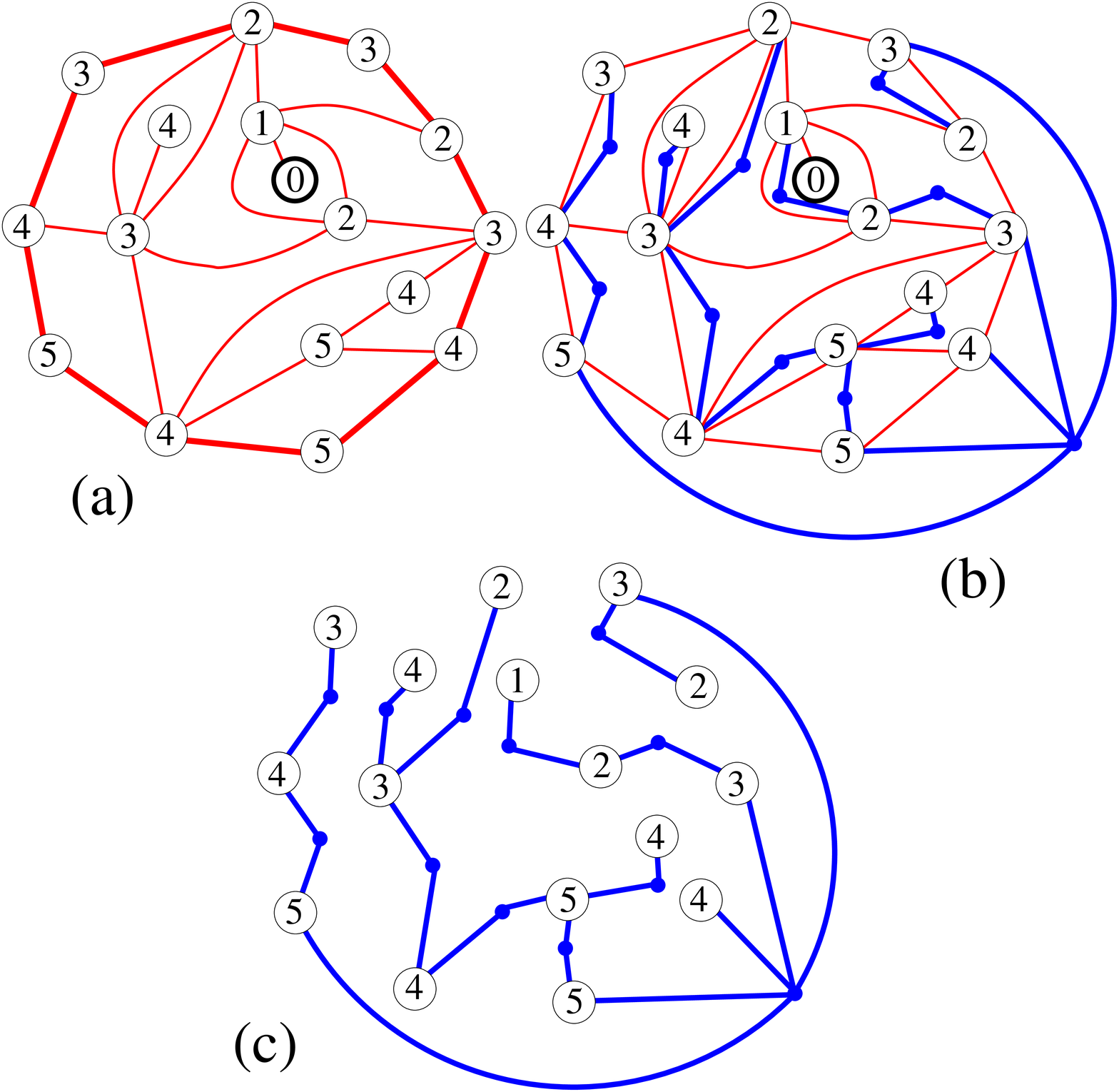}{12.cm}
\figlabel\quadmobone
A quadrangulation with a boundary is a particular instance of a bipartite
planar map. As such, it may be coded by a so-called {\it well-labeled mobile},
as explained in the section 2 of Ref.~\MOB\ (see Fig.~\quadmobone). 
More precisely, the coding is for a pointed map (i.e a map with a 
chosen origin vertex). The
associated mobile is a plane tree with alternating labeled and unlabeled 
vertices. The labeled vertices correspond to the original vertices of the map 
and they carry an integer label equal to the graph distance in the map from 
the corresponding vertex to the origin. The unlabeled 
vertices of the mobile correspond to the faces of the map and their degree 
is half the degree of the corresponding face in the map. Around each unlabeled 
vertex $v$, we have the following property (P): reading the sequence of labels 
of vertices adjacent to $v$ clockwise around $v$, any label $\ell$ is followed 
by a label larger than or equal to $\ell-1$. Finally, the mobile has a minimum 
label equal to $1$.
\fig{The well-labeled mobile of Fig.~\quadmobone-(c), where we erased the
bi-valent unlabeled vertices (a), may alternatively be viewed (b) as 
a set of well-labeled trees satisfying the property (P') of the text,
attached at each descending step of a (counterclockwise-oriented) cyclic 
sequence of integers (green outer circle) reproducing precisely the distance 
from the origin of the successive vertices along the contour of the 
quadrangulation (as apparent in Fig.~\quadmobone-(a)).
}{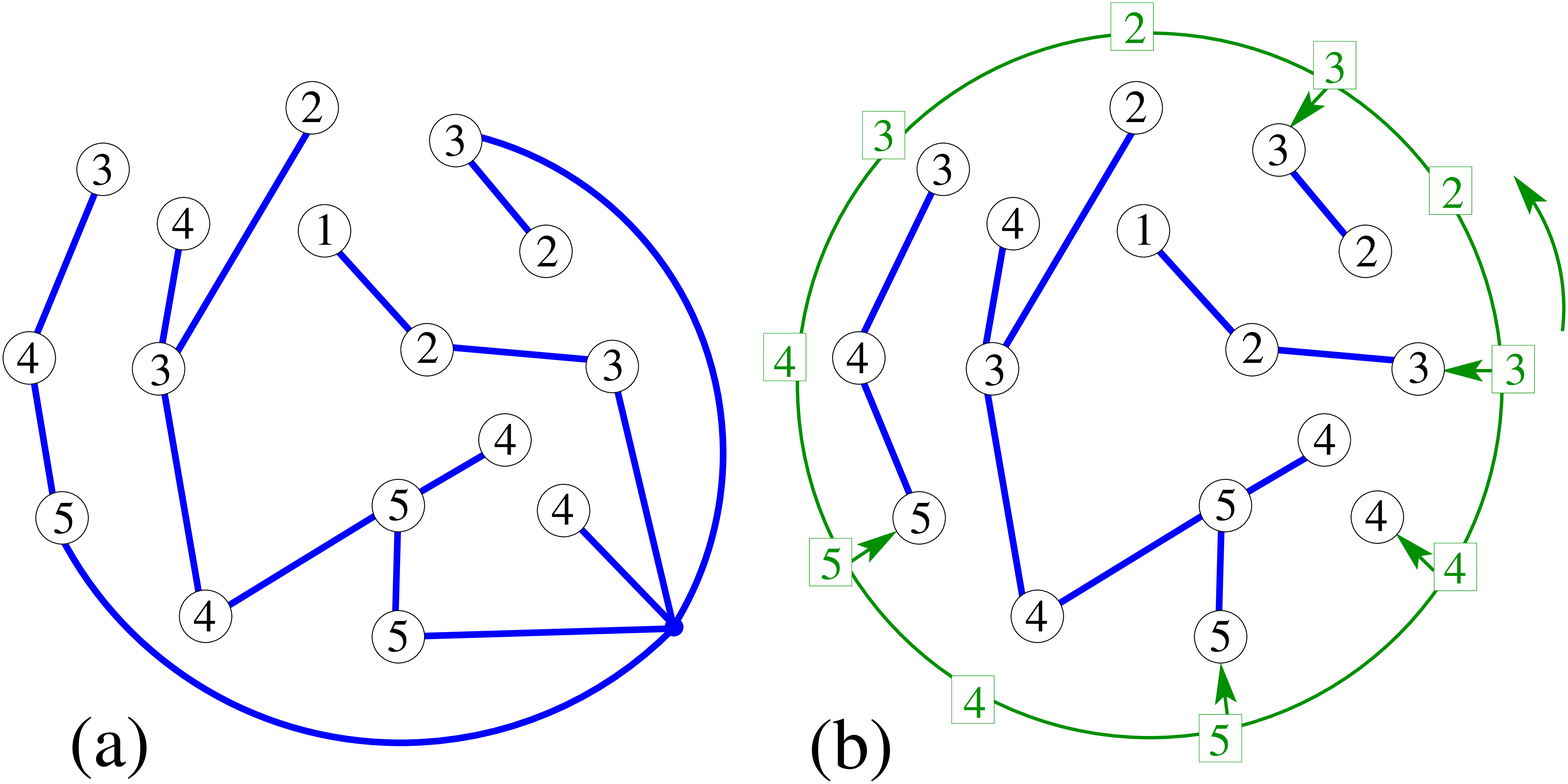}{12.cm}
\figlabel\quadmobtwo
For quadrangulations with a boundary of length $2p$, all the unlabeled 
vertices of the associated mobile necessarily have degree two, except for 
that associated with the external face, which we call the {\it external vertex}
and whose degree is $p$. Those bi-valent unlabeled vertices may be erased, 
giving rise to edges which connect the labeled vertices directly. The property
(P) may then be rephrased into the property (P') that {\it labels on adjacent 
labeled vertices differ by at most 1}. The resulting object is therefore a 
collection of $p$ 
{\it well-labeled trees}, i.e trees with labeled vertices satisfying (P'),
attached to the external vertex by their root vertices, whose clockwise 
sequence of labels satisfies (P) around the external vertex (see 
Fig.~\quadmobtwo-(a)). An equivalent 
but more convenient coding of the sequence of these $p$ root labels around 
the external vertex is via a cyclic sequence of $2p$ non-negative integers 
such that consecutive integers differ by $\pm 1$ (see Fig.~\quadmobtwo-(b)). 
In this coding, the root labels simply correspond to those integers which 
are followed immediately by a smaller integer in the cyclic sequence,
and the property (P) is automatically satisfied. Moreover, 
the cyclic sequence of integers corresponds precisely to the distance
to the origin of the successive boundary vertices along the contour
(see Fig.~\quadmobtwo-(b)).

To summarize, we have a bijection between, on the one hand, pointed 
quadrangulations with a boundary and on the other hand cyclic
sequences of non-negative integers such that consecutive integers 
differ but $\pm 1$,
with a well labeled tree with root label $\ell$ attached to each descending 
step $\ell\to\ell-1$ of the cyclic sequence, and with the requirement
that the global minimum label is $1$. Under this bijection, we have the 
following correspondences: 

\item{-} The total number of edges for all the well-labeled trees
is equal to the area $n$ (number of inner faces) of the quadrangulation.
\item{-} The label of any vertex is equal to the distance to the origin of the 
corresponding vertex on the map.
\item{-} The (even) length of the cyclic sequence is equal to the 
length $2p$ of the boundary. 
\item{-} The successive integers in the cyclic sequence are equal to the 
distance to the origin of the successive boundary vertices along the
contour. In particular, the smallest integer $d$ in this cyclic sequence
is the smallest distance found between the origin and a vertex of the
boundary, i.e the {\it distance from the origin to the boundary}.
\par
\noindent We may transform the cyclic sequence of non-negative
integers into a Dyck path $(0=\ell_0, \ell_1, \cdots,
\ell_{2p-1},\ell_{2p}=0)$ by reading the sequence from one of its
minima, and subtracting $d$ from all the integers.  To each descending
step $\ell_i\to\ell_i-1$ is now attached a well-labeled tree with root
label $\ell_i+d$. Note that the cyclic sequence may have several
minima, and choosing a particular minimum amounts to marking a
boundary edge $d\to d+1$ with, say, the external face to the right.

\subsec{Basic generating functions} 
We now wish to compute the generating function for pointed quadrangulations 
with a boundary, where a map with area $n$ and perimeter $2p$ comes with
a weight $g^n z^p$. Under the bijection, this weight 
simply 
amounts to a weight $g$ per edge of the well-labeled trees and a weight $z$ 
per descending step of the cyclic sequence of integers, or equivalently
of the Dyck path. 

A first ingredient is the generating function $R_\ell$ for 
rooted well-labeled trees with a root label $\ell\geq 1$ and with the 
condition that {\it all the labels on the tree are larger than or equal 
to $1$}. 
This generating function was computed in Ref.~\GEOD\ and reads
\eqn\Rell{R_\ell= R {\q{\ell}\, \q{\ell+3} \over
\q{\ell+1}\, \q{\ell+2}}\ ,}
where we use the notation 
\eqn\defq{\q{\ell}\equiv {1-x^\ell\over 1-x}}
and where the quantities $R$ and $x$ are solutions of 
\eqn\eqnforRx{R=1+3 g R^2\ , \quad \quad x=g R^2 (1+x+x^2)\ ,}
namely 
\eqn\Rxexplicit{\eqalign{R &= {1-\sqrt{1-12g}\over 6 g}\ ,\cr
x &={1-24g-\sqrt{1-12g}+\sqrt{6}\sqrt{72g^2+6g+\sqrt{1-12g}-1}\over
2(6g+\sqrt{1-12g}-1)}\ .\cr}}
\fig{Opening a cyclic sequence of non-negative integers at one of its
minima $d$ results into a Dyck 
path $(0=\ell_0, \ell_1, \cdots,
\ell_{2p-1},\ell_{2p}=0)$ by subtracting $d$ from all integers in the sequence. 
To each descending step $\ell\to\ell-1$ of the Dyck path is attached 
a well-labeled tree with root label $\ell+d$, resulting in a weight
$z R_{d+\ell}$ for the descending step if we demand that all the original 
labels be 
larger than $1$.}{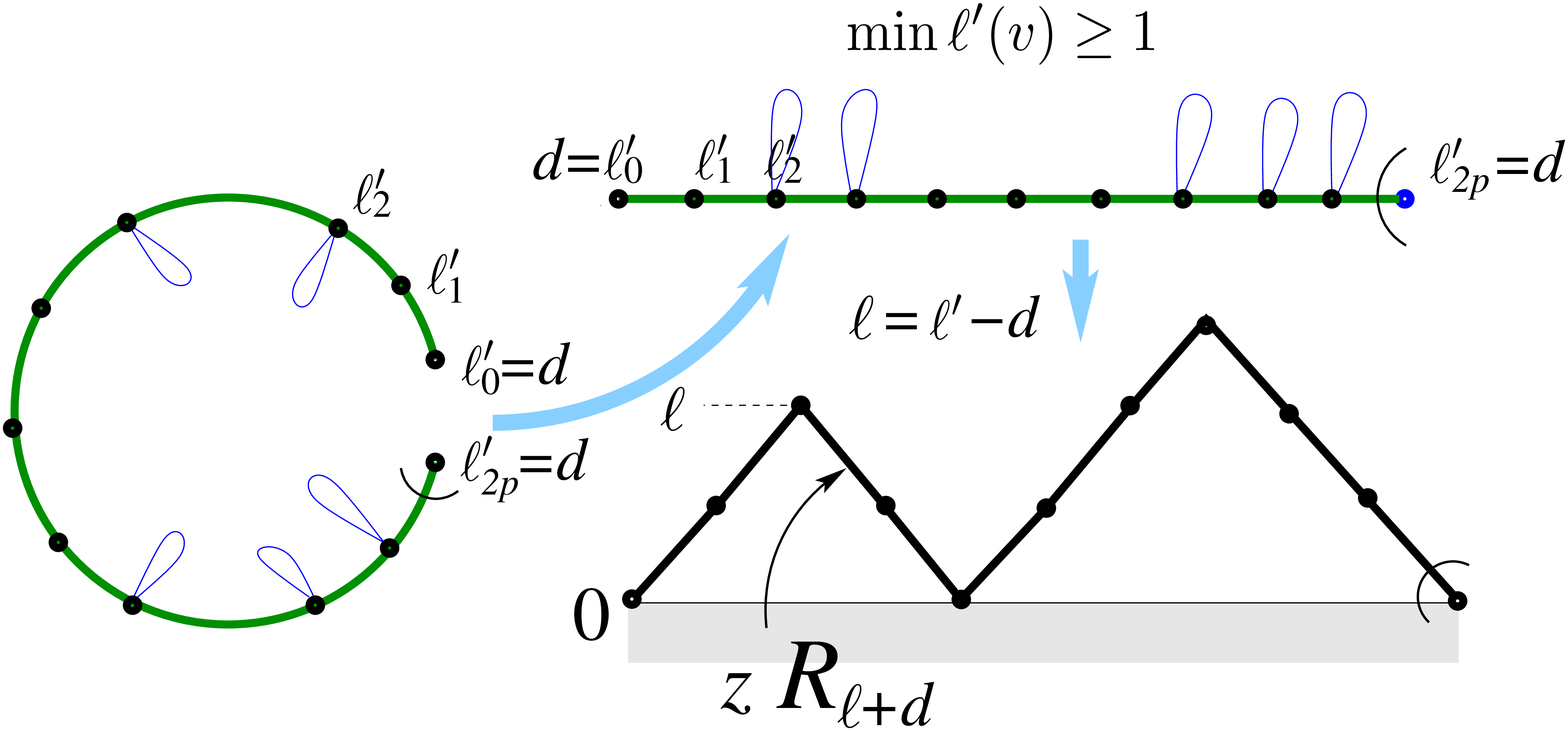}{12.cm}
\figlabel\dyck
We can now express the generating function $W_d$ for Dyck 
paths with a well-labeled tree with root label $\ell_i+d$ attached at each 
descending step $\ell_i\to\ell_i-1$, and with a global minimum
label larger than or equal to $1$. This generating function 
reads
\eqn\dyck{W_d= \sum_{p\geq 0} \quad
\sum_{{\rm Dyck\ paths\ of\ length\ } 2p \atop
(0=\ell_0, \ell_1, \ldots,\ell_{2p}=0)} \quad
\prod_{{\rm descending\ steps}\atop \ell_i \to \ell_i-1} z\, R_{\ell_i+d}}
with a conventional weight $1$ for the trivial Dyck path of length $0$. 
We have the recursion relation
\eqn\recurW{W_d=1+ z R_{d+1} W_{d} W_{d+1}\ ,
}
as obtained by decomposing any non-trivial Dyck path into its first ascending
step $0\to 1$, a path from $1$ to $1$ with all intermediate heights larger 
than or equal to $1$ (weight $W_{d+1}$ as obtained by a simple shift of 
labels), its first descending step $1\to 0$ 
(weight $z R_{d+1}$) and a final Dyck path (weight $W_d$). 

We may look for a solution of this equation in the form
\eqn\formsol{W_d=W {\q{d+2}\over \q{d+3}} {V_{d+1}\over V_d}\ ,}
with $W$ and $V_d$ to be determined. Substituting this particular form 
in Eq.\recurW\ leads to the equation
\eqn\subsform{W \q{d+2} V_{d+1} = \q{d+3} V_d + z R W^2 \q{d+1} V_{d+2}\ .}
This relation is satisfied (for arbitrary $R$ and $x$) upon taking
\eqn\valv{V_d=1+\lambda x^d}
provided we choose $W$ and $\lambda$ such that 
\eqn\eqnforWandlambda{\eqalign{W &=1+ z R W^2 \cr
W(-x^2+\lambda x) &= (-x^3+\lambda)+z RW^2 (-x+\lambda x^2)}}
namely
\eqn\Wlambdaexpl{W={1-\sqrt{1-4 z R}\over 2 z R}, \qquad
\lambda = x {(W-1)-x \over 1-x(W-1)}\ .}
Plugging this last expression back in \valv\ and \formsol, we end up with the 
desired solution
\eqn\Wdexpli{\eqalign{W_d & = W \ {\q{d+2}\over \q{d+3}} \times 
{\q{d+3}-x(W-1) \q{d+1} \over \q{d+2}-x (W-1) \q{d}} \cr
& = W\ {1-(W-1) f_{d+1}\over 1-(W-1) f_d}
\ ,\cr}}
where $W$ is explicitly given in terms of $g$ and $z$ as
\eqn\expliW{W={3g-\sqrt{9g^2-6g(1-\sqrt{1-12 g})z}\over (1-\sqrt{1-12g})z}}
while we introduce the compact notation
\eqn\defwfd{f_d\equiv x {\q{d} \over \q{d+2}}}
with $\q{\cdot}$ defined above. In particular, we have $f_0=0$ and
$f_1=g R^2$.  Note that $W_d$ depends on the variable $z$ only through
the quantity $W$.  In particular, for $d=0$, we find
\eqn\expwzero{W_0=W (1-g\, R^2 (W-1))}
which agrees with the known generating function for rooted 
pseudo-quadrangulations \PDF.

To go from Dyck paths back to cyclic sequences of integers, we simply have
to identify the Dyck paths differing only by the choice of an instance of 
the smallest integer in the sequence. We note that Eq.~\recurW\ may 
be alternatively written as
\eqn\alterrecu{W_d= \sum_{k\geq 0} (z R_{d+1} W_{d+1})^k}
where $(z R_{d+1} W_{d+1})^k$ is nothing but the generating function
for Dyck paths with exactly $k$ returns to $0$. Cyclic sequences are
then enumerated by
\eqn\cyclicenu{\sum_{k\geq 1} {1\over k} (z R_{d+1} W_{d+1})^k = \log{W_d}\ .}
where we use the convention of counting configurations with an inverse
symmetry factor. A given configuration made of a cyclic sequence with
its attached trees may only have a cyclic symmetry group ${\bf Z}_m$
(with $m$ a divisor of the number of minima $k$), and is then counted
with a weight $1/m$.

So far, in the configurations counted by $\log{W_d}$, 
we imposed only that the global minimum label be larger than or equal to $1$. 
To impose that this minimum label be exactly $1$, as required by the bijection,
we must suppress those configurations with a minimum larger than or equal 
to $2$, whose generating function is obtained from the previous one 
by a simple shift of all labels by $-1$, i.e is given by $\log{W_{d-1}}$. 
In particular, we deduce that the generating 
function $G_d$ for quadrangulations with a boundary and a marked 
vertex at distance $d$ from the boundary is simply given by
\eqn\genlog{G_d=\log\left({W_d\over W_{d-1}}\right)\ .}
Again, configurations with an $m$-fold symmetry around the marked vertex
are counted with a factor $1/m$. 

As for the original quantity $W_d$, it is the generating function for 
quadrangulations with a boundary, with a marked vertex at a distance 
less than or equal to $d$ 
from the boundary (the origin) and with a marked ``closest edge'' to 
this origin, i.e a boundary edge incident to a vertex
at distance $d$ from the origin and oriented counterclockwise around
the bulk of the quadrangulation. 
Note that such pointed rooted maps cannot have any non-trivial symmetry.
Finally, we may also interpret $\log W_d$ 
as the generating function for pointed quadrangulations with a boundary 
whose origin is at a distance less than or equal to $d$ from the boundary.

As discussed in Section 2, we are also interested in the 
generating function $T_d$ for 
quadrangulations with a boundary having {\it two marked} (and distinguished)
{\it boundary edges} oriented, say counterclockwise around the bulk
of the quadrangulation, such that the origins of these marked edges
are at a mutual distance $d$ on the map. 
Taking the origin of the first marked edge as the origin of the map, we
get under the bijection a cyclic sequence of integers with minimal value
$0$. The first marked edge defines a first step $0\to 1$ at which
we may start reading the cyclic sequence, leading to a Dyck path 
with a well-labeled tree with root label $\ell_i$ attached to 
each descending step $\ell_i\to\ell_i-1$. Marking the second boundary edge
amounts to choosing a step $d\to d\pm 1$ in the sequence, i.e
to choosing a point of height $d$ in the Dyck path (for $d=0$, 
this point must be different from the last one). Upon decomposing
the Dyck path into a first ascending part from $0$ to the marked point 
$d$ and a second descending step from $d$ to $0$, we get the expression
\eqn\Tdone{\eqalign{T_d & = W_0 W_1 \cdots W_d \times (W_d z R_d) 
(W_{d-1} z R_{d-1}) \cdots (W_1 z R_1) W_0 \cr
& = (W_0 W_1 \cdots W_d)^2 z^d (R_1 R_2 \cdots R_d) \ ,\cr }}
valid for $d>0$. For $d=0$, the actual decomposition yields 
$T_0=W_0^2-W_0$. From now on, we will assume $d>0$ when referring
to $T_d$. Upon substituting the expression \Wdexpli\ for $W_d$, we get
\eqn\Tdtwo{\eqalign{T_d & = W^2 (W-1)^d {\q{1} \over \q{3}} 
{\left(\q{d+3}-x(W-1)\q{d+1}\right)^2
\over \q{d+1}\q{d+3}}\cr &= 
W^2 (W-1)^d f_1 \left({1\over f_{d+1}}-2(W-1)+f_{d+1}(W-1)^2\right)
\ .\cr}}

\subsec{Fixed length generating functions}

As we already noticed, the quantities $W_d$ and $T_d$ depend on
$z$ only via the quantity $W$ as given by Eq.~\eqnforWandlambda, 
which we may rewrite as 
\eqn\ztow{z= {w \over R\, (1+w)^2}}
upon introducing the notation
\eqn\smallwdef{w\equiv W-1\ .}
In this respect, $W_d$ and $T_d$ are so-called Lagrangean generating
functions, i.e for which we can apply the Lagrange inversion theorem
\GJ.
This means that we may extract an explicit expression
for the $z^p$ term of these generating functions. 
More precisely, the $z^p$ term, expressed as a contour integral in the 
variable $z$ around $0$ may be transformed by the change of variable
$z\to w$ into a contour integral around $0$ of the variable $w$,
namely
\eqn\chgvar{\oint {dz \over 2 {\rm i} \pi} {1\over z^{p+1}}
\Big\{ \cdot \Big\} = R^p\, \oint {dw \over 2{\rm i} \pi}
{1\over w^{p+1}} (1-w) (1+w)^{2p-1} \Big\{ \cdot \Big\}\ .}
Upon expanding \Wdexpli\ in $w=W-1$, we get the expression
\eqn\Wdtwo{W_d=(1+w) \left(1-w\, (f_{d+1}-f_d) \sum_{k\geq 1} 
(w\, f_d)^{k-1} \right)}
Taking the contour integral \chgvar\ of this quantity, we immediately 
get
\eqn\Wdzp{\eqalign{W_d\vert_{z^p}& =
R^p \left\{\!
{2p \choose p}\!-\!{2p \choose p\!-\!1}\!-\!
(f_{d+1}\!-\!f_d)\!\sum_{k\geq 1}\left(\!{2p \choose p\!-\!k}\!-\!{2p
\choose p\!-\!1\!-\!k}\!\right)\!(f_d)^{k-1}\!\right\}\cr
&= 
R^p \left\{
{(2p)!\over p! (p\!+\!1)!}-
(f_{d+1}\!-\!f_d)\sum_{k=1}^p{(2p)! \over (p\!-\!k)!(p\!+\!k\!+\!1)!} (2k+1)
(f_d)^{k-1}\right\}
\ .\cr}}
By a similar calculation, we easily obtain
\eqn\logWdzp{(\log W_d)\vert_{z^p} =
R^p \left\{
{(2p-1)!\over (p!)^2}-2
\sum_{k=1}^p{(2p-1)! \over (p\!-\!k)!(p\!+\!k\!)!}
\left((f_{d+1})^k-(f_d)^k\right)\right\}
}
for $p\geq 1$. 
For $d=0$, we get in particular
\eqn\Wzerozp{W_0\vert_{z^p}= R^p
{(2p)! \over p!(p+2)!}(p+2-3 p f_1) = R^p 
{(2p)! \over p!(p+2)!}\left(2+p\left(2-R\right)\right) 
} 
with $R$ given by \Rxexplicit. A second application of the Lagrange
inversion formula, now for the variable $g$, yields
\eqn\Wzerozpgn{W_0\vert_{g^n z^p}= {3^n (2p)! \over p!(p-1)!}
{(2n+p-1)!\over n!(n+p+1)!}} 
for $p\geq 1$.
On the other hand, for $d\to \infty$, we have $W_d\to W$, with 
\eqn\Winfzp{W\vert_{z^p}= R^p {(2p)! \over p!(p+1)!} }
so that
\eqn\Winfzpgn{W\vert_{g^n z^p}= {3^n (2p)! \over (p-1)!(p+1)!}
{(2n+p-1)!\over n!(n+p)!}} 
for $p\geq 1$.
The quantity $W_0\vert_{g^n z^p}$ is the number
of quadrangulations with area $n$ and boundary 
of length $2p$, where we have marked one of the $2p$ boundary 
edges on the contour. The quantity 
$W\vert_{g^n z^p}$ is the number of 
the same quadrangulations with additional markings of one of their 
$n+p+1$ vertices (at some arbitrary distance from the boundary) 
as well as of a particular boundary edge closest to this vertex. 
The quantity 
\eqn\Wratio{{W\vert_{g^n z^p} \over W_0\vert_{g^n z^p}}
{2p \over  n+p+1}= {2p\over p+1}}
measures therefore the average number of boundary edges closest to a 
uniformly chosen random vertex, for the ensemble of quadrangulations 
with area $n$ and perimeter $2p$. Note that, surprisingly,
 this average number is 
independent of $n$ (and its value thus matches that obtained for $n=0$, 
where quadrangulations with a boundary reduce to plane trees
with $p$ edges).

Finally, we may write \Tdtwo\ as
\eqn\Tdtwo{T_d= (1+w)^2 f_1 \left({w^d \over f_{d+1}}
-2 w^{d+1} + w^{d+2} f_{d+1}\right)
}
so that we get
\eqn\Tdzp{\eqalign{T_d\vert_{z^p}= 
& R^p f_1 \left\{
{1\over f_{d+1}}\!\left(\!{2p+1 \choose p\!-\!d}\!-\!{2p+1 \choose p\!-\!1\!-d}
\!\right)\right.\cr & \left.\ \ \  -\!2\left(\!{2p+1 \choose p\!-\!1\!-d}\!-\!
{2p+1 \choose p\!-\!2\!-d}\!\right)\! 
+\!f_{d+1}\!\left(\!{2p+1 
\choose p\!-\!2\!-d}\!-\!{2p+1 \choose p\!-\!3\!-d}\!\right)\right\}\cr
& = 2 R^p f_1 \left\{
{(d+1)\over f_{d+1}}\!{(2p+1)!\over (p\!-\!d)!(p\!+\!d\!+2)!}
\right.\cr & \left.\ \ \  -\!2 (d+2)\!{(2p+1)!\over (p\!-\!d\!-1)!(p\!
+\!d\!+3)!}\!+\!(d+3)\!f_{d+1}\!{(2p+1)! \over
(p\!-\!d\!-2)!(p\!+\!d\!+4)!}\!\right\}\cr
}}

By a slight refinement of this calculation, we can compute 
the generating function $T_d(s,s')$ for quadrangulations with a boundary
of length $2p=s+s'$ with two marked boundary edges whose origins 
are separated by $s$ steps counterclockwise along the contour and
$s'$ steps clockwise, and are at a mutual distance $d$ on the map.
We find
\eqn\Tddes{\eqalign{T_d(s;s')= R^{s+s'} f_1
& \Big\{ {1\over f_{d+1}} C(s,d)C(s',d)
- C(s,d)C(s',d+2)\cr & -C(s,d+2)C(s',d)
+f_{d+1} C(s,d+2)C(s',d+2) \Big\}\cr}}
where
\eqn\Asd{C(s,d)\equiv {s \choose {s-d\over 2}}-{s \choose {s-d\over 2}-1}\ .}
This results holds when $d$, $s$ and $s'$ have the same parity, while
$T_d(s,s')$ vanishes otherwise.

\newsec{Quadrangulations with a boundary: asymptotics}

\subsec{Critical lines}

In order to describe the statistics of distances in large
quadrangulations, we have to analyze the singular behavior of the
various generating functions above. More precisely, the
asymptotics for a large area $n$ is encoded in the singularity 
reached at the radius of
convergence in $g$ of these generating functions.  It is simpler to
work first with a {\it fixed value of $z$}, corresponding to 
the fixed $z$ ensemble mentioned in Section 2.  A first singularity is
associated with the singular behavior of $R$ and $x$, as given by
Eqs.~\eqnforRx\ or \Rxexplicit, when $g$ approaches the critical value
\eqn\gcritone{g_{\rm crit}^{(1)}={1\over 12}} irrespectively of the
value of $z$. This is the dominant singularity for $z<1/8$ while, for
$z\geq 1/8$, another singularity comes from the singular behavior of
$W$, as given by Eqs.~\eqnforWandlambda\ or \Wlambdaexpl, when $z R$
approaches the value $1/4$, which defines the critical line
\eqn\gcrittwo{g_{\rm crit}^{(2)}(z)={1\over 3}\ 4z\, (1-4z)\quad
(z\geq 1/8)\ .}
Since $g_{\rm crit}^{(2)}(z)\leq 1/12$, this second
singularity is dominant (i.e determines the radius of convergence)
whenever $z>1/8$.
 
The radius of convergence therefore changes determination at
\eqn\zcrit{z_{\rm crit}={1\over 8}\ .} 
As we shall see below, the generating functions have very different 
scaling behaviors when $z$ is smaller or
larger than $z_{\rm crit}$. As discussed in Section 2, this is the 
manifestation of a drastic
change in the geometry of large quadrangulations with a boundary
at this critical value. 

\subsec{Scaling limit: the $z<1/8$ regime}

Let us first discuss the case $z<1/8$ for which the generating functions
have radius of convergence $g_{\rm crit}^{(1)}=1/12$. We may analyze the
associated singularity by setting 
\eqn\gscal{g={1\over 12}(1-\mu \epsilon)}
with $\epsilon \to 0$. A sensible scaling limit for $W_d$ is obtained 
by considering large distances of the form
\eqn\scald{d=D\, \epsilon^{-1/4}\ ,}
with $D$ finite. We then have the following small $\epsilon$ expansions
\eqn\expan{\eqalign{
f_{d} &= 1-2 f(D;\mu)\, \epsilon^{1/4} + (4 f^2(D;\mu)-3 \sqrt{\mu})\,  
\epsilon^{1/2} +\cdots \cr
f_{d+1} &= 1-2 f(D;\mu)\, \epsilon^{1/4} + (6 f^2(D;\mu)-6 \sqrt{\mu})\,  
\epsilon^{1/2} +\cdots \cr}}
where we define
\eqn\deffDmu{f(D;\mu)\equiv \sqrt{{3\over 2}}\, \mu^{1/4}\ \coth \left(
\sqrt{{3\over 2}}\, \mu^{1/4} D \right)\ .}
We also have the expansion
\eqn\expantwo{W = A(z) - z\, A'(z)\, \sqrt{\mu}\, \epsilon^{1/2} + \cdots}
where we define
\eqn\defAz{A(z)= {1-\sqrt{1-8z}\over 4z}\ .}
Plugging these expressions in \Wdexpli, we obtain the expansion
\eqn\expanthree{W_d
= A(z) - {\cal F}(D;\mu)\, z\, A'(z)\, \epsilon^{1/2} + \cdots}
where we introduce the notation
\eqn\defcalF{{\cal F}(D;\mu) \equiv 2 (f^2(D;\mu) -\sqrt{\mu}) 
= \sqrt{\mu}\left( 1+{3\over \sinh^2 
\left(\sqrt{{3\over 2}}\, \mu^{1/4} D \right)
}\right)\ .}
We may alternatively derive \expanthree\ by using the explicit
form \Wdzp\ of $W_d\vert_{z^p}$ and expanding it at small $\epsilon$.
Using the expansion
\eqn\expanR{R=2(1-\sqrt{\mu}\, \epsilon^{1/2}+\cdots)\ ,}
we obtain that 
\eqn\Wdzpexpan{\eqalign{W_d\vert_{z^p}& =
2^p \left\{ {(2p)!\over p! (p\!+\!1)!}
- \sqrt{\mu}\, \epsilon^{1/2}
{(2p)!\over (p\!-\!1)! (p\!+\!1)!}\right.
\cr 
\ \ \ \ \ \ 
& \left. \!-\!
(2f^2(D;\mu)\!-\!3\sqrt{\mu})\, \epsilon^{1/2}\, 
\sum_{k=1}^p{(2p)! \over (p\!-\!k)!(p\!+\!k\!+\!1)!} (2k+1)
\right\} +\cdots \cr
&= 2^p 
{(2p)!\over p! (p\!+\!1)!}\left\{1-p\ {\cal F}(D;\mu)\, \epsilon^{1/2} 
+\cdots\right\}\ .\cr}}
Upon summing over $p$ with a weight $z^p$, this reproduces precisely the
expression \expanthree. Note that the ``kernel'' ${\cal F}(D;\mu)$ 
occurring in \Wdzpexpan\ is independent of $p$. For $p=1$, quadrangulations 
with a boundary of length $2$ are equivalent, upon closing the boundary, 
to {\it rooted quadrangulations}, i.e quadrangulations with a marked edge. 
The kernel ${\cal F}(D;\mu)$ is therefore the same as that encountered in 
Ref.~\GEOD\ when deriving the continuous two-point function of planar 
quadrangulations from the generating function of rooted quadrangulations. 

We may repeat this analysis for the quantity $\log(W_d)$. We have the expansion 
\eqn\expanlog{\log W_d
= \log A(z) - {\cal F(D;\mu})\, {z\, A'(z)\over A(z)}\, \epsilon^{1/2} + \cdots}
or the equivalent expansion
\eqn\logWdzpexpan{(\log W_d)\vert_{z^p} =
2^p {(2p-1)!\over (p!)^2}
\left\{1-p\ {\cal F}(D;\mu)\, \epsilon^{1/2} 
+\cdots\right\}\ . }

From the above singularity analysis, we may deduce the bulk-boundary
distance statistics when $z<1/8$ in the ensemble of quadrangulations
(with a boundary) with a fixed area $n$, in the
limit where $n\to \infty$. Indeed, we may extract the contribution to
$W_d$ or $\log W_d$ of 
these quadrangulations by a contour integral around $0$ in
the variable $g$. At large $n$, this translates into an integral over
a real variable $\xi$ upon setting (see Ref.~\GEOD\ for a more detailed
discussion)
\eqn\gtoxi{g={1\over 12}\left(1+{\xi^2\over n}\right)\ .}
We may indeed write at large $n$
\eqn\gtoxiint{\oint {dg \over 2{\rm i}\pi}{1\over g^{n+1}}\{\cdot\}
\sim {12^n \over \pi n} \int_{-\infty}^\infty d\xi\ {\xi \over {\rm
i}}\ \ e^{-\xi^2}\{\cdot\} \left(1+{\cal O}\left({\xi^2 \over
n}\right)\right) \ .}

Setting \eqn\dD{d=D\ n^{1/4}} with $D$ finite, we may use the
expansion \expanthree\ above with $\epsilon=1/n$ and $\mu=-\xi^2$ and
deduce that, at large $n$
\eqn\largenWd{W_d\vert_{g^n} \sim 
{12^n \over \pi n^{3/2}} z A'(z)\ \int_{-\infty}^\infty
d\xi\ {\rm i}\xi\ \ e^{-\xi^2}{\cal F}(D;-\xi^2)} 
since the first (regular) term $A(z)$ in the expansion \expanthree\ leads to 
a vanishing integral in $\xi$ by parity. In particular, for $D\to \infty$,
we have ${\cal F}(D,-\xi^2)\to -{\rm i}\xi$ and we get
\eqn\largenWdbis{W\vert_{g^n} \sim 
{12^n \over 2 \sqrt{\pi} n^{3/2}} z A'(z)\ ,}
which may alternatively be obtained directly from the general
expression \Winfzpgn.

Similarly, we obtain from \logWdzpexpan\ that
\eqn\largenlogWd{\eqalign{\log W_d\vert_{g^n} & \sim 
{12^n \over \pi n^{3/2}} {z A'(z)\over A(z)}\ \int_{-\infty}^\infty
d\xi\ {\rm i}\xi\ \ e^{-\xi^2}{\cal F}(D;-\xi^2)\cr   
\log W\vert_{g^n} & \sim 
{12^n \over 2 \sqrt{\pi} n^{3/2}} {z A'(z)\over A(z)}\ .\cr}}
The ratio of these quantities tends at large 
$n$ to a finite quantity 
\eqn\ratiolim{\Phi(D)={2\over \sqrt{\pi}}  \int_{-\infty}^\infty
d\xi\ {\rm i} \xi\ e^{-\xi^2}{\cal F}(D;-\xi^2)}
which is the {\it (cumulative) distribution function} for $D$ giving, 
in the ensemble of pointed quadrangulations with a boundary, the 
probability that the distance to the boundary of the marked vertex be 
less than $D$. In the regime $z<1/8$, this 
distribution function is independent of $z$ and identical to the 
two-point function of the Brownian map [\xref\AW,\xref\AJW,\xref\GEOD]. 
This follows from the fact that, 
in the regime $z<1/8$, the length of the boundary does not scale with $n$ 
but remains finite 
at large $n$. This property can be measured as follows: as we already noticed, 
the generating function $W_0$ counts quadrangulations with one marked 
edge along the boundary. To remove this marking, we must consider
instead the generating function $\int_0^z dz' (W_0(z')-1)/(2z')$ , so that
the average half-perimeter reads
\eqn\avebl{\langle p\rangle_n(z)={(W_0(z)-1)\vert_{g^n} \over
\int_0^z dz' {W_0(z')-1 \over z'}\vert_{g^n}}\ .}
From \Wzerozpgn, we immediately get that, at large $n$,
\eqn\asymwo{(W_0(z)-1)\vert_{g^n} \sim {12^n \over 2 \sqrt{\pi} n^{3/2}}
z(z\ A(z))''}
so that
$\langle p\rangle_n(z)$ tends to the finite value
\eqn\valpave{\langle p\rangle_n(z) \to {z (z\, A(z))'' \over (z\, A(z))'-1}
={4 z \over (1-8z)(1-\sqrt{1-8z})} \quad\ (z<1/8)\ .}

Let us now consider the boundary-boundary distance statistics for $z<1/8$.
When $g\to 1/12$, a sensible scaling limit is now obtained by 
{\it keeping $d$ finite} in $T_d$. This is consistent with 
the fact that the perimeter itself remains finite
and $d\leq p$ obviously. For finite $d$, we now have the expansion 
\eqn\fdlim{f_{d+1}=
{d+1\over d+3}- {(d+1)(d+2)\over d+3} \sqrt{\mu}\, \epsilon^{1/2} 
+{\cal O}(\epsilon)}
while the expansion \expantwo\ still holds. Using \Tdtwo, 
we get
\eqn\Tdlim{T_d={A(z)^2 \big(2(d+2)-(d+1)A(z)\big)^2
\over 3(d+1)(d+3)}\ \big(A(z)-1\big)^d\ \big(1+ {\cal O}(\epsilon)\big)\ .}
In particular, $T_d$ decays exponentially with $d$ as $\exp{(-d/\xi(z))}$
with a correlation length
\eqn\valxi{\xi(z)=-{1\over \log (A(z)-1)}\ .} 
If we now wish to compute the large $n$ behavior of the 
term $T_d\vert_{g^n}$, we have to extract the singular part of 
$T_d$, which requires continuing the expansion \Tdlim\ up to
order $\epsilon^{3/2}$. This yields a rather complicated and non-universal
expression which exhibits the same exponential decay in the distance $d$.

To conclude, the regime $z<1/8$ is characterized by a perimeter 
which remains finite at large $n$ and by boundary-boundary distances 
which also remain finite and are governed by a non-universal probability
law with an exponential decay. This is to be contrasted with the bulk-boundary 
distances, which scale as $n^{1/4}$ and are governed by the universal two-point
function of the Brownian map.
 
\subsec{Scaling limit: the $z>1/8$ regime}

Let us now discuss the situation $z>1/8$, and more precisely
$1/8<z<1/4$. In this regime, we have a dominant singularity of the 
generating functions at $g_{\rm crit}^{(2)}(z)$ and we 
set
\eqn\devgtwo{g=g_{\rm crit}^{(2)}(z)(1-\nu\, \epsilon)}
with $\epsilon \to 0$.
We then have the expansions
\eqn\expanfour{\eqalign{W &= 2 \left(
1- \sqrt{{1-4z \over 8z-1}}\sqrt{\nu}\, \epsilon^{1/2} + \cdots \right)\cr
x & =x_{\rm crit}(z)\ +{\cal O}(\epsilon) 
\quad  \hbox{with}\quad x_{\rm crit}(z)={16 z-1-\sqrt{3((8z)^2-1)} \over
2(1-4z)}\ .\cr}}
Here $x_{\rm crit}(z)$ is the value of $x$, as given by \Rxexplicit, for
$g=g_{\rm crit}^{(2)}(z)$. {\it Keeping $d$ finite}, 
the generating function $W_d$ has the expansion
\eqn\expWd{W_d=2 {(1-x_{\rm crit}^{d+2})(1+x_{\rm crit}^{d+2})
\over (1-x_{\rm crit}^{d+3})
(1+x_{\rm crit}^{d+1})}-2 \sqrt{{1-4z \over 8z-1}} {(1-x_{\rm crit}^{d+1})
(1-x_{\rm crit}^{d+2})^2\over (1-x_{\rm crit}^{d+3})
(1+x_{\rm crit}^{d+1})^2} 
\sqrt{\nu}\, \epsilon^{1/2} +\cdots}
with $x_{\rm crit}=x_{\rm crit}(z)$ as above. We therefore find in this case
a simple square root singularity in $(g_{\rm crit}^{(2)}-g)$. This singularity
(i.e the term proportional to $\sqrt{\nu}\, \epsilon^{1/2}$ in \expWd) 
may be obtained alternatively from the behavior of $W_d\vert_{z^p}$ 
at large $p$. Indeed, from \Wdzp, we have at large $p$
\eqn\Wdzplarge{
\eqalign{W_d\vert_{z^p}& \sim 
{(4\, R)^p \over \sqrt{\pi}\, p^{3/2}}
\left\{\!1\!-\!
(f_{d+1}\!-\!f_d)\!\sum_{k\geq 1}(2k+1)\!(f_d)^{k-1}\!\right\}\cr
&= 
{(4\, R)^p \over \sqrt{\pi}\, p^{3/2}}
{(1-x^{d+1})
(1-x^{d+2})^2\over (1-x^{d+3})
(1+x^{d+1})^2}\ .
\cr}}
Upon summing over $p$ with a weight $z^p$, the $p$-dependent prefactor
gives rise when $4\, R\, z \to 1$ (which happens precisely
when $g$ approaches $g_{\rm crit}^{(2)}(z)$ as in \devgtwo), 
to a square root singularity $-2 \sqrt{1-4\, R\, z}
\sim -2 \sqrt{(1-4z)/(8z-1)}\sqrt{\nu}\, \epsilon^{1/2}$, 
while $x$ simply tends to its value $x_{\rm crit}(z)$ at
$g=g_{\rm crit}^{(2)}(z)$. We therefore recover
the singular behavior in \expWd\ from the contribution of
large values of $p$. 

As in the previous Section, we can consider quadrangulations with a fixed and large number $n$ 
of inner faces. This is done again by performing a contour integral in
$g$ and setting
\eqn\gtozeta{g=g_{\rm crit}^{(2)}(z)\left(1+{\zeta^2\over n}\right)\ .}
We may indeed write at large $n$
\eqn\gtozetaint{\oint {dg \over 2{\rm i}\pi}{1\over g^{n+1}}\{\cdot\} \sim
{(g_{\rm crit}^{(2)}(z))^{-n} \over \pi n} 
\int_{-\infty}^\infty
d\zeta\ {\zeta \over {\rm i}}\ \ e^{-\zeta^2}\{\cdot\} 
\left(1+{\cal O}\left({\zeta^2 \over n}\right)\right)
\ .} 
Using \expWd\ with $\epsilon=1/n$ and $\nu=-\zeta^2$, we obtain that
\eqn\largeWdbis{
W_d\vert_{g^n} \sim {(g_{\rm crit}^{(2)}(z))^{-n} 
\over \sqrt{\pi} n^{3/2}}
\sqrt{{1-4z \over 8z-1}} {(1\!-\!x_{\rm crit}^{d+1})
(1\!-\!x_{\rm crit}^{d+2})^2\over (1\!-\!x_{\rm crit}^{d+3})
(1\!+\!x_{\rm crit}^{d+1})^2} \ , \quad
W\vert_{g^n}\sim {(g_{\rm crit}^{(2)}(z))^{-n} 
\over \sqrt{\pi}
 n^{3/2}}
\sqrt{{1-4z \over 8z-1}}\ .}
As for $\log W_d$, we have the expansion 
\eqn\explogWd{\log W_d=\log\left(2 {(1-x_{\rm crit}^{d+2})(1+x_{\rm crit}^{d+2})
\over (1-x_{\rm crit}^{d+3})
(1+x_{\rm crit}^{d+1})}\right) -\sqrt{{1-4z \over 8z-1}} {(1-x_{\rm crit}^{d+1})
(1-x_{\rm crit}^{d+2})\over (1+x_{\rm crit}^{d+1})
(1+x_{\rm crit}^{d+2})} 
\sqrt{\nu}\, \epsilon^{1/2} +\cdots}
with a singular part which can be alternatively read off the large $p$
behavior 
\eqn\logWdzplarge{
\eqalign{\left(\log W_d\right)\vert_{z^p}& \sim 
{(4\, R)^p \over 2 \sqrt{\pi}\, p^{3/2}}
\left\{\!1\!-\!
2\!\sum_{k\geq 1}\!\left((f_{d+1})^k-(f_d)^k\right)\!\right\}\cr
&= 
{(4\, R)^p \over 2 \sqrt{\pi}\, p^{3/2}}
{(1-x^{d+1})
(1-x^{d+2})\over (1+x^{d+1})
(1+x^{d+2})}\ .
\cr}}
We immediately deduce the leading behaviors
\eqn\largeWdbisbis{\eqalign{ 
\log W_d\vert_{g^n} & \sim {(g_{\rm crit}^{(2)}(z))^{-n} 
\over 2 \sqrt{\pi} n^{3/2}}
\sqrt{{1-4z \over 8z-1}} {(1\!-\!x_{\rm crit}^{d+1})
(1\!-\!x_{\rm crit}^{d+2})\over (1\!+\!x_{\rm crit}^{d+1})
(1\!+\!x_{\rm crit}^{d+2})} \cr
\log W\vert_{g^n} & \sim {(g_{\rm crit}^{(2)}(z))^{-n} 
\over  2 \sqrt{\pi}
 n^{3/2}}
\sqrt{{1-4z \over 8z-1}} \ . \cr } 
}
\fig{
Plots of the (non-universal) cumulative distribution function 
$\phi_z(d)$ for $z$ approaching the critical value $1/8$ from
above, namely $z=0.13$, $0.126$, $0.1255$ and $0.1251$ (dotted plots 
from left to right). 
Beside each plot, we display the corresponding (universal) 
limiting scaling form (solid line) of Eq.~(4.40).
}{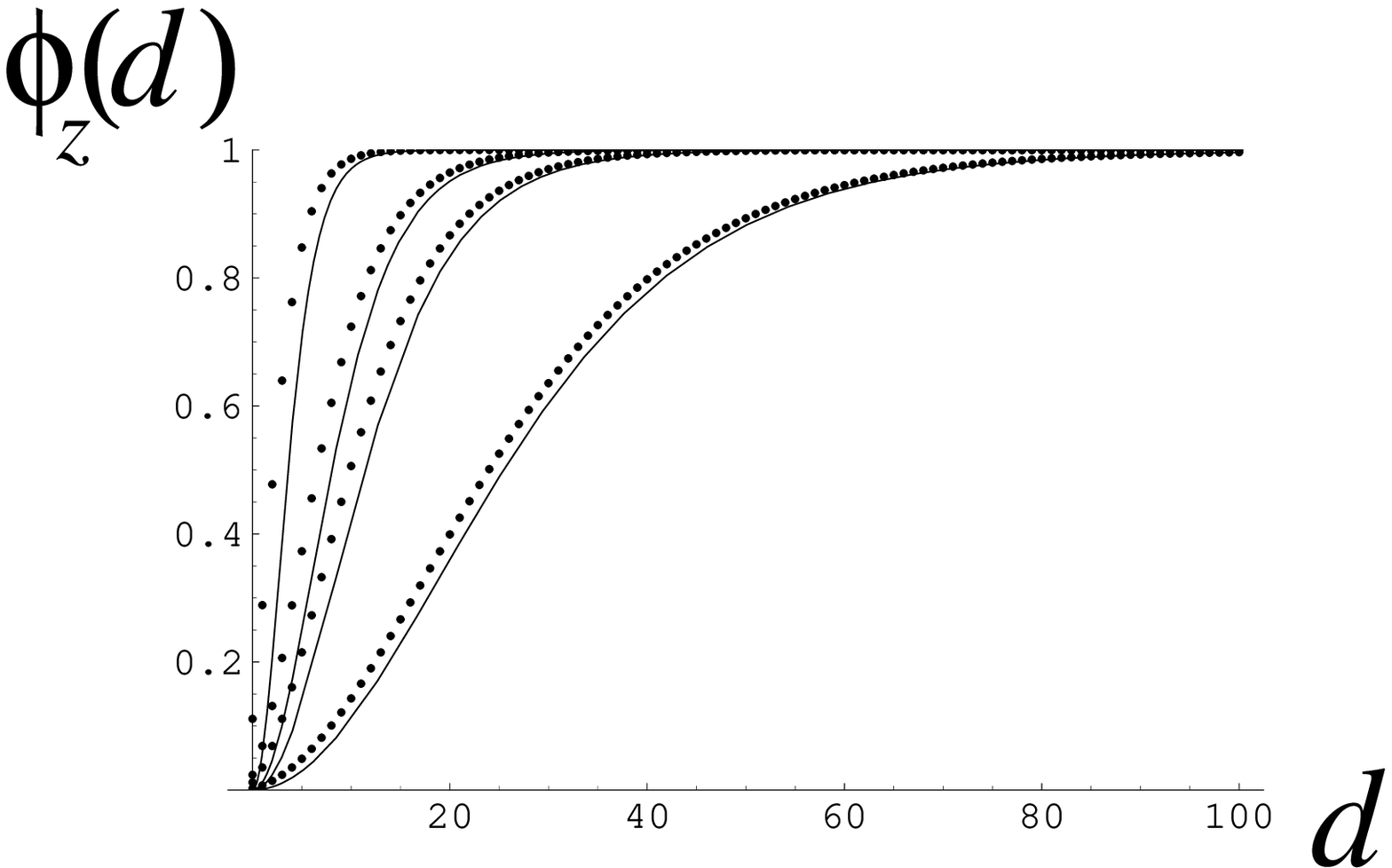}{9.cm}
\figlabel\phiofd
Taking the ratio $\log W_d\vert_{g^n}/\log W_d\vert_{g^n}$, we deduce 
the large $n$ asymptotic expression $\phi_z(d)$ for the probability 
that the distance to the boundary of the marked vertex be smaller 
than or equal to $d$ in the ensemble of pointed quadrangulations
with a boundary, namely:
\eqn\probsmaller{\phi_z(d)= 
{(1-(x_{\rm crit}(z))^{d+1})
(1-(x_{\rm crit}(z))^{d+2})\over (1+(x_{\rm crit}(z))^{d+1})
(1+(x_{\rm crit}(z))^{d+2})}\ ,}
with $d$ finite and $x_{\rm crit}(z)$ as in \expanfour.
This function is expected to be non-universal. However, 
when $z$ approaches the critical value $1/8$, we have
\eqn\appro{\phi(d) \sim \tanh^2\left(d\, \beta(z)\right)
\quad \hbox{with}\ \beta(z)=2\sqrt{3}\sqrt{z-{1\over 8}}\ ,}
and we expect that, except for the precise value of $\beta$, 
the above scaling form is universal. The function 
$\phi_z(d)$ is plotted against its scaling form \appro\ 
for $z=0.13$, $0.126$, $0.1255$ and $0.1251$ in Fig.~\phiofd.

From the exponential growth with $n$ of $W_0\vert_{g^n}$, 
and from the general formula \avebl, we now have at large $n$ 
\eqn\valavebl{\eqalign{\langle p \rangle_n(z) &  \sim 
n\times g_{\rm crit}^{(2)}(z) 
{d\ \over dz} (g_{\rm crit}^{(2)}(z))^{-1}\cr  & = n \times {8z-1 \over 1-4z}
\quad (1/8<z<1/4)\ .\cr}}

More precisely, the probability to have a prescribed value of $p$ is
proportional to $W_0\vert_{g^n z^p} z^p/(2p)$. Using the explicit
form \Wzerozpgn, and expanding it at large $n$ with $p \propto n$,
we find that, asymptotically, this probability tends to a Gaussian 
distribution peaked at $p=\langle p \rangle_n(z)$ as above, and with 
width $\sqrt{n}\ (1-4z)/\sqrt{4z}$.

Let us now consider the boundary-boundary distance statistics in
the regime $z>1/8$. From the expression \Tdtwo\ and the expansion 
\expanfour, we see that a sensible scaling limit is now obtained by taking 
$d$ large as
\eqn\newsacld{d=D \epsilon^{-1/2}}
with $D$ finite.
Using $f_1=g R^2 \sim (1-4z)/(12 z)$ at leading order in $\epsilon$
and $f_{d+1}\to x_{\rm crit}(z)$ for large $d$, we obtain that, 
in the scaling limit \newsacld, 
\eqn\Tddev{T_d \sim \left(8-{1\over z}\right) e^{
-2 \sqrt{{1-4z\over 8z-1}}\sqrt{\nu}\, D}\ .}

We can again consider the fixed $n$ ensemble by performing a contour 
integral in $g$. Taking $d=D n^{1/2}$, we now get
\eqn\Tdlargen{\eqalign{T_d\vert_{g^n} & \sim
{(g_{\rm crit}^{(2)}(z))^{-n} \over \pi n} 
\left(8-{1\over z}\right)
\int_{-\infty}^\infty
d\zeta\ {\zeta \over {\rm i}}\ \ e^{-\zeta^2
+2 {\rm i}\, \zeta\, \sqrt{{1-4z\over 8z-1}}\, D}\
\cr 
&= 
{(g_{\rm crit}^{(2)}(z))^{-n} \over \sqrt{\pi} n}
\left(8-{1\over z}\right) 
\times
\sqrt{{1-4z\over 8z-1}}\, D e^{-{1-4z\over 8z-1}\ D^2} 
\ .\cr}} 
To have a proper probability density, we must multiply this quantity
by the infinitesimal step $n^{1/2} dD$ and normalize it by
the generating function of quadrangulations with two
marked edges on the boundary, given at large $n$ by
\eqn\normade{\eqalign{2z\, {d\ \over dz} W_0\vert_{g^n} & 
\sim 2z\, {d\ \over dz} 
\left\{{(g_{\rm crit}^{(2)}(z))^{-n}
\over \sqrt{\pi} n^{3/2}}\sqrt{{1-4z\over 8z-1}}\left(1-{1\over 
2 z}\right) \right\} \cr & \sim 
{(g_{\rm crit}^{(2)}(z))^{-n}
\over \sqrt{\pi} n^{1/2}} 
\sqrt{{8z-1\over 1-4z}}\left(4-{1\over 
2 z}\right)\ .\cr }}
We obtain finally the probability density
\eqn\bounprob{\rho_{\rm bound.}(D)
= 2 D \ {1-4z\over 8z-1} \ e^{-D^2 {1-4z\over 8z-1}}\ . }
Here $\rho_{\rm bound.}(D) dD$ measures the probability that the two marked
edges on the boundary be at a distance in the quadrangulation in the 
range $[D,d+dD]$, in the ensemble of quadrangulations with two marked
edges on the boundary.
It is natural to measure the boundary-boundary distance $d$ in 
units of the square root of the average half-perimeter 
$\langle p\rangle_n(z)$, as given by \valavebl. This is done by introducing the variable
\eqn\natd{\delta\equiv {d\over \sqrt{n\, {8z-1\over 1-4z}}}
= D \sqrt{{1-4z\over 8z-1}}\ , } 
which remains a finite quantity in the scaling limit. 
The probability density for the variable $\delta$ follows simply
from \bounprob\ and reads
\eqn\prodelta{{\tilde \rho}_{\rm bound.}(\delta) = \sqrt{{8z-1\over 1-4z}}\ 
\rho_{\rm bound.} \left(\delta \sqrt{{8z-1\over 1-4z}}\right)\  
=\  2 \delta \  e^{-\delta^2}\ .}
This probability density is independent of $z$ and, as announced in Section 2,
is identical to the two-point function of the Brownian Continuum Random 
Tree \ALDOUS, i.e a simple Rayleigh law. 

To conclude, the regime $z>1/8$ is characterized by a perimeter 
which is proportional to $n$ at large $n$ and governed by 
a Gaussian law peaked at its average value \valavebl. The distance of 
a point in the bulk to this boundary remains finite and governed
by the non-universal distribution $\phi_z(d)$ above. The boundary-boundary
distance is of order $n^{1/2}$ and, when measured in natural units
given by the square root of the average perimeter, is characterized
by the universal two-point function of the Brownian Continuum Random
Tree.

\subsec{Scaling limit: the critical regime}
Let us finally discuss the vicinity of the transition point 
$z=z_{\rm crit}=1/8$. A sensible scaling limit is now obtained by 
setting
\eqn\critexp{\eqalign{ g & ={1\over 12}(1-\mu\, \epsilon) \cr
z & ={1\over 8}(1-\mu_B\, \epsilon^{1/2}) \cr
d &= D\, \epsilon^{-1/4}\cr }}
where $d$ may now stand for both the bulk-boundary distance (as in $W_d$) 
and the boundary-boundary distance (as in $T_d$). 
Using the expansions \expan\ and the expansion
\eqn\expanW{W=2\left( 1 - \sqrt{\mu_B+\sqrt{\mu}}\, \epsilon^{1/4} +\cdots \right)\ ,}
we get
\eqn\expanWd{W_d=2\left( 1- {\cal H}(D;\mu,\mu_B)\, \epsilon^{1/4} +\cdots \right)}
where
\eqn\valcalH{{\cal H}(D;\mu,\mu_B)= f(D;\mu)+{\mu_B-\sqrt{\mu}/2\over
\sqrt{\mu_B+\sqrt{\mu}}+f(D;\mu)} }
with $f(D;\mu)$ given by \deffDmu.
The scaling function ${\cal H}(D;\mu,\mu_B)$, obtained here by a direct scaling
limit of the discrete expression \Wdexpli\ for $W_d$, can be obtained 
alternatively
as the solution of a non-linear differential equation as follows:
using the expansion for $R_d$ (as obtained for instance via \Wdzpexpan\
for $p=1$ since $R_d=W_d\vert_{z^1}$)
\eqn\valcalF{R_d=2\left(1-{\cal F}(D;\mu)\, \epsilon^{1/2} +\cdots \right)}
with ${\cal F}(D;\mu)$ given by \defcalF, and expanding the recursion
relation \recurW\ at order $\epsilon^{1/2}$ with $W_d$ as in
\expanWd, we get the equation 
\eqn\recurexpan{\partial_D{\cal H}(D;\mu,\mu_B)-
{\cal H}^2(D;\mu,\mu_B)+{\cal F}(D;\mu)+\mu_B=0\ .}
The expression \valcalH\ above for ${\cal H}(D;\mu,\mu_B)$ is then 
the unique solution of this equation 
satisfying ${\cal H}(\infty;\mu,\mu_B)=\sqrt{\mu_B+\sqrt{\mu}}$, 
as required by \expanW.
We may also relate our expression to the result of Refs.~\AW\ and \AJW, 
by introducing the quantity 
\eqn\calG{{\cal G}^*(D;\mu,\mu_B)\equiv 
\partial_D\partial_{\mu_B}{\cal H}(D;\mu,\mu_B)\ .}
This is indeed the continuous counterpart of the generating function considered 
in Refs.~\AW\ and \AJW\ (in the slightly different context of triangulations)
corresponding in our language to pointed rooted maps with a boundary where 
the origin-boundary distance has a fixed value $D\epsilon^{-1/4}$ 
(hence the operator $\partial_D$) and where the root edge lies
anywhere on the boundary (hence the operator $\partial_{\mu_B}$). With
our explicit expression for ${\cal H}$, it is easy to check that ${\cal G}^*$
satisfies:
\eqn\checkAJW{\partial_D{\cal G}^*=-2\partial_{\mu_B}\left({\cal K}{\cal G}^*
\right)
\quad\hbox{with}\ {\cal K}={\cal K}(\mu,\mu_B)\equiv \left(\mu_B-{\sqrt{\mu}
\over 2}\right) \sqrt{\mu_B+\sqrt{\mu}}\ .}
This is precisely the equation used in Refs.~\AW\ and \AJW\ to determine
${\cal G}^*$ and the two-point function. 

We have finally the expansion
\eqn\expanlogWd{\log W_d=\log(2)-{\cal H}(D;\mu,\mu_B)\, \epsilon^{1/4} +\cdots }

The regime \critexp\ corresponds to typical values of the 
perimeter of order $\epsilon^{-1/2}$. Rather than fixing $\mu_B$,
we may alternatively work directly with a fixed value of the half-perimeter 
$p$ being of order $\epsilon^{-1/2}$, i.e consider 
our fixed length generating functions of Section 3.3 and set
\eqn\scalfixp{\eqalign{g& ={1\over 12}(1-\mu \epsilon)
\cr p &= P\, \epsilon^{-1/2}\cr d &=D\, \epsilon^{-1/4}
\ . \cr}}
Upon setting $k=K\, \epsilon^{-1/4}$, 
the expression \Wdzp\ translates into 
\eqn\Wdzpexp{\eqalign{& {W_d\vert_{z^p}\over 8^p}
\sim\ \epsilon^{3/4} {\bar{\cal H}}(D,P;\mu) \quad \hbox{with} 
\cr &  {\bar{\cal H}}(D,P;\mu)  =
{e^{-\sqrt{\mu} P}\over \sqrt{\pi} P^{3/2}}
\left\{1+
\left(3\sqrt{\mu}-2 f^2(D;\mu)\right)\ 
\int_0^\infty dK\, e^{-{K^2 \over P}-2 f(D;\mu) K} 2K
\right\}\ . \cr}}
The scaling functions ${\cal H}$ 
and ${\bar{\cal H}}$ are then simply related by 
\eqn\HHrel{{\cal H}(D;\mu,\mu_B)=\sqrt{\mu_B+\sqrt{\mu}}+ 
{1\over 2}\int_0^\infty dP\ e^{-\mu_B P}
\left( {e^{-\sqrt{\mu} P} \over \sqrt{\pi}P^{3/2}} -{\bar{\cal H}}(D,P;\mu)
\right)\ .}
For small $P$, we have in particular
\eqn\Wdzpexpsmallp{\eqalign{{\bar{\cal H}}(D,P;\mu) &
{\buildrel {P\to 0} \over \sim}\ {1 \over \sqrt{\pi} P^{3/2}}
(1-\sqrt{\mu}\, P+\cdots)\left(1+ P (3\sqrt{\mu}-2 f^2(D;\mu)) 
+\cdots \right) 
\cr & \sim {1 \over \sqrt{\pi} P^{3/2}}(1- P {\cal F}(D;\mu) +\cdots )
\cr}}
Note that, as could be expected, this small $P$ behavior matches precisely the 
large $p$ behavior of $W_d\vert_{z^p}$ obtained in the regime $z<1/8$ for
finite $p$, as given by \Wdzpexpan.

\noindent For large $P$, we get instead
\eqn\Wdzpexplargep{\eqalign{{\bar{\cal H}}(D,P;\mu) &
{\buildrel {P\to \infty} \over \sim} {e^{-\sqrt{\mu}\, P} 
\over \sqrt{\pi} P^{3/2}}
\left(1+ {3\sqrt{\mu}-2 f^2(D;\mu)\over 2 f^2(D;\mu)}\right)
\cr & = {e^{-\sqrt{\mu}\, P} \over \sqrt{\pi} P^{3/2}}
\tanh^2\left(\sqrt{3\over 2}\, \mu^{1/4}\, D\right)\ .
\cr}}
Again, as expected, this expression matches precisely
that obtained for $W_d\vert_{z^p}$ in the regime $z>1/8$, 
as given by \Wdzplarge, upon considering the scaling
limit $p=P \epsilon^{-1/2}$, $R=2(1-\sqrt{\mu} \epsilon^{1/2})$, 
$d=D\epsilon^{-1/4}$ and $x=1- \sqrt{6} \mu^{1/4} \epsilon^{1/4}$.
As for $\log W_d$, since, up to a factor $1/2$, it has the same singularity 
than $W_d$, we have 
\eqn\logWdzpexp{{\log W_d\vert_{z^p}\over 8^p}
\sim\ \epsilon^{3/4} {1\over 2}{\bar{\cal H}}(D,P;\mu)\ .}
This behavior can alternatively be obtained by taking directly 
the scaling limit of \logWdzp.

We can now turn to the fixed $n$ ensemble, with $n$ large  
and in the critical scaling regime 
\eqn\scalnp{p= P\ n^{1/2}}
with $P$ finite. We then get a cumulative distribution function
\eqn\cumdi{{\bar \Phi}(D,P)=2\sqrt{P}\ e^{P^2/4}\int_{-\infty}^{\infty}
d\xi\ {\xi\over {\rm i}}\ e^{-\xi^2}\ {\bar{\cal H}}(D,P;-\xi^2)\ ,}
which measures the probability that a vertex chosen uniformly at random
in the quadrangulation be at a rescaled distance less than $D$ 
from the boundary. 
This distribution function is plotted in Fig~\PhiofP\ for
$P=0.01$, $0.1$, $0.5$, $1.0$, $2.0$ and $5.0$. 
For fixed $P$ and small $D$, we have the expansion
\eqn\smallDbar{{\bar \Phi}(D,P)={3\over 4} P\, D^2-
{3\over 8}(P^2-1) D^4 +\cdots}
\bigskip
For fixed $D$ and small $P$, 
using \Wdzpexpsmallp, we immediately see that ${\bar \Phi}(D,P)\to \Phi(D)$
as expected. This property is illustrated in Fig.~\PhiofP.
Note that the small $D$ and small $P$ limits do not commute. 

On the other hand, when $P$ is large, using \Wdzpexplargep\
and evaluating the integral over $\xi$ by a saddle point estimate, 
we deduce that 
\fig{
Plots of the cumulative distribution function ${\bar{\Phi}}(D,P)$ as
a function of the scaling variable $D\sqrt{P}$ in the regime of
large values of $P$ (thin solid lines) and their comparison with
the limiting scaling form of Eq.~(4.67) (thick blue line). 
The plots represented here are for $P=1.$, $2.$
and $5.$, from left to right.}{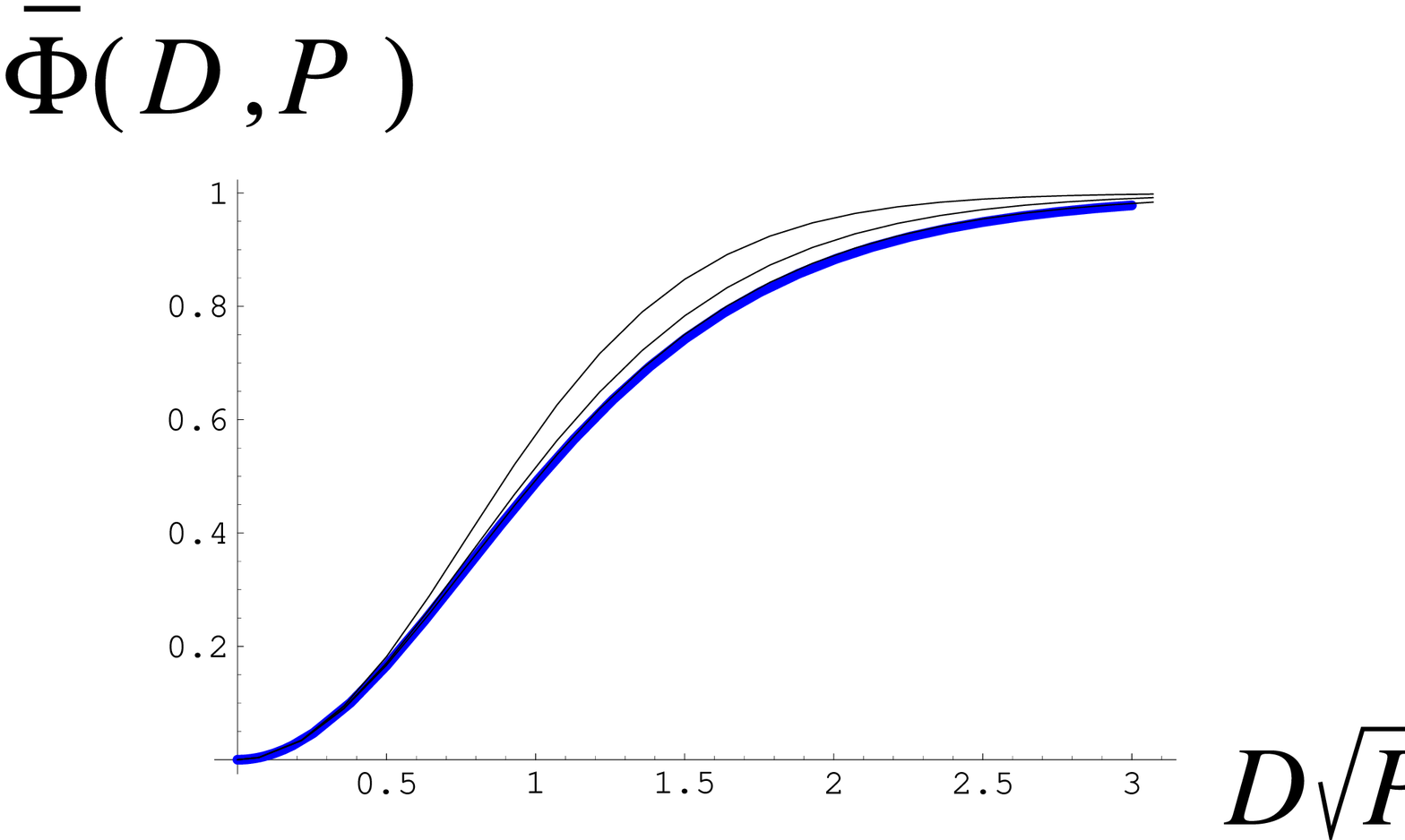}{10.cm}
\figlabel\PhiofPscal
\eqn\PhilimlargeP{{\bar{\Phi}}(D,P){\buildrel {P\to \infty}
\over \rightarrow} \tanh^2\left({\sqrt{3} \over 2} D \ \sqrt{P}\right) \ .}
This property is illustrated on Fig.~\PhiofPscal\ for $P=1.$, $2.$ 
and $5.$.

Concerning the boundary-boundary distance, using the expression 
\Tdzp, we get in the scaling regime \scalfixp\ the limiting expression 
for $T_d\vert_{z^p}$:
\eqn\limitTd{{T_d\vert_{z^p} \over 8^p} \sim
{8\over 3} \epsilon {e^{-\sqrt{\mu}\, P-{D^2\over P}} \over \sqrt{\pi}
P^{7/2}} \left\{ (2 D^3\!-\!3 D\, P)\!+\!(4 D^2 P\!-\!2 P^2) f(D;\mu)\!+\!
2 D P^2 f^2(D;\mu) \right\}\ .}
It is natural to measure the boundary-boundary distances in units
of $\sqrt{p}$, i.e write $d=\delta \sqrt{p}$ or equivalently
\eqn\defdelta{D=\delta \sqrt{P}\ .}
In the variable $\delta$, Eq.~\limitTd\ translates into
\eqn\limitTd{{T_d\vert_{z^p} \over 8^p} \sim
{8\over 3} \epsilon {e^{-\sqrt{\mu}\, P-\delta^2} \over \sqrt{\pi}
P^2} \left\{ (2 \delta^3\!-\!3 \delta)\!+\!(4 \delta^2 \!-\!2 ) \sqrt{P} 
f(\sqrt{P}\delta ;\mu)\!+\!
2 \delta \left(\sqrt{P} f(\sqrt{P}\delta;\mu)\right)^2 \right\}\ .}
At small $P$, we have the expansion
\eqn\TdzpsmallP{{T_d\vert_{z^p}\over 8^p}
{\buildrel {P\to 0} \over \sim} 
{8\over 3}\epsilon {e^{-\delta^2} \over \sqrt{\pi}
P^2}\left\{(\delta\!+\!2\delta^3)\!-\!{5\delta\!+\!6\delta^3\!+\!
2\delta^5\over 10}
\mu P^2\!+\!{35\delta\!+\!28 \delta^3\!+\!12\delta^5\!+\!3\delta^7
 \over 105}\mu^{3/2}P^3+\cdots
\right\}}
while at large $P$, we have 
\eqn\TdzplargeP{{T_d\vert_{z^p}\over 8^p}
{\buildrel {P\to \infty} \over \sim} 4 \sqrt{\mu}\, \epsilon\, {e^{-\sqrt{\mu}\, P} \over \sqrt{\pi}
P}\left\{2\delta \ e^{-\delta^2}+ \cdots \right\}}

We turn finally to fixed values of $n$ and $p$, 
in the critical scaling regime $n\to\infty$ with the ratio $P=p/n^{1/2}$
fixed.
Using \limitTd\ with $\epsilon=1/n$ and $\mu=-\xi^2$, multiplying 
by the elementary step $n^{1/4}dD$ and normalizing
by $2p W_0\vert_{g^n z^p}$, which behaves as
\eqn\scalwo{{2p W_0\vert_{g^n z^p}\over 8^p} \sim 
{12^n \over \pi n^{7/4}}\ e^{-P^2/4}\ P^{3/2}\ ,}
we get a critical probability density for the rescaled 
boundary-boundary distance $D=d\cdot n^{-1/4}$:

\eqn\prdfixP{\eqalign{{\bar \rho}_{\rm bound.}(D,P)& ={4\over 3 P^4}
e^{-D^2/P} 
\left\{(2D^3\!-\!3DP)\!+\!(4D^2P\!-2\!P^2) \sigma_1(D,P)\!+2\!DP^2
\sigma_2(D,P) \right\}\cr
{\rm with} \ \ &\sigma_1(D,P)=
{2 e^{P^2/4} \over \sqrt{\pi} P}
\int_{-\infty}^{\infty}d\xi\ {\xi\over {\rm i}}\ e^{-\xi^2+{\rm i}\, \xi\, P}
f(D;-\xi^2)
 \cr
&\sigma_2(D,P)= 
{2 e^{P^2/4} \over \sqrt{\pi} P}
\int_{-\infty}^{\infty}d\xi\ {\xi\over {\rm i}}\ e^{-\xi^2+{\rm i}\, \xi\, P}
f^2(D;-\xi^2)
\ .\cr}}
As before, it is natural to measure the boundary distances in units
of $\sqrt{p}$, i.e write $d=\delta \sqrt{p}$ or equivalently
$D=\delta \sqrt{P}$. We find for $\delta$ a probability density
\eqn\prdeltafixP{\eqalign{{\tilde \rho}_{\rm bound.}(\delta,P)& 
={4\over 3 P^2}
e^{-\delta^2} 
\left\{(2\delta^3\!-\!3\delta)\!+\!(4\delta^2\!-\!2) 
{\tilde\sigma}_1(\delta,P)\!+\!2\delta
{\tilde \sigma}_2(\delta,P) \right\}\cr
{\rm with} \ \ &{\tilde \sigma}_1(\delta,P)=
{2 e^{P^2/4} \over \sqrt{\pi}\sqrt{P}}
\int_{-\infty}^{\infty}d\xi\ {\xi\over {\rm i}}\ e^{-\xi^2+{\rm i}\, \xi\, P}
f(\delta\sqrt{P};-\xi^2)
 \cr
&{\tilde \sigma}_2(\delta,P)= 
{2 e^{P^2/4} \over \sqrt{\pi}}
\int_{-\infty}^{\infty}d\xi\ {\xi\over {\rm i}}\ e^{-\xi^2+{\rm i}\, \xi\, P}
f^2(\delta\sqrt{P};-\xi^2)
\ .\cr}}
This probability density is plotted in
Fig.~\rhotildeboundP\ for $P=0.5$, $1.0$, $1.5$, $2.0$, $3.0$, $5.0$
and $10.0$.

At small $P$, we find in particular from \TdzpsmallP\ that 
\eqn\prdeltasmallP{{\tilde \rho}_{\rm bound.}(\delta,P)
{\buildrel {P\to 0} \over \to} {2\over 105}
e^{-\delta^2} (35\delta+28 \delta^3+12 \delta^5+3 \delta^7)\ .} 
At large $P$, we have the simple result
\eqn\prdeltalargeP{{\tilde \rho}_{\rm bound.}(\delta,P)
{\buildrel {P\to \infty} \over \to} 2 \ \delta\ 
e^{-\delta^2} \ .}

Repeating the above analysis for the refined generating 
function $T_d(s,s')$ of Eq.~\Tddes, we have access to the
refined probability ${\tilde \rho}_{\rm bound.}(\delta,u,P)$
that the vertex of the boundary at rescaled distance 
$2 P u$ ($0\geq u\geq 1$) {\it along the boundary} from a
given vertex (chosen uniformly at random on the boundary) be at
rescaled distance $\delta \sqrt{P}$ 
{\it in the quadrangulation} from this vertex. 
This probability density reads
\eqn\prdeltaufixP{\eqalign{{\tilde \rho}_{\rm bound.}(\delta,u,P)
& ={1\over 6\sqrt{\pi} P^2}{
e^{-{\delta^2\over 4u(1\!-\!u)}}\over u^{5/2}(1\!-\!u)^{5/2}}
\Big\{(\delta^2\!-\!2u)(\delta^2\!-\!2(1\!-\!u))\cr
&\ \ \ +\!2\delta(\delta^2\!-\!4u(1\!-\!u))
{\tilde\sigma}_1(\delta,P)\!+\!4\delta^2 u(1\!-\!u)
{\tilde \sigma}_2(\delta,P) \Big\}\ .\cr}}
We have in particular, for small $P$,
\eqn\prdeltausmallP{{\tilde \rho}_{\rm bound.}(\delta,u,P)
{\buildrel {P\to 0} \over \to} {1\over 6\sqrt{\pi}}{
e^{-{\delta^2\over 4u(1\!-\!u)}}\over u^{5/2}(1\!-\!u)^{5/2}}
{\delta^4 \over 140}
\Big\{70\!+\!21\delta^2
\!+\!3\delta^4\!-\!42u(1\!-\!u)(\delta^2\!+\!5)
\Big\}}
while, for large $P$, 
\eqn\prdeltaulargeP{{\tilde \rho}_{\rm bound.}(\delta,u,P)
{\buildrel {P\to \infty} \over \to} {1\over 2\sqrt{\pi}}{\delta^2\ 
e^{-{\delta^2\over 4u(1\!-\!u)}}\over u^{3/2}(1\!-\!u)^{3/2}}\ .}
\fig{The average value $\langle \delta(u) \rangle$ (measured in units
of $\sqrt{P}$) for the distance in the quadrangulation of two boundary 
points at rescaled distance $u$ (measured in units of $2P$) along the
boundary, here for $P=0.5$, $1.0$, $1.5$, $2.0$, $3.0$, $5.0$ and $10.0$ 
(thin lines from top to bottom). The plots interpolate between two
limiting laws (thick lines): the non-trivial (but universal) law of
Eq.~(4.82) for $P\to 0$ and the simple semi-circle law of Eq.~(4.83) 
for $P\to \infty$.
}{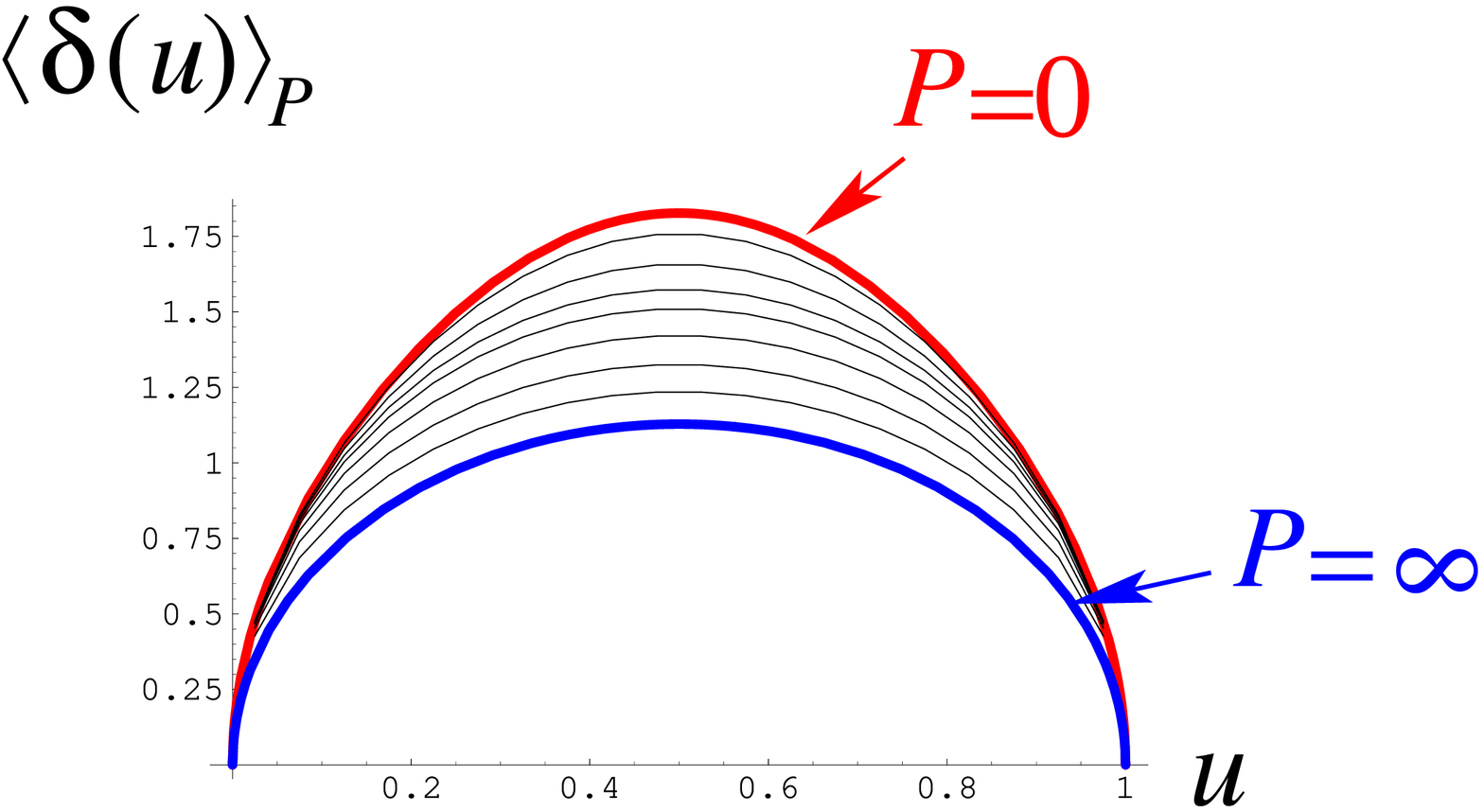}{10.cm}
\figlabel\avedetu
A simpler characterization of the distance in the quadrangulation
of two boundary vertices at distance $2 P u$ along the
boundary is through the corresponding average 
value 
\eqn\profil{\langle \delta(u) \rangle_P 
\equiv \int_0^\infty\ d\delta\ \delta\ {\tilde \rho}_{\rm bound.}(\delta,u,P)
\ .}
This average value is plotted in Fig.~\avedetu\ for
$P=0.5$, $1.0$, $1.5$, $2.0$, $3.0$, $5.0$ and $10.0$.
For $P$ small, we get in particular
\eqn\smallPavedelt{\langle \delta(u) \rangle_P
{\buildrel {P\to 0} \over \to}
{16\over 105} \sqrt{{u(1-u)\over \pi}}\Big\{
35+21u(1-u)+36 u^2(1-u)^2\Big\}}
while, for $P$ large, we have
\eqn\largePavedelt{\langle \delta(u) \rangle_P
{\buildrel {P\to \infty} \over \to}
4 \sqrt{{u(1-u)\over \pi}}\ .}
\bigskip
To conclude, quadrangulations with a boundary display an interesting 
scaling behavior when both the number $n$ of inner faces and
the length $2p$ of the boundary become large, keeping the ratio 
$P=p/n^{1/2}$ fixed. In this regime, the bulk-boundary distances
scale as $n^{1/4}$ and are characterized by a {\it universal distribution 
function} ${\bar \Phi}(D,P)$ which interpolates between the (cumulative)
two-point function of $\Phi(D)$ the Brownian map for small $P$
to a simple $\tanh^2$ function in the variable $D\sqrt{P}$ at large $P$.
As for the boundary-boundary distances, they scale as the 
square root of the perimeter and are characterized by 
a {\it universal probability density} ${\tilde \rho}_{\rm bound.}(\delta,P)$
which interpolates between the two-point function of the 
Brownian Continuum Random Tree when $P$ is large to the small
$P$ expression \prdeltasmallP. Note that this latter formula, even if it 
looks rather involved, is expected to be universal, and so are the formulas 
\prdeltausmallP\ and \smallPavedelt. 

\newsec{Self-avoiding boundary}

\subsec{Generating functions}
So far, we considered quadrangulations whose boundary may contain 
separating vertices or edges, corresponding to vertices or edges
encountered several times along the contour. We may instead 
consider quadrangulations {\it with a self-avoiding boundary},
i.e demand that the $2p$ vertices (and consequently the $2p$ edges)
along the contour be all distinct. We shall denote by 
${\tilde W}_d$ the generating function for quadrangulations with a 
self-avoiding boundary having a marked vertex at distance 
{\it smaller than or equal to $d$} from the boundary, and as before with 
a marked "closest edge", i.e a boundary edge incident to a vertex
at minimal distance from the marked vertex and oriented 
counterclockwise around the bulk of the quadrangulation. As before,
when $d=0$, these markings reduce to the choice of a boundary edge oriented
clockwise around the bulk.  
\fig{A schematic picture of the Eq.~(5.1). In any rooted 
quadrangulation with a boundary, the root edge selects a particular
irreducible component (in magenta) which may be either a single
edge (a) or an irreducible component with non-zero area and 
self-avoiding boundary of perimeter $2p_0$. The other irreducible components
may be reassembled into (possibly empty) quadrangulations with
a generic boundary attached to each of the two endpoints of the
edge (case (a)) or to each of the $2p_0$ boundary vertices of the selected 
irreducible component (case (b)). These attached quadrangulations are 
naturally rooted and each of them gives rise to a factor
$W_0(g,z)$.}{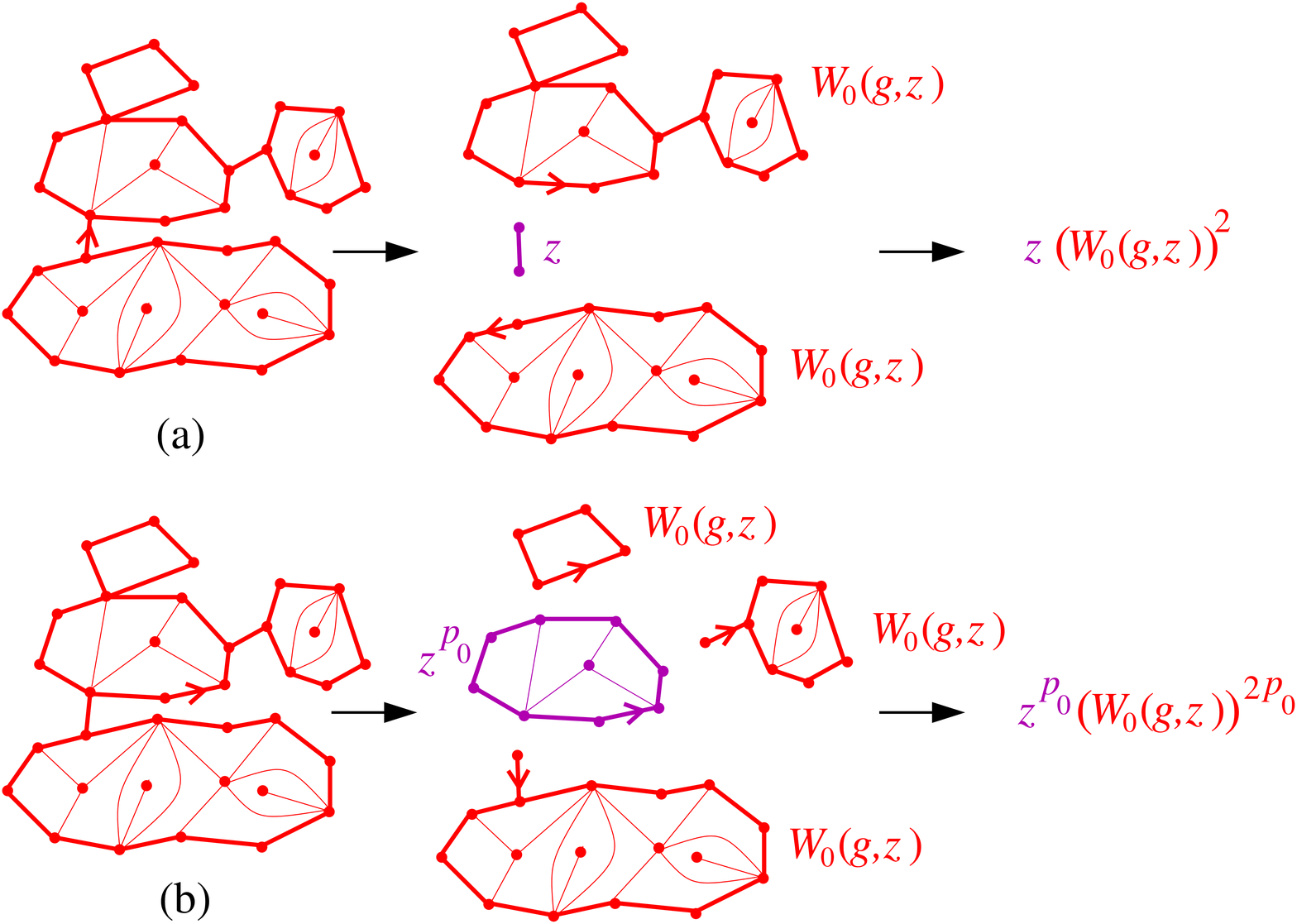}{12.cm}
\figlabel\irreducible
In ${\tilde W}_d$,
we weight each configuration by a factor $g$ per inner face and a factor
$\sqrt{Z}$ per edge of the boundary. For convenience, we decide to
also add to the configurations counted by ${\tilde W}_d$ the empty
configuration (weight $1$) and the configuration with $p=1$ and 
$n=0$ (weight $Z$) corresponding to a single edge embedded in the
external face, although the boundary is not 
self-avoiding in this case. In other words, we take the convention that
${\tilde W}_d\vert_{g^0}=1+Z$ in the following. 
With this convention, it is easy 
to check that we have the combinatorial identity
\eqn\wwtilde{W_0(g,z)= {\tilde W}_0(g,Z)\quad \hbox{with}
\quad Z=z W_0^2(g,z)\ .}
Indeed, considering the root edge in any (non-empty) 
configuration counted by $W_0$, with the external face on its 
right, it belongs to some irreducible component which is either a 
quadrangulation
of non-zero area, with a self-avoiding boundary of length, say $2p_0$ 
with $p_0\geq 1$, or a single edge ($p_0=1$) if the root edge
is a separating edge (see Fig.~\irreducible\ for an illustration). 
The configurations in the latter case are counted by $z W_0^2(g,z)$, 
while those in the former case are counted by 
$({\tilde W}_0-1-Z)\vert_{Z^{p_0}}\times  \left(z W_0^2\left(g,z\right)
\right)^{p_0}$ since we may attach to each vertex of the self-avoiding boundary 
of the irreducible component a configuration counted by
$W_0(g,z)$. Summing over $p_0$, we end up with the relation \wwtilde.

\fig{A schematic picture of the Eq.~(5.2). In any pointed-rooted 
quadrangulation with a boundary counted by $W_d-W_0$, the origin
vertex lies strictly in the bulk of some particular irreducible 
component (in magenta), with self-avoiding boundary of perimeter
$2p_0$. The other irreducible components may be reassembled into $2p_0$
naturally rooted quadrangulations with a generic boundary attached 
to the $2p_0$ boundary vertices of the selected irreducible component.
One of these quadrangulations contains the originally marked closest
edge, inducing a natural splitting in two parts (in dark red) 
which results in an extra factor $W_0(g,z)$.
}{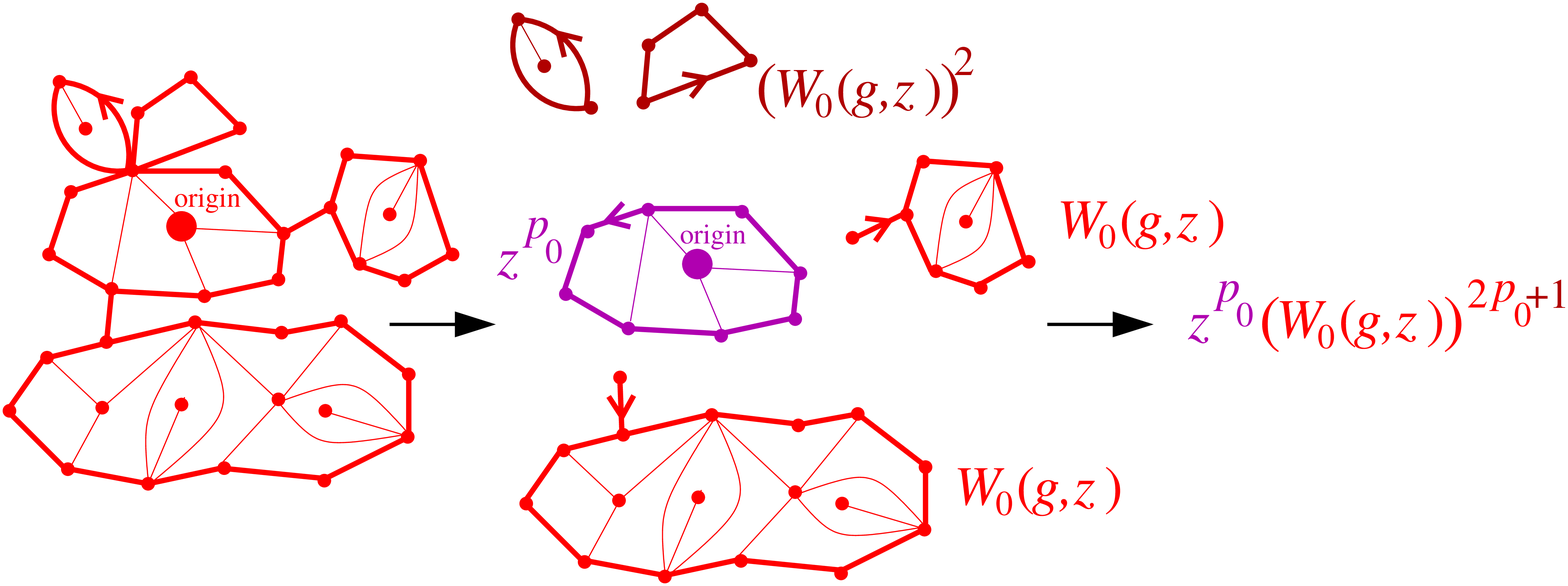}{13.cm}
\figlabel\irreduciblebis
For $d>0$, we have the slightly different relation
\eqn\wwtilded{W_d(g,z)-W_0(g,z)=W_0(g,z)\left({\tilde W}_d\left(
g,Z\right)-{\tilde W}_0\left(
g,Z\right)\right) \quad \hbox{with}\quad Z=z W_0^2(g,z)\ .}
Indeed, $W_d-W_0$ counts configurations with a marked vertex strictly
in the bulk of the quadrangulation, at distance less than or equal to $d$
and with a marked closest edge. The marked vertex belongs to
one particular irreducible component and its distance to the whole boundary
is equal to its distance to the self-avoiding boundary of this 
component. The most general configuration is then obtained
by attaching to each vertex of the self-avoiding boundary
of the irreducible component a configuration counted by
$W_0(g,z)$, leading again to the effective weight $Z=z W_0^2$ in
the generating function for the irreducible component (see 
Fig.~\irreduciblebis\ for an illustration). 
Finally, the marked closest edge on the original 
configuration starts form a particular closest {\it vertex} on the
self-avoiding boundary which in turns selects a closest edge
on the self-avoiding boundary, leading eventually to ${\tilde W}_d
-{\tilde W}_0$. The marking of the original closest edge
induces a splitting in two parts of the contour of the configuration 
attached to the marked closest vertex on the self-avoiding boundary, 
resulting in an extra multiplicative factor $W_0(g,z)$ in \wwtilded.

\fig{A schematic picture of the decomposition of a pointed quadrangulation
with a self-avoiding boundary, whose origin lies strictly in the bulk
and with $k$ closest vertices on the boundary (here $k=7$), into
$k$ slices, as defined in the text. From each closest vertex on the boundary 
(red dots), we draw the leftmost geodesic path to the origin (blue circle).
Note that these geodesics cannot cross but may merge before reaching the origin
(as in $v_3$). This splits the quadrangulation into $k$ triangular slices
whose depths are generally different, with the maximal depth being equal
to the origin-boundary distance.
}{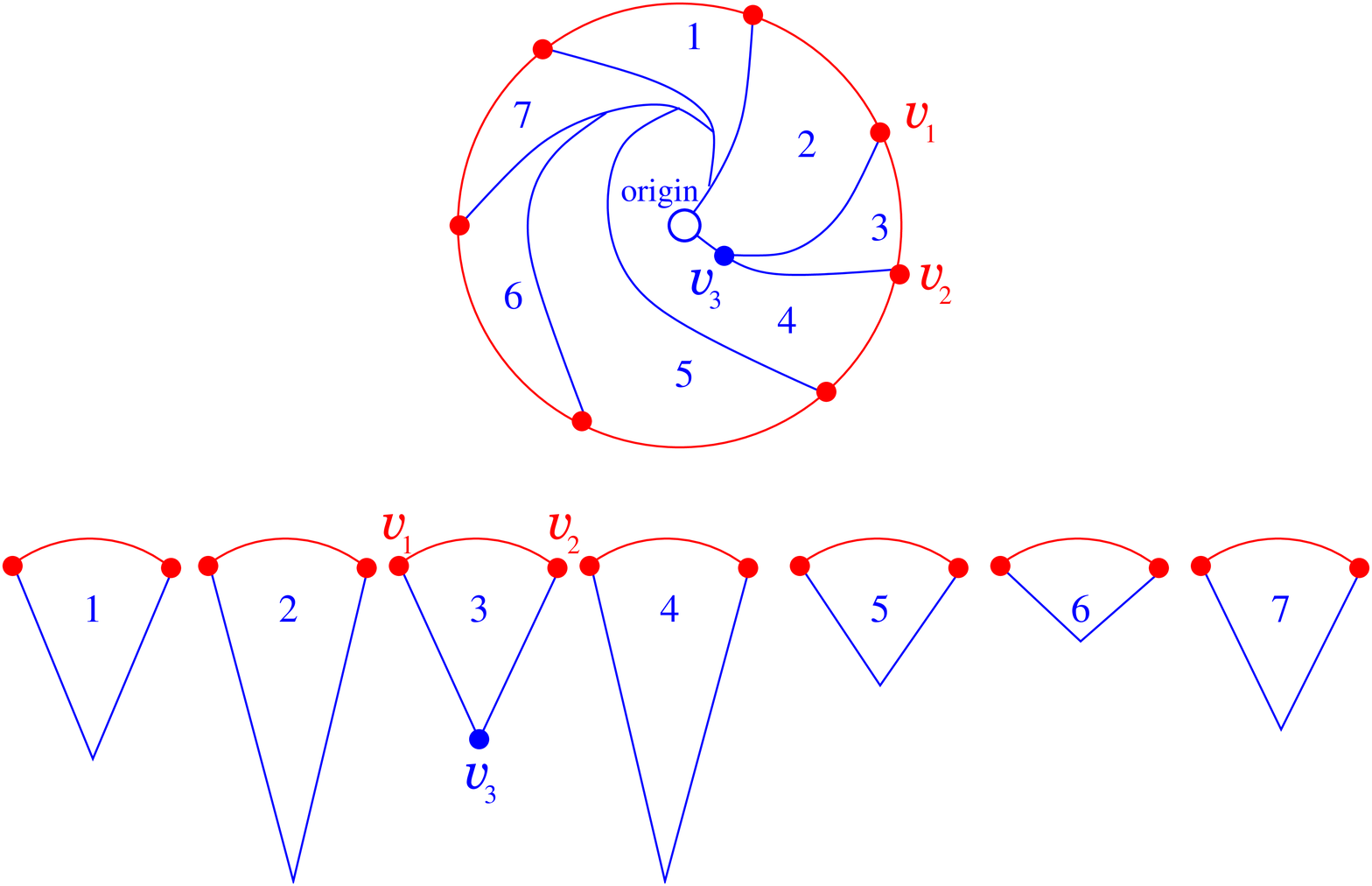}{12.cm}
\figlabel\slicing

It is more natural to consider configurations without a marked closest
edge. In the case of a non self-avoiding boundary, this was done by
changing $W_d$ into $\log W_d$. Here, we find that suppressing the
marking of the closest edge results in replacing the generating
function ${\tilde W}_d$ by a modified generating function
\eqn\noclosest{{\tilde W}_d'\equiv
{\tilde W_0}-1+\log\left({\tilde W}_d-({\tilde W}_0-1)\right)\ .}
A proof of this relation is sketched as follows. It is equivalent to
exhibit a generating function $S_d$ such that:
\eqn\sdst{\eqalign{{\tilde W}_d - (\tilde{W}_0 - 1) &=
\sum_{k \geq 0} (S_d)^k \cr {\tilde W}'_d - (\tilde{W}_0 - 1) &=
\sum_{k \geq 1} {1 \over k} (S_d)^k. \cr}}
Combinatorially, both left hand sides are generating functions for
pointed quadrangulations with a self-avoiding boundary, such that the
origin-loop distance is between $1$ and $d$, with an additional
marking of a closest boundary vertex in the first case (note that
the marking of an edge or of a vertex are equivalent for self-avoiding
boundaries). Clearly, we shall look for a decomposition according 
to the number $k$ of such
closest vertices.  A possible way is to decompose the map into $k$
``slices'' by cutting along each {\it leftmost geodesic} from a
closest vertex to the origin, see Fig.~\slicing. Each slice is itself
a quadrangulation with a self-avoiding boundary, and has three
distinguished vertices on the boundary: two of them, say $v_1$ and
$v_2$, correspond to two clockwise consecutive closest boundary
vertices in the original map and the third, say $v_3$, is the point
where their outgoing leftmost geodesics merge (note that $v_3$ is not
necessarily the origin). Furthermore, the slice contour consists of
three segments $(v_1v_2)$, $(v_2v_3)$ and $(v_3v_1)$ of non-zero
length, and the following properties hold within the slice:
\item{-} $(v_3v_1)$ and $(v_2v_3)$ are geodesics of the same length
$\ell$, hereafter called the depth of the slice, with $1 \leq \ell \leq d$,
\item{-} there is no other geodesic from $v_1$ to $v_3$ (due to our
convention of cutting along leftmost geodesics),
\item{-} all vertices on $(v_1v_2)$ distinct from $v_1$ and $v_2$ are
at distance strictly larger than $\ell$ from $v_3$.
\par
\noindent Let $S_d$ be the generating function for slices satisfying
the above characterization, with a weight $g$ per face and $\sqrt{Z}$
per edge of $(v_1v_2)$. Then, we have a bijective decomposition of the
original map into a linear or cyclic sequence of slices, depending
whether a closest boundary vertex is marked or not. 
Note that the slices obtained in the decomposition do not necessarily have
the same depth, and that the maximal depth is equal to the origin-loop
distance. Omitting further details of the
proof, the identities \sdst\ are established.

The generating function for pointed quadrangulations with
a self-avoiding boundary with origin-boundary distance equal to
$d$ is
\eqn\tildeGW{{\tilde G}_d={\tilde W}'_d -{\tilde W}'_{d-1}, \quad
{\tilde G}_0={\tilde W}'_0={\tilde W}_0-1\ ,}
which yields the formula \sabulkbouncorr\ announced in Section 2.

Let us now derive explicit expressions for ${\tilde W}_d$ and
${\tilde W}'_d$. 
From \wwtilde\ and from the general expression \expwzero\ for $W_0$, 
we immediately deduce that
\eqn\expwzerotild{{\tilde W}_0(g,Z)={\tilde W}(1-f_1  \ 
({\tilde W}-1))\ ,}
where
\eqn\tildwZ{\eqalign{& {\tilde W}\equiv 
{\tilde W}(g,Z) =  W(g,z(g,Z))\quad\hbox{with 
$z(g,Z)$ given implicitly by} \cr &
Z=z(g,Z)W_0^2(g,z(g,Z))=z(g,Z) {\tilde W}_0^2(g,Z)\cr}}
Writing these equations as
\eqn\tildwZbis{\eqalign{&{\tilde W}=1+z(g,Z)\,R\,{\tilde W}^2\cr
& Z=z(g,Z)\left({\tilde W} \left(1-f_1 
({\tilde W}-1)\right)\right)^2\ ,\cr}}
and eliminating $z(g,Z)$, we deduce the relation 
\eqn\ZWtilderel{Z={{\tilde W}-1\over R}(1-f_1
({\tilde W}-1))^2}
which determines ${\tilde W}$ as a function of $g$ and $Z$.
As for ${\tilde W}_d$ for $d>0$, we deduce from \wwtilded\ and from the general 
expression \Wdexpli\ for $W_d$ that
\eqn\Wdtildeexpli{{\tilde W}_d(g,Z) ={1\over 
1-f_1 ({\tilde W}-1)}\times{1-({\tilde W}-1) f_{d+1}\over 1-({\tilde W}-1) f_d}
+{\tilde W}_0-1}
with ${\tilde W}$ related to $Z$ as in \ZWtilderel\ above.
Note that the quantity ${\tilde W}_d$ depends on $Z$ only via 
${\tilde W}$ but that, unlike generic boundaries, ${\tilde W}$ is
not the limit of ${\tilde W}_d$ for $d\to \infty$. Finally, we have
\eqn\Wdprim{{\tilde W}_d'(g,Z) =\log\left({1\over 
1-f_1 ({\tilde W}-1)}\times{1-({\tilde W}-1) f_{d+1}\over 1-({\tilde W}-1) f_d}
\right)
+{\tilde W}_0-1\ .}

As before, ${\tilde W}_d$ and ${\tilde W}'_d$ are Lagrangean generating
functions and we may extract an explicit expression
for their $Z^p$ term. Writing \ZWtilderel\ as 
\eqn\Ztowtilde{Z= {{\tilde w}\over R}(1-f_1 {\tilde w})^2}
upon introducing the quantity 
\eqn\smallwtildedef{{\tilde w}\equiv {\tilde W}-1\ ,}
we may easily transform any contour integral in the 
variable $Z$ into a contour integral in the 
variable ${\tilde w}$, namely
\eqn\chgvartilde{\oint {dZ \over 2 {\rm i} \pi} {1\over Z^{p+1}}
\Big\{ \cdot \Big\} = R^p\, \oint {d{\tilde w} \over 2{\rm i} \pi}
{1\over {\tilde w}^{p+1}} {1-3 f_1 {\tilde w}\over 
(1-f_1{\tilde w})^{2p+1}} \Big\{ \cdot \Big\}\ .}
Using \expwzerotild, which we write as  
\eqn\expwzerotildbis{{\tilde W}_0=(1+{\tilde w})(1-f_1 
{\tilde w})\ ,}
and $f_1=g\, R^2$, we obtain
\eqn\expwzerotilzp{{\tilde W}_0\vert_{Z^p}=(g R^3)^p {(3p-3)!\over p!(2p-1)!}
({p\over  g\, R^2}+2-3p)} 
for $p\geq 1$, and ${\tilde W}_0\vert_{Z^0}=1$.
Extracting the $g^n$ term of the right hand side, we deduce the equivalent
formula
\eqn\expzerotildezpgn{{\tilde W}_0\vert_{g^n Z^p}=3^{n-p} {(3p)!\over p!(2p-1)!}{(2n+p-1)!\over (n-p+1)!(n+2p)!}\ ,}
valid for all $n\geq 0$, $1\leq p\leq n+1$.

As for ${\tilde W}_d$, we get upon expanding \Wdtildeexpli\ 
in ${\tilde w}={\tilde W}-1$ the expression
\eqn\Wdtildetwo{{\tilde W}_d={1\over 1-f_1 {\tilde w}} 
\left(1-{\tilde w}\, (f_{d+1}-f_d) \sum_{k\geq 1} 
({\tilde w}\, f_d)^{k-1} \right)+{\tilde W}_0-1}
from which we deduce 
\eqn\WdtildeZp{\eqalign{{\tilde W}_d\vert_{Z^p} & =
(g R^3)^p \left\{\!
{3p \choose p}\!-\!2{3p \choose p\!-\!1}\!-\!
(f_{d+1}\!-\!f_d)\!\sum_{k\geq 1}\!{{3p-k \choose p-k}\!-\!2{3p-k
\choose p-1-k}\over (g\, R^2)^k}\!(f_d)^{k-1}\!\right\}\cr
&\ \  +{\tilde W}_0\vert_{Z^p}-\delta_{p,0}\cr
&= 
(g R^3)^p \left\{
{(3p)!\over p! (2p\!+\!1)!}-
(f_{d+1}\!-\!f_d)\sum_{k=1}^p{(3p-k)! \over (p\!-\!k)!(2p\!+\!1)!} {(2k+1)
\over (g\,R^2)^k}
(f_d)^{k-1}\right\}\cr
&\ \  +{\tilde W}_0\vert_{Z^p}-\delta_{p,0}
\ .\cr}}
By a similar argument, we have
\eqn\logWdtildeZp{\eqalign{{\tilde W}_d'\vert_{Z^p} & =
(g R^3)^p \left\{
{(3p-1)!\over p! (2p)!}-2
\sum_{k=1}^p{(3p-k-1)! \over (p\!-\!k)!(2p)!} {1
\over (g\,R^2)^k}
\left((f_{d+1})^{k}-(f_d)^k\right)\right\}\cr
&\ \  +{\tilde W}_0\vert_{Z^p}
\cr}}
for $p\geq 1$.

\subsec{Critical behavior and scaling limits}

As before, the generating functions ${\tilde W}_d$ and
${\tilde W}'_d$ have a first singularity
at $g=1/12$, irrespectively of the value of $Z$. A second singularity
comes from ${\tilde W}$ which, from \ZWtilderel, is singular when 
\eqn\newcrit{{27\over 4} g\, R^3\, Z = 1\ .}
This equality may occur only when $Z\geq 2/9$ and it defines a critical 
line $g={\tilde g}_{\rm crit}(Z)$, with
\eqn\gtildecrit{{\tilde g}_{\rm crit}(Z)= 
{1\over 32}\left((4+9Z)\sqrt{16Z+9 Z^2}-9 Z(4+3 Z)\right)\qquad
(Z\geq 2/9)\ .}
For $Z>2/9$, we have ${\tilde g}_{\rm crit}(Z)<1/12$ so that 
this new value determines the radius of convergence in $g$ of
${\tilde W}_d$.
As before, we have a change of determination of the radius of 
convergence at the critical value
\eqn\Zcrit{Z_{\rm crit}={2\over 9}\ ,}
corresponding to a transition between two differently behaved regimes.

To study the regime $Z<2/9$, we set as before
\eqn\regimeone{\eqalign{g&={1\over 12}(1-\mu\, \epsilon)\cr
d&= D\, \epsilon^{-1/4}\cr}}
and we use the expansion \WdtildeZp\ to get
\eqn\Wdzpexpantilde{\eqalign{
{\tilde W}_d\vert_{Z^p}& =
{\tilde W}_0\vert_{Z^p}\!-\!\delta_{p,0}\!+\!
\left({2\over 3}\right)^p \Big\{ {(3p)!\over p! (2p\!+\!1)!}
\left(1-3p\, \sqrt{\mu}\, \epsilon^{1/2}\right)
\cr 
&\ \ \  \!-\!
(2f^2(D;\mu)\!-\!3\sqrt{\mu})\, \epsilon^{1/2}\, 
\sum_{k=1}^p 3^k {(3p-k)! \over (p\!-\!k)!(2p\!+\!1)!} (2k+1)
\Big\} +\cdots \cr
&= 
{\tilde W}_0\vert_{Z^p}\!-\!\delta_{p,0}\!+\!
\left({2\over 3}\right)^p 
{(3p)!\over p! (2p\!+\!1)!}\left\{1-3 p\ {\cal F}(D;\mu)\, \epsilon^{1/2} 
+\cdots\right\}\ .\cr}}
Here we have used the identity
\eqn\identr{\sum_{k=1}^p 3^k {(3p-k)! \over (p\!-\!k)!(2p\!+\!1)!} (2k+1)
= 3\ {(3p)!\over (p\!-\!1)! (2p\!+\!1)!}}
obtained by writing the summand in the left hand side as
$\alpha_k-\alpha_{k+1}$ with $\alpha_k\equiv 3^k {3p-k-1 \choose p-1}$.
Similarly, we have 
\eqn\logWdzpexpantilde{
{\tilde W}_d'\vert_{Z^p} =
{\tilde W}_0\vert_{Z^p}\!+\!
\left({2\over 3}\right)^p 
{(3p-1)!\over p! (2p)!}\left\{1-3 p\ {\cal F}(D;\mu)\, \epsilon^{1/2} 
+\cdots\right\}\ }
for $p\geq 1$.

Finally, from \expwzerotilzp, we have:
\eqn\expantildeWz{{\tilde W}_0\vert_{Z^p}= 2
\left({2\over 3}\right)^p {(3p-3)!\over p!(2p-1)!}(1+{\cal O}(\epsilon))}
for $p>0$ and ${\tilde W}_0\vert_{Z^0}= 1$. 
Summing \Wdzpexpantilde\ over $p$ with a weight $Z^p$, we obtain
\eqn\expti{\eqalign{& {\tilde W}_d 
={\tilde A_0}(Z)+{\tilde A(Z)}-3Z{\tilde A}'(Z)\, {\cal F}(D;\mu)\, 
\epsilon^{1/2}+\cdots \cr \hbox{with}\quad &
{\tilde A_0}(Z)\!=\!2 \sum_{p\geq 1}\left({2\over 3}\right)^p Z^p
{(3p\!-\!3)!\over p!(2p\!-\!1)!} 
= {2\over 3}\left(-1\!+{}_2F_1\left(\left\{-\!{2\over 3},-\!{1\over 3}
\right\},
\left\{{1\over 2}\right\},{9Z\over 2}\right)\right)
\cr & 
{\tilde A}(Z)\!=\!\sum_{p\geq 0} 
\left({2\over 3}\right)^p Z^p  {(3p)!\over p!(2p\!+\!1)!} 
\!=\!\sqrt{2\over Z}
\sin\left({1\over 3}\arcsin\left(3\sqrt{{Z\over 2}}\right)\right)\ ,\cr}}
while
\eqn\logexpti{{\tilde W}_d' 
={\tilde A_0}(Z)+\log {\tilde A(Z)}-{3Z{\tilde A}'(Z)\over
\tilde A(Z)}\, {\cal F}(D;\mu)\, 
\epsilon^{1/2}+\cdots} 
Going to a fixed $n$ ensemble, we deduce from this expansion that
\eqn\largenlogWdtilde{\eqalign{{\tilde W}_d'\vert_{g^n} 
& \sim 
{12^n \over \pi n^{3/2}} {3Z{\tilde A}'(Z)\over {\tilde A}(Z)}
\ \int_{-\infty}^\infty
d\xi\ {\rm i}\xi\ \ e^{-\xi^2}{\cal F}(D;-\xi^2)\cr   
{\tilde W}_\infty'\vert_{g^n}
& \sim 
{12^n \over 2 \sqrt{\pi} n^{3/2}} {3Z{\tilde A}'(Z)\over {\tilde A}(Z)}
\quad \hbox{where} \ 
{\tilde W}'_\infty\equiv \lim_{d\to \infty} {\tilde W}'_d  
\cr}}
at large $n$ so that, taking the ratio of these quantities, we 
get {\it the same asymptotic distribution function} $\Phi(D)$ as in 
Section 3 for the rescaled distance $D$. In the $Z<2/9$ regime, 
the length of the boundary does not scale with $n$ and 
quadrangulations with a self-avoiding boundary stay, in the scaling limit,
in the universality class of the Brownian map.

To study the regime $Z>2/9$, we must now set
\eqn\resimedeux{Z={\tilde g}_{\rm crit}(Z)(1-{\tilde \nu}\, \epsilon)}
and keep $d$ finite. We introduce the notations
\eqn\valcrits{\eqalign{& {\tilde x}_{\rm crit}(Z) \equiv
{1\over 16}\big(27Z\!-\!8\!+\!9\sqrt{Z (16\!+\!9Z)}\cr
&\ \ \ \ \ \ \ \ \ \ \ \ \ \ \ -\!\sqrt{6}\sqrt{243 Z^2\!+\!144 Z\!-\!32\!+\!
3(27 Z\!-\!8)\sqrt{Z(16\!+\!9Z)}}\big)
\cr &
{\tilde f}_d^{\rm crit} \equiv {\tilde x}_{\rm crit}(Z){1-\left({\tilde x}_{\rm crit}(Z)\right)^d 
\over 1-\left({\tilde x}_{\rm crit}(Z)\right)^{d+2}}\ .\cr}}
Here ${\tilde x}_{\rm crit}(Z)$ is simply the value of $x$ when 
$g={\tilde g}_{\rm crit}(Z)$. With these notations, we have the expansion
\eqn\newexpti{{\tilde W}= \left(1+{3\over {\tilde f}_1^{\rm crit}}\right)
-{\tilde C}(Z)\sqrt{{\tilde \nu}}\, \epsilon^{1/2}}
where ${\tilde C}(Z)$ is some function of $Z$ (with no singularity 
for $Z>2/9$) which we do not make explicit. We then find the expansion
\eqn\lastexpti{\eqalign{{\tilde W}_d & =\left\{{4+21 {\tilde f}_1^{\rm crit}
\over 18 {\tilde f}_1^{\rm crit}}+{3\over 2}{({\tilde f}_d^{\rm crit} -
{\tilde f}_{d+1}^{\rm crit})\over ( 3 {\tilde f}_1^{\rm crit}-
{\tilde f}_d^{\rm crit})}\right\}\cr
& \ \ \ -{\tilde C}(Z)\sqrt{{\tilde \nu}}\, \epsilon^{1/2} \left\{{4+15 {\tilde f}_1^{\rm crit}\over 12}
+{9 {\tilde f}_1^{\rm crit}\over 4}{({\tilde f}_d^{\rm crit}
-{\tilde f}_{d+1}^{\rm crit})(9 {\tilde f}_1^{\rm crit}-{\tilde f}_d^{\rm crit})
\over (3 {\tilde f}_1^{\rm crit}-{\tilde f}_d^{\rm crit})^2} \right\}\cr}} 
and the similar expansion 
\eqn\lastexpti{\eqalign{{\tilde W}_d' & =\left\{{2-3 {\tilde f}_1^{\rm crit}
\over 9 {\tilde f}_1^{\rm crit}}+\log\left({3\over 2}{(3
{\tilde f}_1^{\rm crit} -
{\tilde f}_{d+1}^{\rm crit})\over ( 3 {\tilde f}_1^{\rm crit}-
{\tilde f}_d^{\rm crit})}\right)\right\}\cr
& \ \ \ -{\tilde C}(Z)\sqrt{{\tilde \nu}}\, \epsilon^{1/2} \left\{{2+3 {\tilde f}_1^{\rm crit}\over 6}
+9 ({\tilde f}_1^{\rm crit})^2{({\tilde f}_d^{\rm crit}
-{\tilde f}_{d+1}^{\rm crit})
\over (3 {\tilde f}_1^{\rm crit}-{\tilde f}_d^{\rm crit})
(3 {\tilde f}_1^{\rm crit}-{\tilde f}_{d+1}^{\rm crit})
} \right\}\ .\cr}} 
\fig{
Plots of the (non-universal) cumulative distribution function 
${\tilde \phi}_Z(d)$ for $Z$ approaching the critical value $2/9$ from
above, namely $Z=0.23$, $0.225$, $0.223$ and $0.2227$ (dotted plots 
from left to right). 
Beside each plot, we display the corresponding (universal) 
limiting scaling form (solid line) of Eq.~(5.38).
}{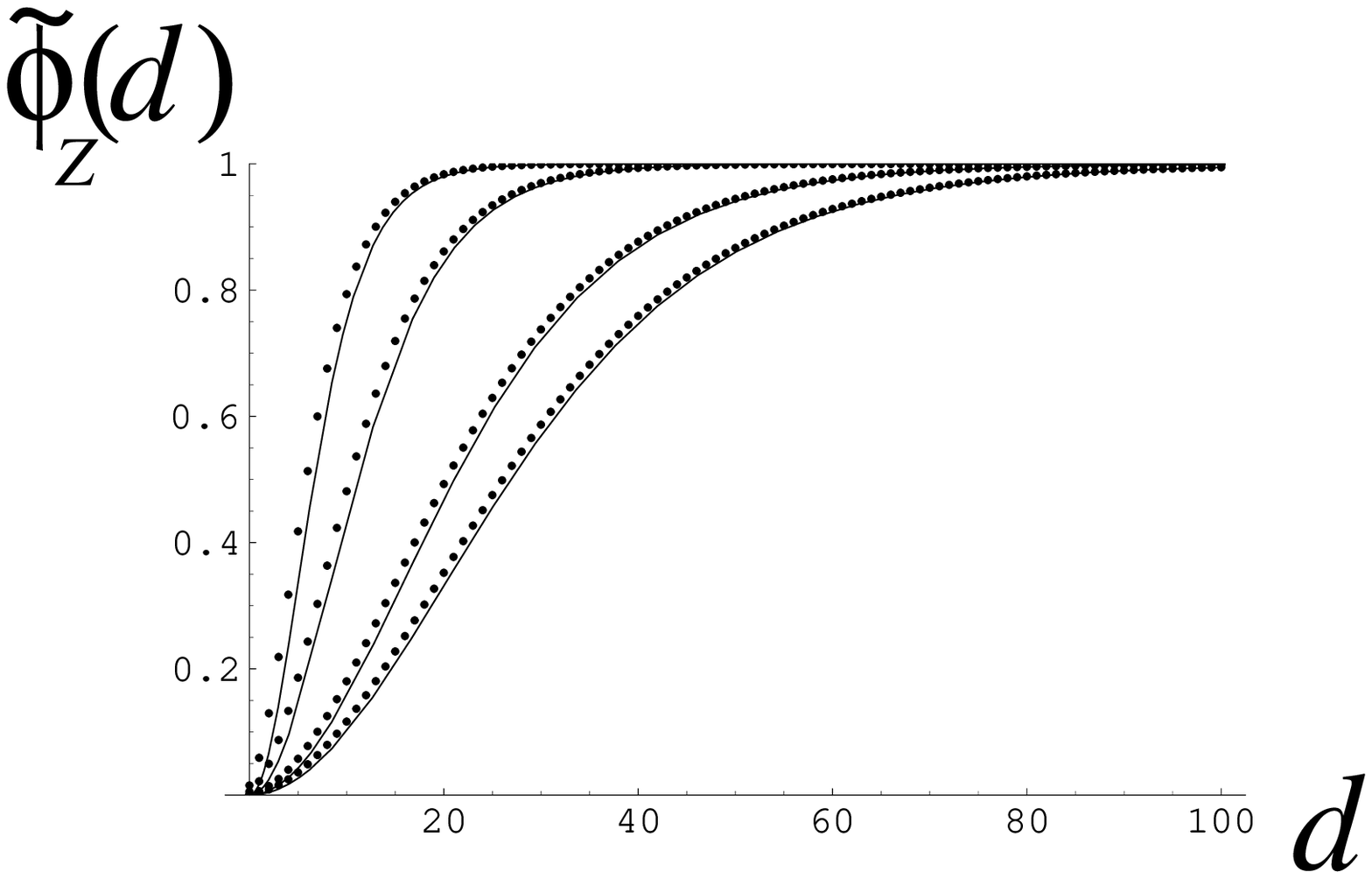}{9.cm}
\figlabel\tildephiofd
\noindent In the fixed $n$ ensemble, we deduce that the ratio
${\tilde W}_d'\vert_{g^n}/{\tilde W}_\infty'\vert_{g^n}$ tends
at large $n$ to the cumulative distribution function for $d$:
\eqn\phitilde{\eqalign{{\tilde \phi}_Z(d)& = 1+{
54 ({\tilde f}_1^{\rm crit})^2 \over 2+3 {\tilde f}_1^{\rm crit}}
{({\tilde f}_d^{\rm crit}-{\tilde f}_{d+1}^{\rm crit})\over 
(3{\tilde f}_1^{\rm crit}- {\tilde f}_d^{\rm crit})
(3{\tilde f}_1^{\rm crit}- {\tilde f}_{d+1}^{\rm crit})
}\cr
&= 1-{27 x(1+x+x^2) \over (2+x)^2(1+2x)^2}\left\{1-
{({2+x\over 1+2x}-x^d)({2+x\over 1+2x}-x^{d+1})\over 
({2+x\over 1+2x}+x^d)({2+x\over 1+2x}+x^{d+1})}\right\}
\quad \hbox{with}\ x={\tilde x}_{\rm crit}(Z)\ . 
\cr}}
Note that this distribution is different from the distribution
$\phi_z(d)$ of Eq.~\probsmaller\ obtained for a non self-avoiding boundary.
When $Z$ approaches the critical value $2/9$ however, we have
\eqn\approtilde{{\tilde \phi}_Z(d) \sim \tanh^2\left(d\, {\tilde \beta}(Z)\right)
\quad \hbox{with}\ {\tilde \beta}(Z)={3\over 2}\sqrt{Z-{2\over 9}}\ ,}
and we recover the universal scaling form of Eq.~\appro.  
The function ${\tilde \phi}_Z(d)$ is plotted against its scaling form 
\approtilde\ for $Z=0.23$, $0.225$, $0.223$ and $0.2227$ in Fig.~\tildephiofd.

Finally, to study the regime $Z \sim 2/9$, we can set
\eqn\critique{\eqalign{g& ={1\over 12}(1- \mu\, \epsilon)
\cr Z& ={2\over 9}(1-{\tilde \mu}_B\, \epsilon^{1/2})\cr
d &= D\, \epsilon^{-1/4}\cr}}
if we work with a fixed value of $Z$, or alternatively replace the
second equation by 
\eqn\setp{p=P\, \epsilon^{-1/2}}
if we work in a fixed $p$ ensemble. 
In the first ensemble, we have the expansion
\eqn\firstex{\eqalign{
{\tilde W}_d -({\tilde W}_0-1) & = {3\over 2}\left(1-\epsilon^{1/4}
{\cal H}\left(D;\mu,{\tilde \mu}_B/3\right)+\cdots \right)\cr
{\tilde W}_d' -({\tilde W}_0-1) & = \log\left({3\over 2}\right)-\epsilon^{1/4}
{\cal H}\left(D;\mu,{\tilde \mu}_B/3\right)+\cdots \cr}}
with ${\cal H}$ as in \valcalH, 
while ${\tilde W}_0=4/3+{\cal O}(\epsilon^{1/2})$.
We thus recover the {\it same scaling function} as that obtained for
non self-avoiding boundaries, with a simple renormalization of the 
``boundary cosmological constant'' by $1/3$.  

In the fixed length ensemble, we obtain
from \WdtildeZp\ the scaling behavior
\eqn\Wdzpexptilde{
{{\tilde W}_d\vert_{Z^p}\over (9/2)^p}
\sim\ 
\epsilon^{3/4} 
{9\over 4}{\bar {\cal H}}(D;3P;\mu)
}
with ${\bar {\cal H}}$  as in \Wdzpexp.
Except for the trivial factor $9/4$, we recover again the same scaling 
function as that obtained for non self-avoiding boundaries, with now
a renormalization of the perimeter by a factor of $3$. 
As for ${\tilde W}_d'$, since it has the same singular behavior
as ${\tilde W}_d$ up to a factor $3/2$, we deduce
\eqn\Wdzpprimeexptilde{
{{\tilde W}_d'\vert_{Z^p}\over (9/2)^p}
\sim\ \epsilon^{3/4} \ 
{3\over 2}{\bar {\cal H}}(D;3P;\mu)\ .
}
This behavior can also be obtained directly by taking the scaling limit
of \logWdtildeZp.

After going to the fixed $n$ ensemble 
and taking the appropriate ratio ${\tilde W}_d'\vert_{g^n Z^p}/
{\tilde W}_\infty'\vert_{g^n Z^p}$, we now 
find that 
\eqn\Renorone{{\bar \Phi}_{{\rm self-avoiding}\atop
{\rm boundary}}(D,P) = {\bar \Phi}(D,3P)}
for the distribution function for $D$ in the fixed $p$ ensemble.

Note that the ratio ${\tilde W}_d \vert_{Z^P}/{\tilde W}'_d\vert_{Z^P}$
tends to $3/2$ which may be interpreted as the asymptotic average number 
of boundary vertices closest to the origin in the case of a self-avoiding 
boundary. This number is independent of $D$ in the scaling regime.

To conclude, large quadrangulations with a self-avoiding boundary behave
essentially as large quadrangulations with a non-self-avoiding boundary, 
as far as the bulk-boundary distance is concerned. For boundary
lengths of the order $n^{1/2}$, quadrangulations with a self-avoiding
boundary of length $P$ behave as quadrangulations with a non-self-avoiding
boundary of length $3P$. This suggests that, in quadrangulations
with a non-self-avoiding boundary, only one of the irreducible
components is macroscopic and has a boundary of 
length equal to $1/3$ of the total perimeter, while
all the other components are microscopic but altogether, the lengths
of their boundaries represent $2/3$ of the total perimeter.

\newsec{Self-avoiding loops} 

We end this paper by a study of the statistics of distances 
in {\it quadrangulations
with a self-avoiding loop}. As already discussed in Section 2, 
a self-avoiding loop is
a closed path made of consecutive edges of the quadrangulation, which 
is simple, i.e visits any vertex at most once. For convenience,
we shall suppose that the loop is oriented. Upon cutting 
along the loop, we obtain two quadrangulations with a self-avoiding
boundary, constrained to have the same perimeter. The orientation
allows to distinguish this two pieces as left and right. We can now 
express a number of generating functions for this problem in terms of the
generating functions found in Section 4.

A first simple generating function is that of quadrangulations 
with a self-avoiding loop with a marked vertex on the loop.
It reads immediately
\eqn\rootdloop{\Omega_0(g,y)=\sum_{p\geq 1}y^p
\left({\tilde W}_0(g,Z)\vert_{Z^p}\right)^2}
where $g$ is the weight per face while $\sqrt{y}$ is a weight per
edge of the loop (the length of the loop is necessarily even).

Another interesting quantity involving the distance is the
generating function for quadrangulations with a self-avoiding loop
with a marked vertex at distance {\it less than or equal to $d$},
lying to the right of the loop (or possibly on the loop itself). 
It reads
\eqn\pointloop{\Omega_d(g,y)=\sum_{p\geq 1}y^p
{\tilde W}_0(g,Z)\vert_{Z^p}
{\tilde W}_d'(g,Z)\vert_{Z^p}}
if the configurations are counted with their usual inverse
symmetry factor.
This can be seen by first noting that the generating function for the
same objects with an additional marking of a closest vertex on the loop
is obviously given by $\sum_{p\geq 1}y^p {\tilde W}_0(g,Z)\vert_{Z^p}
{\tilde W}_d(g,Z)\vert_{Z^p}$ upon gluing the two boundaries in such
a way that their marked vertices coincide. Removing the marked
closest vertex, we see that a given configuration 
is over-counted by a factor of $k$ equal to the number of closest
vertices on the loop. Dividing by this factor $k$ amounts to     
replacing ${\tilde W}_d$ by ${\tilde W}_d'$ as in Section 4.
Note that, clearly, the quantity $\Omega_d$ is related to the 
quantity $\Gamma_d$ of Section 2 by
\eqn\OmegaGamma{\Gamma_d=\Omega_d-\Omega_{d-1}, \quad \Gamma_0=\Omega_0.}

From the exponential growth ${\tilde W}_0 \sim (9/2)^p$ and 
${\tilde W}_d'\sim (9/2)^p$, we immediately deduce that a transition
occurs now at $y=y_{\rm crit}$ with
\eqn\ycrit{y_{\rm crit}= \left({2\over 9}\right)^2={4\over 81}\ .}
For $y<4/81$, the typical values of $p$ contributing to $\Omega_d(g,y)$
remain finite when $g$ tends to its critical value $1/12$. 
Setting
\eqn\againgcrit{\eqalign{g&={1\over 12}(1-\mu\, \epsilon)
\cr d & =D\, \epsilon^{-1/4}\ ,\cr} }
we may use \logWdzpexpantilde\ and \expantildeWz\ to write
\eqn\expanOmegad{\Omega_d\vert_{y^p}\!=\!
4\!\left({4\over 9}\right)^p\!\left({(3p-3)!\over p!(2p-1)!}\right)^2\!
+\!2\!\left({4\over 9}\right)^p\!
{(3p\!-\!1)!\over p! (2p)!}{(3p\!-\!3)!\over p!(2p\!-\!1)!}\!
\left\{1\!-\!3 p\, {\cal F}(D;\mu)\, \epsilon^{1/2} 
\!+\!\cdots\right\}}
for $p\geq 1$.
Summing over $p$ with a weight $y^p$, we obtain
\eqn\exptiomega{\eqalign{& \Omega_d 
=a(y)-b(y) {\cal F}(D;\mu)\, 
\epsilon^{1/2}+\cdots \cr \hbox{with}\quad &
a(y)\!=\!2 \sum_{p\geq 1}\left({4\over 9}\right)^p y^p
{(3p\!-\!3)!\over p!(2p\!-\!1)!}\left( 2{(3p\!-\!3)!\over p!(2p\!-\!1)!}
+{(3p-1)!\over p! (2p)!}\right)
\cr & 
b(y)\!=\!6 \sum_{p\geq 1} 
\left({4\over 9}\right)^p y^p  
{(3p-3)!\over (p-1)!(2p\!-\!1)!} 
{(3p-1)!\over p!(2p)!} 
\ ,\cr}}
which are functions of $y$ with no singularity for $y<4/81$.
Going to a fixed $n$ ensemble and considering large values of $n$, 
we immediately deduce from the singular behavior \exptiomega\
that the cumulative distribution function for the rescaled distance 
$D=d/n^{1/4}$ from the marked vertex in the bulk to the 
self-avoiding loop is again, in
the regime $y<y_{\rm crit}$, equal to the universal two-point
function $\Phi(D)$ of the Brownian map, as given by Eq.~\ratiolim.
In this regime, the size of the loop does not scale
with $n$ and becomes negligible in the large $n$ limit.

In the regime $y>4/81$, we expect that the dominant singularity
corresponds to large values of $p$. In this case, we may use \expwzerotilzp\
and \logWdtildeZp\ to write the large $p$ behavior
\eqn\equivOmega{\eqalign{\Omega_d\vert_{y^p} & 
\sim \left({27g R^3\over 4}\right)^{2p}\!{1\over 
\pi\, (3p)^3}\!
\left({1\over 3 f_1}\!-\!1\right) \left\{
\left({1\over 3 f_1}\!-\!1\right)\!+\!
{3\over 2} \left\{1\!-\!2
\sum_{k=1}^\infty
{(f_{d+1})^{k}\!-\!(f_d)^k \over (3 f_1)^k}
\right\}\right\}\cr 
&= 
\left({27g R^3\over 4}\right)^{2p}\!{1\over
\pi\, (3p)^3}\!
\left({(1\!-\!3 f_1)(2\!+\!3 f_1)\over
18 f_1^2}\right)
\left\{1+{54 f_1^2\over 2+3 f_1}
{f_d-f_{d+1}\over (3f_1-f_d)(3f_1-f_{d+1})}\right\}
\ .\cr}}
The line $y (27 g R^3/4)^2=1$ defines the critical 
value ${\hat g}_{\rm crit}(y)$ of $g$ for $y>y_{\rm crit}$, 
namely
\eqn\hatgcrit{{\hat g}_{\rm crit}(y)= {\tilde g}_{\rm crit}
(\sqrt{y})}
with ${\tilde g}_{\rm crit}(Z)$ as in Eq.~\gtildecrit. 
Upon summing over $p$ with a weight $y^p$, the $p$-dependent prefactor
gives rise when $((27/4) g\, R^3)^2\, y \to 1$
to a singularity
\eqn\newsing{-{1\over 2}\left(1\!-\!\left({27 g\, R^3\over 4}\right)^2
\, y \right)^2
\log \left(1\!-\!\left({27 g\, R^3\over 4}\right)^2\, y \right)
\propto \left(1\!-\!{g\over {\hat g}_{\rm crit}(y)}\right)^2
\log\left(1\!-\!{g\over {\hat g}_{\rm crit}(y)}\right)\ .}
This singularity translates into an asymptotic behavior of the form 
$\left({\hat g}_{\rm crit}(y)\right)^{-n} / n^3$ for 
$\Omega_d\vert_{g^n}$ at large $n$, with a multiplicative factor
proportional to the $d$-dependent coefficient in \equivOmega,
taken at $x=x({\hat g}_{\rm crit}(y))={\tilde x}(\sqrt{y})$.
Note that this $d$-dependent coefficient comes from ${\tilde W}'_d$ only
and we therefore recover in the fixed $n$ ensemble (and in the regime 
$y>4/81$) the same form \phitilde\ for the cumulative distribution 
function ${\tilde \phi}_Z(d)$ for $d$ provided we identify 
$Z=\sqrt{y}$ in this formula.

The most interesting situation is when $y$ is in the vicinity 
of $4/81$. Alternatively, we may work in the fixed length ensemble
and study the scaling limit \againgcrit\ with moreover
\eqn\againscalp{p=P\, \epsilon^{-1/2}\ .}
In this limit, we may use the relation \Wdzpprimeexptilde\ and the relation
\eqn\scalpwzerotilzp{{{\tilde W}_0\vert_{Z^p}
\over (9/2)^p} \sim \epsilon^{5/4}\ 2 {e^{-3 \sqrt{\mu} P} 
\over \sqrt{\pi} (3P)^{5/2}} (1+ 3P\sqrt{\mu})} 
inherited from \expwzerotilzp\ to deduce that
\eqn\omegafixl{\eqalign{{\Omega_d\vert_{Z^p}\over (81/4)^p}& \sim\ \epsilon^2
\ 3 {e^{-3 \sqrt{\mu} P}             
\over \sqrt{\pi} (3P)^{5/2}} (1\!+\!3P\sqrt{\mu}){\bar {\cal H}}(D;3P;\mu)
\cr 
&=\epsilon^2 {\hat {\cal H}}(D,P;\mu) \ \  \hbox{with} \cr 
{\hat {\cal H}}(D,P;\mu)
&=
3 {e^{-6 \sqrt{\mu} P}
\over \pi (3P)^4} (1\!+\!3P\sqrt{\mu})
\left\{1\!+\!
\left(3\sqrt{\mu}\!-\!2 f^2(D;\mu)\right)\ 
\int_0^\infty dK\, e^{-{K^2 \over 3P}-2 f(D;\mu) K} 2K
\right\}\ ,\cr}}
involving a {\it new universal scaling function} 
${\hat {\cal H}}(D,P;\mu)$.

\noindent For small $P$, we have 
\eqn\omegafixlsmallp{\eqalign{{\hat {\cal H}}(D,P;\mu) &
{\buildrel {P\to 0} \over \sim}\ 
{3 \over \pi (3P)^4}}(1- (3P) {\cal F}(D;\mu) +\cdots) }
which matches the 
large $p$ behavior of $\Omega_d\vert_{y^p}$ in the regime $y<4/81$.
\noindent For large $P$, we have instead
\eqn\omegafixllargep{{\hat {\cal H}}(D,P;\mu) 
{\buildrel {P\to \infty} \over \sim} 3 {e^{- 6 \sqrt{\mu}\, P} 
\over \pi (3P)^3} \sqrt{\mu}
\tanh^2\left(\sqrt{3\over 2}\, \mu^{1/4}\, D\right)\ .}
If we now turn to the fixed $n$ ensemble, in the limit of large $n$, 
with the scaling
\eqn\scalnp{p= P\ n^{1/2}\ ,}
we get a cumulative distribution function
\eqn\cumdihat{{\hat \Phi}(D,P)={18 P^3 \sqrt{\pi} \over
1+18 P^2}\ e^{9 P^2}\int_{-\infty}^{\infty}
d\xi\ {\xi\over {\rm i}}\ e^{-\xi^2}\ {\hat{\cal H}}(D,P;-\xi^2)\ ,}
which measures the probability that a vertex chosen uniformly at random
in the quadrangulation with a self-avoiding loop of rescaled length
$2P$ be at a rescaled distance less than $D$ from this loop. 
This is a new universal distribution which cannot be reduced 
to the distribution ${\bar \Phi}(D,P)$ 
of previous sections by a simple rescaling
of $P$. It is plotted in Fig~\hatPhiofP\ for
$P=0.01$, $0.1$, $0.2$, $0.5$ and $1.0$. 
For small $D$ and fixed $P$, we have the expansion
\eqn\smallDhat{{\hat \Phi}(D,P)={9\over 2} P\, D^2 -{3\over 2} 
{1+9P^2+162 P^4 \over 1+18 P^2}\, D^4 +\cdots}
\bigskip
From \omegafixlsmallp, we see that ${\hat \Phi}(D,P)\to \Phi(D)$ at small $P$
for fixed $D$,
as expected. From \omegafixllargep, and via a saddle point estimate of
the integral over $\xi$, we find that, for $P\to \infty$, the typical value of 
$D$ is of order $P^{-1/2}$, with the scaling form
\eqn\hatPhilimlargeP{{\hat{\Phi}}(D,P) \sim 
\tanh^2\left(\sqrt{9 \over 2} D \ \sqrt{P}\right) \ ,}
i.e we recover the general form found in previous Sections.

\newsec{Discussion and conclusion}

To briefly summarize our results, we have been able to find discrete exact
expressions for the bulk-boundary and boundary-boundary correlators for
quadrangulations with a generic boundary. In the case of a self-avoiding 
boundary, we have been only able to express the bulk-boundary correlator,
leading eventually to results for a model of self-avoiding loop. From our
discrete expressions, we identified three scaling regimes for which we gave 
the asymptotic behaviors. 

Most of the above results follow from the discovery of the ``master formula''
\Wdexplit\ for $W_d$, which is remarkable in itself. This is yet another 
manifestation of the still mysterious integrability of the equations 
governing distance statistics in maps, as already observed for the
two-point and three-point functions. Here two levels of integrability
are involved, first for the equation determining $R_d$, involving the
parameter $g$ only, and then for the equation \recurW\ determining $W_d$, 
involving $z$ and $R_d$ as an ``external potential''. In this respect, 
it is worth mentioning that we have a continued fraction expansion
\eqn\contfrac{W_d={1\over \displaystyle{1- {z R_{d+1}\over \displaystyle{
1-{z R_{d+2}\over \cdots}}}}}}
and furthermore, we have the conserved quantity 
\eqn\conserve{W_d-g\, z\, R_dR_{d+1}R_{d+2} W_dW_{d+1}W_{d+2}=
W-g\, z\, R^3W^3 \quad \hbox{for all}\ d\ .}
This identity may be checked directly from the explicit expression \Wdexpli\ 
for $W_d$. Alternatively, using \dyck\ to expand both sides of this equation
in powers of $z$, it is equivalent to an infinite number of conserved 
quantities involving $R_d$ and $g$ only, and it is easy to check that those 
correspond precisely to the conserved quantities found in Ref.~\DFG.  

Another remarkable fact is the strong similarity, already apparent at the 
discrete level, between the bulk-boundary correlators for generic and 
self-avoiding boundaries. Unfortunately, we have not been able to 
exhibit the same phenomenon for boundary-boundary correlators. Furthermore, 
had we gone over this problem, we are still far from understanding the
distance statistics between two points lying on a self-avoiding loop:
the loop and geodesics may cross each other, which prevents the decoupling
of both sides observed in \pointloop\ for the bulk-loop correlator. 

Our universal expressions are properties of what could be called the 
Brownian map with a boundary. In this respect, we may wonder whether
these results may be re-obtained in a purely continuous formalism, which
would likely require a proper probabilistic definition of this object.
At a more physical level, we notice that the expression \prdeltaufixP\ 
may be re-derived from the solution ${\cal P}(D,U;\mu)$ of some
diffusion equation in a potential ${\cal F}(D;\mu)$ as in Eq.~\defcalF:
\eqn\diffuse{{\partial\ \over \partial U} {\cal P}= {\partial^2 \over
\partial D^2} {\cal P} - {\cal F}{\cal P}\ .}
Here ${\cal P}$ describes the law of the position $D$ at time $U$ of a 
particle diffusing in the potential ${\cal F}$ in one spatial dimension. 
Discarding normalization factors, it reads:
\eqn\soldif{{\cal P}(D,U;\mu) \propto {e^{-\sqrt{\mu} U}\over U^{5/2}}
e^{-{D^2\over 4U}}
\left(D^2-2 U+ 2 U D f(D;\mu)\right)\ ,}
with $f(D;\mu)$ as in \deffDmu. The expression \prdeltaufixP\ 
can be obtained by considering the quantity ${\cal P}(D,U;\mu)
{\cal P}(D,P-U;\mu)$, setting $\mu=-\xi^2$ and performing the usual
appropriate integral over $\xi$ to go to a fixed area ensemble, 
and finally introducing the rescaled variables $\delta=D/\sqrt{P}$ and
$u=U/P$. Heuristically, ${\cal P}(D,U;\mu) {\cal P}(D,P-U;\mu)$ is
the continuous counterpart of the generating function for Dyck paths 
of fixed length, constrained to reach some prescribed height at some given 
step.  The potential ${\cal F}(D;\mu)$ is the continuous counterpart
of the weight $R_d$ attached to each descent $d\to d-1$ of the Dyck path. 

To conclude, let us list a few possible generalizations of our results.
We expect integrability to survive when considering maps with faces
of degrees other than $4$, as found for the two-point function in \GEOD.
We may also consider maps of higher genus and/or with several boundaries. 
For instance, Refs.~\AW\ and \AJW\ give continuous 
results for surfaces with two 
boundaries at a prescribed mutual distance, for which discrete formulas
may as well exist. Finally, introducing multiple boundaries may pave the 
way towards understanding the distance statistics in the general 
${\bf O}({\cal N})$ loop model on dynamical random lattices.  
\listrefs
\end